\documentclass[12pt]{article}
\pdfoutput=1

\usepackage{textcomp}
\usepackage{color}

\usepackage{putex}
\usepackage{autobreak}
\usepackage{graphicx}
\graphicspath{{plots/}}
\usepackage{tabularx}
\usepackage{caption}
\usepackage{amsmath}
\usepackage{bbm}
\usepackage{array}
\usepackage{subcaption}
\usepackage{epstopdf}
\usepackage{enumerate}
\usepackage{cite}
\usepackage{youngtab}
\usepackage{booktabs}
\usepackage{tensor}
\usepackage{slashed}
\usepackage[aligntableaux=center]{ytableau}
\usepackage[utf8]{inputenc}
\usepackage{rotating}
\usepackage{makecell}
\usepackage{multirow}
\usepackage{braket}
\usepackage{tikz}
\usepackage[
colorlinks=true,
linkcolor=blue,
urlcolor=blue,
filecolor=black,
citecolor=red,
]{hyperref}
\usepackage{bm}

\newcommand{\abs}[1]{\left\lvert #1 \right\rvert}

\newcommand{\bbbbZ}{(\partial^4_b-15 \partial^2_b)\log Z}
\newcommand{\bbmmZ}{\partial^2_b \partial^2_{m} \log Z}
\newcommand{\mmmmZ}{\partial^4_m \log Z}
\newcommand{\bbttZ}{\partial_{\tau}\partial_{\bar{\tau}} \partial^2_b \log Z}
\newcommand{\mmttZ}{\partial_{\tau}\partial_{\bar{\tau}} \partial^2_m \log Z}

\newcommand{\mmmmZT}{\left(\partial^4_m \log Z\right)_{\mathcal{T}}}
\newcommand{\bbbbZT}{\left((\partial^4_b-15 \partial^2_b)\log Z\right)_{\mathcal{T}}}
\newcommand{\bbmmZT}{\left(\partial^2_b \partial^2_{m} \log Z\right)_{\mathcal{T}}}

\newcommand{\calallO}{\mathcal{O}_1\mathcal{O}_2\mathcal{O}_3\mathcal{O}_4}

\newcommand{\vecx}{\vec{x}}

\newcommand{\vecy}{\vec{y}}

\newcommand {\be} {\begin {equation}}
\newcommand {\ee} {\end {equation}}

\newcommand {\bes} {\begin {equation*}}
\newcommand {\ees} {\end {equation*}}

\newcommand{\es}[2] {\begin{equation} \label{#1} \begin{split} #2 \end{split} \end{equation}}

\newcommand{\R}{\mathbb{R}}

\newcommand{\beq}{\begin{equation}}
	\newcommand{\eeq}{\end{equation}}

\def\ie{\begin{equation}\begin{aligned}}
		\def\fe{\end{aligned}\end{equation}}

\numberwithin{equation}{section}

\def\<{\langle}
\def\>{\rangle}

\newcommand{\bareps}{\bar{\epsilon}}
\newcommand{\deltau}{\partial_{\tau}}
\newcommand{\delbartau}{\partial_{\bar{\tau}}}
\newcommand{\calT}{\mathcal{T}}
\newcommand{\Dbar}{\bar{D}}

\DeclareMathOperator{\SU}{SU}
\DeclareMathOperator{\U}{U}
\DeclareMathOperator{\SO}{SO}

\makeatletter
\DeclareFontFamily{OMX}{MnSymbolE}{}
\DeclareSymbolFont{MnLargeSymbols}{OMX}{MnSymbolE}{m}{n}
\SetSymbolFont{MnLargeSymbols}{bold}{OMX}{MnSymbolE}{b}{n}
\DeclareFontShape{OMX}{MnSymbolE}{m}{n}{
	<-6>  MnSymbolE5
	<6-7>  MnSymbolE6
	<7-8>  MnSymbolE7
	<8-9>  MnSymbolE8
	<9-10> MnSymbolE9
	<10-12> MnSymbolE10
	<12->   MnSymbolE12
}{}
\DeclareFontShape{OMX}{MnSymbolE}{b}{n}{
	<-6>  MnSymbolE-Bold5
	<6-7>  MnSymbolE-Bold6
	<7-8>  MnSymbolE-Bold7
	<8-9>  MnSymbolE-Bold8
	<9-10> MnSymbolE-Bold9
	<10-12> MnSymbolE-Bold10
	<12->   MnSymbolE-Bold12
}{}

\let\llangle\@undefined
\let\rrangle\@undefined
\DeclareMathDelimiter{\llangle}{\mathopen}%
{MnLargeSymbols}{'164}{MnLargeSymbols}{'164}
\DeclareMathDelimiter{\rrangle}{\mathclose}%
{MnLargeSymbols}{'171}{MnLargeSymbols}{'171}
\newcommand{\avg}[1]{\left\llangle #1 \right\rrangle}
\makeatother
\allowdisplaybreaks

\let\bar\relax
\newcommand{\bar}[1]{\overline{#1}}

\begin{document}
	
	\preprint{PUPT-2658\\MIT-CTP/5974}
	
	\institution{Imp}{Abdus Salam Centre for Theoretical Physics, Imperial College London, London SW7 2AZ, UK}
	\institution{MIT}{Center for Theoretical Physics -- a Leinweber Institute, Massachusetts Institute of Technology, Cambridge, MA 02139, USA}
	\institution{Pu}{Joseph Henry Laboratories, Princeton University, Princeton, NJ 08544, USA}
	\institution{PCTS}{Princeton Center for Theoretical Science, Princeton University, Princeton, NJ 08544, USA}
	
	\title{Integral constraints for $\mathcal{N}=4$\\super-Yang-Mills from a squashed sphere}

	\authors{Shai M.~Chester,\worksat{\Imp} Ross Dempsey,\worksat{\MIT} 
		Debaditya Pramanik,\worksat{\Pu}\\[10pt] and Silviu S. Pufu\worksat{\Pu,\PCTS}  }

	\abstract{
	Supersymmetric localization and Ward identities have been used in the past several years to derive two integral constraints on the four-point function of the stress-tensor multiplet in $\mathcal{N} = 4$ super-Yang-Mills theory. These constraints are powerful tools for studying the theory, especially when used in tandem with analytic and/or numerical bootstrap techniques. In this paper, we consider three additional integral constraints that can be derived starting from the ${\cal N} = 4$ super-Yang-Mills theory placed on a squashed four-sphere. These constraints are technically much more challenging to derive than the ones in the literature, and much of the paper is devoted to developing  techniques to make this computation tractable. Our end result is that these three constraints are implied by the two constraints already appearing in the literature.
	}
	\date{}

	\maketitle
	
	\tableofcontents
	
	\pagebreak
	
	\section{Introduction and Summary}
	\label{intro}

$\mathcal{N} = 4$ super-Yang-Mills (SYM) theory \cite{Osborn:1979tq} is one of the most-studied and best-understood four-dimensional gauge theories. Our understanding of it stems from the overlapping successes of many of the most powerful tools in non-perturbative quantum field theory:  the numerical conformal bootstrap \cite{Rattazzi:2008pe} (see \cite{Chester:2019wfx,Poland:2018epd,Poland:2022qrs,Rychkov:2023wsd} for reviews) was applied to ${\cal N} = 4$ superconformal field theories (SCFTs) in \cite{Beem:2013hha,Beem:2013qxa,Beem:2016wfs,Chester:2021aun,Chester:2023ehi,Caron-Huot:2024tzr}; the planar limit was found to be integrable \cite{Minahan:2002ve,Bena:2003wd,Beisert:2003tq,Kazakov:2004qf,Beisert:2005bm,Staudacher:2004tk,Arutyunov:2004vx,Beisert:2005tm,Beisert:2006ib,Gromov:2014bva,Gromov:2014caa} (see \cite{Beisert:2010jr,Gromov:2017blm} for reviews); S-duality \cite{Montonen:1977sn} was tested extensively in a long series of papers starting with \cite{Sen:1994yi,Vafa:1994tf}; the gauge/gravity duality \cite{Maldacena:1997re,Gubser:1998bc,Witten:1998qj} was used to explore the strong-coupling properties; analytic bootstrap methods were used to study the ${\cal N} = 4$ SYM theory at and beyond the tree-level supergravity approximation \cite{Alday:2013opa,Rastelli:2016nze,Aharony:2016dwx,Rastelli:2017udc,Rastelli:2017ymc,Caron-Huot:2022sdy,Alday:2014qfa,Alday:2017xua,Alday:2018pdi,Aprile:2017bgs,Aprile:2019rep,Aprile:2017qoy,Aprile:2018efk,Alday:2017vkk,Alday:2019nin,Binder:2019jwn,Chester:2020dja,Chester:2019jas,Chester:2020vyz};  and, of course, supersymmetry has been the key ingredient that made many non-perturbative computations possible.  One way in which supersymmetry has been used in ${\cal N} = 4$ SYM theory is through exact computations of supersymmetry-protected observables.  Using the technique of supersymmetric localization~\cite{Witten:1988xj} (see \cite{Pestun:2016zxk} for a collection of reviews), one can calculate the partition function $Z$, or equivalently the free energy $F\equiv - \log Z$, of the  ${\cal N} = 4$ SYM theory placed on a four-sphere and deformed by ${\cal N} = 2$-preserving parameters \cite{Pestun:2007rz, Hama:2012bg}.  The deformations we consider involve a mass parameter $m$ and a parameter $b$ that corresponds to a particular squashing of the sphere. Thus, the free energy $F = F(\tau, \bar \tau, b, m)$ depends on $m$, $b$, and the complexified gauge coupling $\tau = \frac{\theta}{2 \pi} + \frac{4 \pi i}{g_\text{YM}^2}$.  It can be computed exactly in terms of a matrix model for any values of these parameters \cite{Pestun:2007rz, Hama:2012bg,Nekrasov:2002qd,Nekrasov:2003rj}.

In this work, we will investigate the various fourth derivatives of $F$ evaluated at the point $(m, b) = (0, 1)$ that corresponds to vanishing mass and a round $S^4$.  Not all fourth derivatives are non-trivial however;  the non-trivial ones are \cite{Chester:2020vyz} 
 \begin{equation}
  \begin{aligned}
     &\partial_\tau \partial_{\bar \tau} \partial_m^2 F \Big|_{(m, b) = (0, 1)} \,, \qquad
     \partial_m^4 F \Big|_{(m, b) = (0, 1)} \,, 
     \\
    & \partial_\tau \partial_{\bar \tau} \partial_b^2 F \Big|_{(m, b) = (0, 1)}  \,, \qquad 
    \partial_b^2 \partial_m^2 F \Big|_{(m, b) = (0, 1)} \,, \qquad
    (\partial_b^4 - 15 \partial_b^2) F \Big|_{(m, b) = (0, 1)} \,,
    \end{aligned} \label{eq:independent}
 \end{equation}
where in the first line we included the non-trivial derivatives that can be computed on a round sphere, and in the second line we included the cases where at least one derivative is with respect to the squashing parameter $b$.  The last quantity in \eqref{eq:independent} includes a subtraction of a second derivative with respect to $b$ in order to remove dependence on the renormalization scheme.\footnote{While the free energy $F(\tau, \bar \tau, m, b)$ suffers from regularization-scheme ambiguities, all the combinations in \eqref{eq:independent} are scheme-independent.}  The explicit supersymmetric localization results of \cite{Pestun:2007rz, Hama:2012bg} imply the following relations  \cite{Chester:2020vyz}:
\begin{equation}\label{eq:relation1}
\begin{split}
    0&=(\partial_\tau\partial_{\bar\tau}\partial_m^2-\partial_\tau\partial_{\bar\tau}\partial_b^2)F\vert_{m=0,b=1}\,,\\
    16c&=(3\partial_b^2\partial_m^2-\partial_m^4-16\tau_2^2\partial_\tau\partial_{\bar\tau}\partial_m^2)F\vert_{m=0,b=1}\,,\\
    0&=(-6\partial_b^2\partial_m^2+\partial_m^4+\partial_b^4-15\partial_b^2)F\big\vert_{m=0,b=1}\,,
\end{split}
\end{equation}
from which one can solve for the derivatives in the second line of \eqref{eq:independent} in terms of those in the first line.  The goal of this paper is to investigate whether the relations \eqref{eq:relation1} are specific to the ${\cal N} = 4$ SYM theory (as obtained from supersymmetric localization) or whether they follow more generally from the Ward identities of ${\cal N} = 4$ superconformal symmetry and may also hold for other ${\cal N} = 4$ SCFTs that do not necessarily belong to the ${\cal N} = 4$ SYM family.\footnote{At present, the only known ${\cal N} = 4$ SCFTs are the ${\cal N} = 4$ SYM theories with various gauge groups, but it is not clear whether this is an exhaustive list.}

This question is particularly relevant for conformal bootstrap studies of ${\cal N} = 4$ SYM theory.  Indeed, the derivatives in \eqref{eq:independent} correspond to four-point integrated correlators \cite{Binder:2019jwn, Chester:2020dja} of the operators that couple to the parameters $(\tau, \bar \tau, b, m)$ at linear order in an expansion around $(m, b) = (0, 1)$ and arbitrary $(\tau, \bar \tau)$ (plus any lower-point integrated correlators that are required by supersymmetry).  In recent years, the integral constraints corresponding to the derivatives in the first line of \eqref{eq:independent} have been used with great success in both analytical \cite{Binder:2019jwn,Chester:2020dja,Chester:2019jas,Chester:2020vyz,Alday:2023pet,Alday:2021peq,Alday:2022uxp, Alday:2023mvu} and numerical \cite{Chester:2021aun,Chester:2023ehi, Caron-Huot:2024tzr} conformal bootstrap studies.  Such studies rely on crossing symmetry and superconformal symmetry, with limited information about the specific ${\cal N} = 4$ SYM theory one is targeting.  In particular, the rank of the gauge group has been inputted through the $c$ anomaly coefficient, which is related to an operator product expansion (OPE) coefficient \cite{Beem:2013qxa}, and the dependence on the Yang-Mills coupling has been inputted precisely through the correlators in the first line of \eqref{eq:independent} \cite{Chester:2021aun,Chester:2023ehi}.    If the relations \eqref{eq:relation1} do not follow from the superconformal Ward identities, then the integral constraints corresponding to the derivatives in the {\em second} line of \eqref{eq:independent} can be used in the bootstrap analysis to further constrain the space of ${\cal N} = 4$ SCFTs.  On the other hand, if the relations \eqref{eq:relation1} do follow from the superconformal Ward identities, then there is nothing to gain from the quantities in the second line of \eqref{eq:independent} as far as the bootstrap studies are concerned.

In more detail, all the derivatives in \eqref{eq:independent} can be expressed in terms of the so-called reduced correlator ${\cal T}(u, v)$, which is a function of the conformally-invariant cross-ratios $(u, v)$ defined in \eqref{eq:defuv} below.   The ${\cal N} = 4$ superconformal Ward identities imply that the correlator of any four operators that belong to the ${\cal N} = 4$ stress-tensor multiplet can be written in terms of ${\cal T}(u, v)$ \cite{Dolan:2001tt}.  For the derivatives with respect to the mass and the complexified coupling in the first line of \eqref{eq:independent}, the ${\cal N} = 4$ superconformal Ward identities yield the two relations \cite{Binder:2019jwn,Chester:2020dja}:
\begin{equation}\label{eq:knownintcorr}
\begin{split}
    \tau^2_2 \deltau \delbartau \partial^2_mF\vert_{m=0} &= \frac{8c^2}{\pi}\int dr\, d\theta\, \frac{r^3 \sin^2\theta}{u^2} \mathcal{T}(u,v) \,,\\
    \partial^4_mF\vert_{m=0} &= -48\zeta(3) c - \frac{32c^2}{\pi}\int dr \, d\theta\, \frac{r^3 \sin^2 \theta}{u^2} (1+u+v)\Dbar_{1,1,1,1}(u,v)\calT(u,v)\,.
\end{split}
\end{equation}
Here, $c$ is the central charge, the function $\bar{D}_{1,1,1,1}$ is defined in terms of the dilog function (see \eqref{eq:D1111func}), and the conformal cross ratios are written in terms of radial coordinates $(r, \theta)$ as $u= 1+r^2-2r\cos \theta$ and  $v=r^2$ (see \eqref{eq:uvinrtheta}).   Our end result consists of the following three integral constraints involving the derivatives in the second line of \eqref{eq:independent}:
\begin{equation}\label{eq:bbmmresult}
\begin{split}
    \tau^2_2 \deltau \delbartau \partial^2_bF\vert_{b=1} &= \frac{8c^2}{\pi}\int dr\, d\theta \, \frac{r^3 \sin^2\theta}{u^2}\mathcal{T}(u,v) \,,\\
    \partial^2_b\partial^2_m F\vert_{m=0,b=1} &= \frac{16c}{3}\left(1-3\zeta(3)\right)\\
    &-\frac{32c^2}{3\pi}\int dr\, d\theta\, \frac{r^3\sin^2 \theta}{u^2}((1+u+v)\Dbar_{1,1,1,1}-4)\calT(u,v)\, , \\
    (\partial^4_b - 15\partial^2_b) F\vert_{m=0,b=1} &= 16c(2-3\zeta(3)) \\
    &-\frac{32c^2}{\pi}\int dr\, d\theta\, \frac{r^3\sin^2 \theta}{u^2}((1+u+v)\Dbar_{1,1,1,1}-8)\calT(u,v)\, .
\end{split}
\end{equation}
Together, \eqref{eq:knownintcorr} and \eqref{eq:bbmmresult} show that the relations \eqref{eq:relation1} are consequences of the superconformal Ward identities, and therefore the bootstrap studies cannot be improved by inputting the values of derivatives of $F$ with respect to $b$.

To derive \eqref{eq:knownintcorr}, there are two major steps performed in \cite{Binder:2019jwn,Chester:2020dja} following similar work in three-dimensional ${\cal N} = 8$ SCFTs \cite{Binder:2018yvd}.  The first step is to use the superconformal Ward identities to express the correlator of the four integrated operators obtained by taking derivatives in terms of ${\cal T}(u, v)$. The result of this step is a $16$-dimensional integral over the positions of the four insertions.  The second step is to carry out $14$ out of the $16$ position integrals, leaving behind only the two integrals in each line of \eqref{eq:knownintcorr}.  Performing a similar derivation for the relations in \eqref{eq:bbmmresult} involves two similar steps.  However, since the squashing deformation involves multiple spinning operators \cite{Hama:2012bg}, both steps become much more challenging.  A significant fraction of this paper is spent on the technical complications of calculating the requisite Ward identities and carrying out the requisite integrals.

While we obtain a negative result for the case of the ${\cal N} = 4$ SYM theory, the methods we use can be extended to other cases where derivatives with respect to squashing may provide useful independent pieces of data.  Indeed, integral constraints from supersymmetric localization that do not involve squashing have also been applied in 3D SCFTs with maximal $\mathcal{N}=8$ supersymmetry \cite{Binder:2018yvd,Alday:2021ymb,Alday:2022rly}, in 3D $\mathcal{N}=6$ SCFTs \cite{Binder:2019mpb,Binder:2021cif, Chester:2024esn},  and in 4D SCFTs with half-maximal $\mathcal{N}=2$ supersymmetry\cite{Chester:2022sqb,Behan:2024vwg,Behan:2023fqq,Chester:2025wti,Chester:2025ssu, Alday:2024yax, Alday:2024ksp,Billo:2023kak,Pini:2024uia,Billo:2024ftq,Pini:2024zwi,DeLillo:2025hal,DeSmet:2025mbc,DeLillo:2025eqg}. 
Combined with localization results  (see, for instance, \cite{Chester:2014mea,Dedushenko:2016jxl,Agmon:2017xes, Dedushenko:2017avn, Dedushenko:2018icp, Bomans:2021ldw, Chester:2021gdw}), the methods developed in this paper will be very useful for further deriving constraints from squashing deformations in 3D $\mathcal{N}\geq 6$ theories.
Additional constraints for these theories would likely yield significant improvements in the already-precise numerical bootstrap studies of ABJM theory \cite{Chester:2014fya,Chester:2014mea,Agmon:2017xes,Agmon:2019imm,Chester:2024bij}.  
For 4D $\mathcal{N} = 2$ SCFTs, or 3D $\mathcal{N} = 4$ SCFTs, the mass and squashing deformations couple to different superconformal multiplets. Thus, the constraints involving squashing are guaranteed to be independent, because they constrain independent correlators. See \cite{DeLillo:2025stg} for recent calculations of the localization input in the 4D $\mathcal{N} = 2$ case. In the future, we plan to apply the methods developed in this paper to derive those constraints, and also to assess whether squashing yields independent constraints in 3D $\mathcal{N} \ge 6$ SCFTs.

The rest of this paper is organized as follows. In Section~\ref{sec:background}, we derive the background of a 4D $\mathcal{N}=2$ SCFT with mass and squashing deformations, clearing up various subtleties in the literature. In Section~\ref{4points}, we describe these deformations in a 4D $\mathcal{N}=4$ SCFT, viewing it as a special case of $\mathcal{N}=2$ superconformal symmetry. In particular, we discuss how to identify the operators in the stress-tensor multiplet of a 4D $\mathcal{N}=4$ SCFT that couple to the mass $m$ and squashing parameter $b$, and how to compute the four-point correlators of these operators. In Section~\ref{sec:nosquash}, we rederive the integral constraints \eqref{eq:knownintcorr}, clarifying a technical step that was not justified properly in \cite{Chester:2020dja}.  In Section~\ref{sec:squashintcorr}, we extend this derivation to the constraints \eqref{eq:bbmmresult}.   One of the technical innovations described in Sections~\ref{sec:nosquash} and~\ref{sec:squashintcorr} consists of a new method for computing conformal integrals that also applies to conformal integrals that break rotational symmetry.  Technical details of the calculations are given in the various Appendices.

	\section{Deformations of \texorpdfstring{${\cal N} = 2$}{N = 2} SCFTs on \texorpdfstring{$S^4$}{S\^4}}
	\label{sec:background}
	
	The ${\cal N} = 2$-preserving deformations of ${\cal N} = 4$ SYM theory on $S^4$ mentioned in the Introduction, with parameters $(\tau, \bar \tau, m, b)$, can be described abstractly in terms of deformations of an ${\cal N} = 2$ SCFT placed on a round $S^4$. Let us parametrize the round $S^4$ by coordinates $x \in \R^4$ in which the metric is manifestly conformally flat:
	\es{eq:S4metric}{
	  ds^2_{S^4} = \Omega(x)^2 ds_{\mathbb{R}^4}^2 \,, \qquad \Omega(x) \equiv \frac{2}{1+x^2} \,,
	}
where $x^2 = x \cdot x$, and where, for simplicity, we set the radius of $S^4$ to unity.  Since the round $S^4$ is conformally flat, a CFT can be placed on it by simply rescaling the flat space operators by appropriate powers of $\Omega(x)$.  The operators that we encounter in the rest of this paper are those on a unit-radius round $S^4$. An alternate but equivalent way of describing ${\cal N} = 2$ SCFTs on round $S^4$ by coupling it to a background $\mathcal{N}=2$ Weyl multiplet is explained in Appendix~\ref{sec:SUSYN2def}.

	The operators of ${\cal N} = 4$ SYM theory can be organized into multiplets of the ${\cal N} = 4$ superconformal algebra, $\mathfrak{psu}(2, 2|4)$.  However, for the purpose of studying ${\cal N} = 2$-preserving deformations it is more convenient to organize the operators into multiplets of the ${\cal N} = 2$ superconformal algebra $\mathfrak{su}(2,2|2)$, which is a subalgebra of $\mathfrak{psu}(2, 2|4)$.  The ${\cal N} = 2$ superconformal algebra has as bosonic subalgebras the $\mathfrak{so}(4,2)$ conformal algebra and the R-symmetry algebra $\mathfrak{su}(2)_R\times \mathfrak{u}(1)_R$.  We will thus characterize our operators by the $\mathfrak{so}(4,2)$ quantum numbers (scaling dimension $\Delta$ and $\mathfrak{so}(3, 1)$ spin $(\ell_1, \ell_2)$) and by the $\mathfrak{su}(2)_R$ representation and $\mathfrak{u}(1)_R$ charge.
	
	We use Latin letters from the beginning of the alphabet ($a,b, \ldots$) for frame indices, Greek letters from the middle of the alphabet ($\mu,\nu, \ldots$) for coordinate indices, and Latin letters from the middle of the alphabet ($i, j, \ldots$) for $\mathfrak{su}(2)$ spinor indices.  For the metric in \eqref{eq:S4metric}, the frame $e^a{}_\mu$ is 
	\es{frame}{
		e^a{}_\mu  = \Omega(x) \delta^a_\mu \,.
	} 
		Whenever possible, we will use the conventions set in \cite{Freedman:2012zz,Lauria:2020rhc}, adapted to Euclidean signature. The analytical continuation from Lorentzian signature to Euclidean signature is given by the Wick rotation: $(t,\vec{x}) \to (\vec{x},x_E^4 = i t)$. Objects with raised indices, including the $\gamma^a$ matrices, are adapted similarly. The $\gamma^a$ matrices are given explicitly in \eqref{eq:gammaconv} in terms of the $\sigma^a, \bar{\sigma}^a$\nobreakdash-matrices, given in \eqref{eq:sigmamatrices}. Note that the Levi-Civita tensor $\epsilon^{\mu\nu\rho\sigma}$ changes due to this Wick rotation. As a result, the definition of self-dual and anti-self-dual tensors is altered by this Wick rotation. In Euclidean signature, such tensors satisfy the relation \eqref{eq:defSD}. The explicit components of the $\gamma^*$ matrix remain unchanged under this Wick rotation, as one can see in \eqref{eq:gammaconv}, even though the definition of the  $\gamma^*$ matrix as a product of the four $\gamma^a$ matrices changes. Additional details of the conventions used are presented at the start of Appendix~\ref{sec:SUSYN2def}.

	\subsection{Deformation by chiral and anti-chiral operators}
	\label{sec:taudef}

	As mentioned in the Introduction, in the ${\cal N} = 4$ SYM theory, $\tau$ is the complexified gauge coupling, written in terms of the Yang-Mills coupling $g_\text{YM}$ and the $\theta$-angle as $\tau = \frac{\theta}{2 \pi} + \frac{4 \pi i}{g_\text{YM}^2}$.  The Lagrangian density of the ${\cal N} = 4$ SYM theory can be written as 
	\es{LagAbstract}{
	 {\cal L} = \tau C + \bar \tau \bar C \,,
	}
where $C$ and $\bar C$ are R\nobreakdash-symmetry invariant operators of scaling dimension four.  In ${\cal N} = 2$ notation, the operators $C$ and $\bar C$ belong to a chiral and an anti-chiral multiplet, respectively.  The full operator content of such a chiral multiplet and the quantum numbers of these operators are summarized in Table~\ref{tab:ChiralMultTable}.  
	\begin{table}[!tb]
		\centering
		\renewcommand{\arraystretch}{1.2}
		\begin{tabular}{ccccc}
			\toprule
			Operator & Conformal dimension & Lorentz rep & $\mathfrak{su}(2)_R$ rep & $\mathfrak{u}(1)_R$ charge \\
			\midrule
			$\mathcal{A}$ & 2 & $(0,0)$ & $\mathbf{1}$ & $2$ \\	
			$\Psi_i$ & $\frac{5}{2}$ & $(\frac{1}{2},0)$ & $\mathbf{2}$ & $\frac{3}{2}$ \\
			$N_{ij}$ & 3 & $\left(0,0\right)$ & $\mathbf{3}$ & $1$ \\
			$G^{-}_{ab}$ & 3 & $\left(1,0\right)$ & $\mathbf{1}$ & $1$ \\
			$\Lambda_i$ & $\frac{7}{2}$ & $\left(\frac{1}{2},0\right)$ & $\mathbf{2}$ & $\frac{1}{2}$ \\
			$C$ & 4 & $\left(0,0\right)$ & $\mathbf{1}$ & $0$ \\
			\bottomrule
		\end{tabular}
		\caption{Conformal primary operators in the ${\cal N} = 2$ chiral multiplet whose superconformal primary has scaling dimension 2, and their quantum numbers under the $\mathcal{N}=2$ superconformal algebra.}
		\label{tab:ChiralMultTable}
	\end{table}	
In particular, the superconformal primary is a dimension-two scalar operator $\mathcal{A}$ that is annihilated by all right-handed supersymmetry generators,
\es{eq:DefChiral}{Q^i \mathcal{A} = 0\,, \qquad i = 1,2\,.}
  The other conformal primaries are $\Psi_i$, $N_{ij}$, $G^-_{ab}$, $\Lambda_i$, and $C$;  except for $C$, which appears in \eqref{LagAbstract}, these operators will not be important in the discussion below.  An anti-chiral multiplet with a dimension-two superconformal primary contains operators that are conjugate to those in Table~\ref{tab:ChiralMultTable}.

	One can then consider an infinitesimal change in the couplings $(\tau, \bar \tau)$ under which the Lagrangian density changes by 
	\es{Deform}{
	 \delta {\cal L} = \delta \tau \, C + \delta \bar \tau\,  \bar C \,.
	}
Such a deformation can be phrased more generally in terms of a deformation of an abstract ${\cal N} = 2$ SCFT\@.  If this ${\cal N} = 2$ SCFT contains a chiral and an anti-chiral multiplet whose superconformal primaries have dimension $2$, as in Table~\ref{tab:ChiralMultTable}, then \eqref{Deform} is a supersymmetry-preserving deformation with coupling constants $\delta \tau$ and $\delta \bar \tau$ that do not necessarily have to be interpreted as small changes in the complexified gauge coupling.  On $S^4$, such a deformation was studied in \cite{Gerchkovitz:2014gta,Gomis:2014woa,Gerchkovitz:2016gxx}.

Using a supercharge $\mathbb{Q}$ that belongs to an $\mathfrak{su}(1|1)$ subalgebra of $\mathfrak{su}(2,2|2)$, it can be shown that, up to $\mathbb{Q}$-exact terms, the integrated $C$ operator is equivalent to an insertion at the north pole (the point $\vec{x} = 0$) of the dimension-two chiral superprimary $ \mathcal{A}$  \cite{Gomis:2014woa}.  Likewise, the integrated $\bar C$ operator is equivalent,  also up to $\mathbb{Q}$-exact terms,  to an insertion at the south pole (the point where $\abs{\vec{x}} \to \infty$) of the dimension-two antichiral primary operator $\bar{\mathcal{A}}$. Thus, 
	\es{eq:taupole}{ 
		\delta \tau\int d^4 x \sqrt{g(x)}\, C(x) &= \delta \tau\mathcal{A}(N) + \text{$\mathbb{Q}$-exact terms}\,, 
		   \\ 
		\delta \bar{\tau}\int d^4 x \sqrt{g(x)} \,\overline{C}(x) &= \delta \bar{\tau} \mathcal{\bar{A}}(S) + \text{$\mathbb{Q}$-exact terms} \, .}
In the rest of this paper, we will use the simplified versions of the deformations given on the RHS of \eqref{eq:taupole}.

	\subsection{Mass deformation}
	
	\label{sec:massdeform}

	The second type of deformation we discuss is an ${\cal N} = 2$-preserving mass deformation.  When the ${\cal N} = 4$ SYM theory is viewed as an ${\cal N} = 2$ theory, the ${\cal N} = 4$ vector multiplet decomposes into an ${\cal N} = 2$ vector multiplet and an ${\cal N} = 2$ hypermultiplet transforming in the adjoint representation of the gauge group.  The mass deformation corresponds to giving equal masses $m$ to the scalars and fermions of the hypermultiplet.  In flat space, such a mass deformation breaks superconformal symmetry to Poincar\'e supersymmetry.  On the sphere, the breaking is slightly different, with the superconformal algebra $\mathfrak{su}(2, 2|2)$ being broken to $\mathfrak{osp}(2|4)$, provided that additional terms proportional to $m/r$, where $r$ is the radius of the sphere, accompany the deformation.  In particular, up to linear order in $m$, the deformation on $S^4$ is \cite{Pestun:2007rz}
	\es{DefMass}{
	 \delta {\cal L} = - m {\cal O}_m + O(m^2) \,, \qquad {\cal O}_m \equiv   i (K - \overline{K}) + \frac{2}{r} J_{12}  \,,
	}
where $K$ and $\overline{K}$ are fermion mass terms and $J_{12}$ is a scalar bilinear.  ($J_{12}$ is not the usual scalar mass term;  the latter appears in the mass deformation only at order $m^2$.)  Consequently, $K$ and $\overline{K}$ have scaling dimension $3$ and $J_{12}$ has scaling dimension $2$. The choice of the precise coefficients in \eqref{DefMass} ensures that the $\mathfrak{osp}(2|4)$ algebra that preserves the mass deformation contains the supercharge $\mathbb{Q}$ that was used in \eqref{eq:taupole}.
	
	From the ${\cal N} = 2$ SCFT perspective, the operators $K$, $\overline{K}$, and $J_{12}$ belong to a flavor current multiplet.  (In ${\cal N} = 2$ notation, the ${\cal N} = 4$ SYM theory has an $\mathfrak{su}(2)_F$ flavor symmetry that acts on the adjoint hypermultiplet;  the operators  $K$, $\overline{K}$, and $J_{12}$ belong to a flavor multiplet corresponding to a $\mathfrak{u}(1)_F$ subalgebra of $\mathfrak{su}(2)_F$.)    The operator content of the ${\cal N} = 2$ flavor current multiplet is summarized in Table \ref{tab:FlavMultTable}. 
	\begin{table}[!tb]
		\centering
		\renewcommand{\arraystretch}{1.1}
		\begin{tabular}{ccccc}
			\toprule
			Operator & Conformal dimension & Lorentz rep & $\mathfrak{su}(2)_R$ rep & $\mathfrak{u}(1)_R$ charge \\
			\midrule
			$J_{ij}$ & 2 & $\left(0,0\right)$ & $\mathbf{3}$ & $0$ \\		
			$\xi_i$ & $\frac{5}{2}$& $\left(0,\frac{1}{2}\right)$ & $\mathbf{2}$ & $\frac{1}{2}$ \\
			$\xi^i$ & $\frac{5}{2}$ & $\left(\frac{1}{2},0\right)$ & $\mathbf{2}$ & $-\frac{1}{2}$ \\
			$K$ & 3 & $(0,0)$ & $\mathbf{1}$ & $-1$ \\	
			$\overline{K}$ & 3 & $(0,0)$ & $\mathbf{1}$ & $1$ \\
			$j_{\mu}$ & 3 & $\left(\frac{1}{2},\frac{1}{2}\right)$ & $\mathbf{1}$ & $0$ \\
			\bottomrule
		\end{tabular}
		\caption{Conformal primary operators in the flavor current multiplet and their quantum numbers under the $\mathcal{N}=2$ superconformal algebra.}
		\label{tab:FlavMultTable}
	\end{table}	
The multiplet starts with a superconformal primary $J_{ij}$, which is a scalar transforming as a triplet of $\mathfrak{su}(2)_R$ and obeying $J_{ij} = J_{ji} = \epsilon_{ik}\epsilon_{jl}J^{kl} = (J^{ij})^*$.    Acting with the supercharges on $J_{ij}$, one obtains fermionic operators that will not be important in the discussion below, as well as scalars $K$ and $\overline{K}$ and a conserved current $j_\mu$.  From an abstract point of view, whenever an ${\cal N} = 2$ SCFT has a conserved current multiplet, we can construct a mass deformation of the form \eqref{DefMass}---see Appendix~\ref{sec:massdeformderiv} for more details (see also \cite{Binder:2019jwn}).  After setting $r=1$, this mass deformation takes the form
	\es{eq:masscoupl}{
	 \delta {\cal L} = - m {\cal O}_m  \,, \qquad {\cal O}_m \equiv  i (K - \overline{K}) + 2 J_{12}  \,.
	}
Note that \eqref{DefMass} differs  \eqref{eq:masscoupl} by the presence of the $O(m^2)$ terms. See the end of Section~\ref{sec:nosquash} for a discussion about these subleading terms.

	\subsection{Squashing deformation}
	\label{sec:squashdeform}
	
	We now move on to the deformation by squashing of the $S^4$, which has not been considered before for computing integral constraints.  Let us start from the round sphere, whose line element is obtained by restricting the standard line element $ds$ in flat five-dimensional space.  Using coordinates $y^p$, $p = 1, \ldots, 5$ for $\R^5$, we  have
	\es{RoundSphere}{
	 ds^2 = \sum_{p=1}^5 (dy^p)^2 \,, \qquad \sum_{p=1}^5 (y^p)^2 = 1 \,.
	}
We can solve the constraint in the second equation using stereographic coordinates $x \in \R^4$ related to $y^p$ by 
	\es{eq:varsR4toS4}{
	 y^p = \left( \frac{2 \vec{x}}{1+x^2},  \frac{1-x^2}{1+x^2} \right) \,.
	 }
It is straightforward to check that the first equation in \eqref{RoundSphere} then yields \eqref{eq:S4metric}.  
	The squashed $S^4$ can be described in a similar manner by simply modifying the embedding into $\mathbb{R}^5$.  Instead of \eqref{RoundSphere}, we have
	\es{eq:squashembed}{
	 \widetilde{ds}^2 = \sum_{p=1}^5 (dy^p)^2 \,, \qquad	\frac{(y^1)^2+(y^2)^2}{b^2} + \frac{(y^3)^2+(y^4)^2}{b^{-2}} + (y^5)^2 = 1\,,
	}
where $b$ is the squashing parameter and $\widetilde{ds}$ is the line element on the squashed sphere. For ${b\neq1}$, the squashing breaks the $\mathfrak{so}(5)$ isometry of $S^4$ to an $\mathfrak{so}(2)\times \mathfrak{so}(2)$ isometry corresponding to independent rotations in the $(y^1,y^2)$ and $(y^3,y^4)$ planes. We can parametrize the squashed sphere using modified stereographic coordinates $x \in \R^4$ defined by
	\es{eq:varsR4toSquashedS4}{
	 y^p = \left( \frac{2x_1}{1+x^2} b,\; \frac{2x_2}{1+x^2} b,\; \frac{2x_3}{1+x^2} b^{-1},\; \frac{2x_4}{1+x^2} b^{-1},\;  \frac{1-x^2}{1+x^2} \right) \,.
	 }
The parameterization \eqref{eq:varsR4toSquashedS4} solves the constraint in the second equation of \eqref{eq:squashembed}.  
	 
	 Writing $ds^2 = g_{\mu\nu} dx^\mu dx^\nu$ for the round sphere and $\widetilde{ds}^2 = \widetilde{g}_{\mu\nu} dx^\mu dx^\nu$ for the squashed sphere, we can expand the metric on the squashed sphere as
	 \es{eq:squashmetric}{
	  \widetilde{g}_{\mu\nu} = g_{\mu\nu} + (b-1) \Delta g_{\mu\nu} + O\left((b-1)^2 \right) \,,
	 }
where
 	\es{g1Def}{
	 \Delta g_{\mu\nu} &= \nabla_{\mu}\zeta_{\nu}+ \nabla_{\nu}\zeta_{\mu} + h g_{\mu\nu} \,, \qquad
		h \equiv \frac{8 (x_1^2 + x_2^2 - x_3^2 - x_4^2)}{(1+x^2)^2} \,, \qquad
		\zeta_\mu \equiv \frac 14 \partial_\mu h \,.
	}	
Here, the covariant derivative $\nabla_\mu$ is taken with respect to the round sphere metric \eqref{eq:S4metric}.  The first two terms in the expression for $\Delta g_{\mu\nu}$ correspond to an infinitesimal diffeomorphism, while the last term corresponds to an infinitesimal Weyl rescaling.  Therefore, up to first order in $(b-1)$, the squashed sphere is conformally flat.

	To preserve some supersymmetry, the squashing of the sphere must be accompanied by non-trivial background values for other fields of a background Weyl multiplet \cite{Hama:2012bg,Pestun:2014mja}.  We give the details of this construction in Appendix~\ref{sec:bkgrndsq}, correcting some typos in \cite{Pestun:2014mja}.  In short, the effect of the squashing deformation is to modify the action by $b$-dependent terms involving operators from the $\mathcal{N}=2$ stress-tensor multiplet coupling to the various non-trivial background fields on $S^4$. The operator content of the stress-tensor multiplet is presented in Table~\ref{tab:STMultTable}.
		\begin{table}[!tb]
		\centering
		\renewcommand{\arraystretch}{1.2}
		\begin{tabular}{ccccc}
			\toprule
			Operator & Conformal dimension & Lorentz rep & $\mathfrak{su}(2)_R$ rep & $\mathfrak{u}(1)_R$ charge \\
			\midrule
			$\Phi$ & 2 & $(0,0)$ & $\mathbf{1}$ & 0 \\
			$\zeta_i$ & $\frac{5}{2}$& $\left(\frac{1}{2},0\right)$ & $\mathbf{2}$ & $-\frac{1}{2}$ \\
			$\zeta^i$ & $\frac{5}{2}$ & $\left(0,\frac{1}{2}\right)$ & $\mathbf{2}$ & $\frac{1}{2}$ \\
			$j_{\mu}$ & $3$ & $\left(\frac{1}{2},\frac{1}{2}\right)$ & $\mathbf{1}$ & $0$ \\
			$\tensor{j}{_{\mu}^i_j}$ & $3$ & $\left(\frac{1}{2},\frac{1}{2}\right)$ & $\mathbf{3}$ & $0$ \\
			$Z^{-}_{\mu\nu}$ & $3$ & $\left(1,0\right)$ & $\mathbf{1}$ & $-1$ \\
			$Z^{+}_{\mu\nu}$ & $3$ & $\left(0,1\right)$ & $\mathbf{1}$ & $1$ \\
			$\kappa_{\mu\,i}$ & $\frac{7}{2}$ & $\left(1,\frac{1}{2}\right)$ & $\mathbf{2}$ & $-\frac{1}{2}$ \\
			$\tensor{\kappa}{_{\mu}^i}$ & $\frac{7}{2}$ & $\left(\frac{1}{2},1\right)$ & $\mathbf{2}$ & $\frac{1}{2}$ \\
			$T_{\mu\nu}$&4 & $\left(2,2\right)$& $\mathbf{1}$ & 0 \\
			\bottomrule
		\end{tabular}
		\caption{Conformal primary operators in the $\mathcal{N}=2$  stress-tensor multiplet and their quantum numbers under the $\mathcal{N}=2$ superconformal algebra.}
		\label{tab:STMultTable}
	\end{table}
The operators relevant to the squashing deformation at linear order in $(b-1)$ are the $\mathfrak{su}(2)_R$-symmetry current $\tensor{j}{_{\mu\, i}^j}$, the self-dual and anti-self-dual tensors $Z^{+}_{\mu\nu}$ and $Z^-_{\mu\nu}$, respectively, and the stress-tensor $T^{\mu\nu}$.  The squashing deformation takes the form
 \es{eq:squashcoupl}{
  \delta {\cal L}  &= (b-1) \mathcal{O}_b  + O((b-1)^2)\,, \qquad
    {\cal O}_b \equiv - 2 i \xi^\mu \tensor{{j}}{_{\mu 1}^1} - 4 (\nabla^\mu \xi^\nu) (Z^+_{\mu\nu} - Z^-_{\mu\nu}) 
     + \frac 12 \Delta g_{\mu\nu} T^{\mu\nu} \,,
 }
where $\Delta g_{\mu\nu}$ was given in \eqref{g1Def}, and  $\xi^\mu$ is one of the two Killing vectors on $S^4$ that correspond to the $\mathfrak{so}(2) \times \mathfrak{so}(2)$ isometry preserved by the squashing.  In particular, the one-forms dual to the $10$ Killing vectors can be written as
 \es{KVGen}{
  \xi^{pq} = \frac{1}{2}(y^p dy^q-y^q dy^p) \,,
 }
where $\xi^{pq}$ is anti-symmetric in the $p, q$ indices; the Killing vector $\xi$ appearing in \eqref{eq:squashcoupl} is
 \es{KVExplicit}{
  \xi = \xi^{12} - \xi^{34} \,.
 }
More details can be found in Appendices~\ref{sec:bkgrndsq} and~\ref{sec:stmultN2}. Note that $\Delta g_{\mu\nu}$ in \eqref{g1Def} consists of an infinitesimal diffeomorphism and an infinitesimal Weyl rescaling. This implies that the last term in \eqref{eq:squashcoupl}, $\frac 12 \Delta g_{\mu\nu}T^{\mu\nu}$, vanishes after it is integrated over the entire $S^4$ (as a consequence of the conservation and tracelessness of  $T^{\mu\nu}$).

 The subleading terms in \eqref{eq:squashcoupl} include couplings between the operators of the $\mathcal{N}=2$ stress-tensor multiplet and the subleading terms in the expansion in $(b-1)$ of the Weyl multiplet background fields, given in Appendix~\ref{sec:bkgrndsq}. The subleading terms in \eqref{eq:squashcoupl} might also involve operators that belong to other superconformal multiplets. As we explain at the end of Section~\ref{sec:bbtt}, the contributions of these subleading terms in \eqref{eq:squashcoupl} to squashing integral constraints can be obtained from the relations in \eqref{eq:relation1}, without knowing the specifics of such terms.

	\section{Stress-tensor correlation functions in \texorpdfstring{$\mathcal{N}=4$}{N = 4} SCFTs}
	\label{4points}
	
	In an $\mathcal{N}=4$ SCFT, all the operators coupling to the squashing, mass, and $\tau$ parameters discussed in the previous section belong to the stress-tensor multiplet.
	 The operators of the stress-tensor multiplet and their quantum numbers are given in Table~\ref{tab:N4STOps}.   This is a $\frac{1}{2}$\nobreakdash-BPS multiplet which contains several conserved currents: the $\mathfrak{su}(4)_R$-symmetry current $J_\mu$, the $\mathcal{N}=4$ SUSY currents $\psi_\mu$ and $\bar{\psi}_\mu$, and the stress tensor $T_{\mu\nu}$. The operators of this multiplet that are relevant for describing the mass, squashing, and $\tau$ deformations are the dimension-two superprimary $S$ and the dimension-three operators: the scalars ($P$ and $\bar{P}$), the self-dual and anti-self-dual fields ($B^-_{[\mu\nu]}$ and $B^+_{[\mu\nu]}$), and the $\mathfrak{su}(4)_R$-symmetry current $J_\mu$.
In this section, we briefly discuss the constraints imposed by superconformal symmetry on the four-point correlation functions of operators in the $\mathcal{N}=4$ stress-tensor multiplet.
The method we use to compute these symmetry constraints was previously applied in \cite{Binder:2019jwn,Chester:2020dja} to the correlation functions contributing to the non-squashed integral constraints in \eqref{eq:knownintcorr}. We also discuss how the $\mathcal{N}=2$ operators mentioned in the previous section are embedded in the $\mathcal{N}=4$ stress-tensor multiplet.

	\subsection{Computing \texorpdfstring{$\mathcal{N}=4$}{N = 4} stress-tensor correlation functions}
	\label{WardSolve}
	
	\begin{table}[!tb]
		\centering
		\renewcommand{\arraystretch}{1.2}
		\begin{tabular}{cccc}
			\toprule
			Operator & Conformal dimension & Lorentz rep & $\mathfrak{su}(4)_R$ rep \\
			\midrule
			$S$& $2$ & $\left(0,0\right)$ & $\mathbf{20^{'}}$  \\
			$\chi$& $\frac{5}{2}$ & $\left(\frac{1}{2},0\right)$ & $\mathbf{20}$  \\
			$\overline{\chi}$& $\frac{5}{2}$ & $\left(0,\frac{1}{2}\right)$ & $\overline{\mathbf{20}}$ \\
			$B^{-}_{[\mu\nu]}$& $3$ & $\left(1,0\right)$ & $\mathbf{6}$  \\
			$P$& $3$ & $\left(0,0\right)$ & $\overline{\mathbf{10}}$ \\
			$B^{+}_{[\mu\nu]}$& $3$ & $\left(0,1\right)$ & $\mathbf{6}$ \\
			$\bar{P}$& $3$ & $\left(0,0\right)$ & $\mathbf{10}$ \\
			$J_{\mu}$& $3$ & $\left(\frac{1}{2},\frac{1}{2}\right)$ & $\mathbf{15}$  \\
			$\psi_{\mu}$& $\frac{7}{2}$ & $\left(1,\frac{1}{2}\right)$ & $\mathbf{4}$ \\
			$\overline{\psi}_{\mu}$& $\frac{7}{2}$ & $\left(\frac{1}{2},1\right)$ & $\overline{\mathbf{4}}$  \\
			$\lambda$& $\frac{7}{2}$ & $\left(\frac{1}{2},0\right)$ & $\overline{\mathbf{4}}$  \\
			$\overline{\lambda}$& $\frac{7}{2}$ & $\left(0,\frac{1}{2}\right)$ & $\mathbf{4}$  \\
			$\phi$& $4$ & $\left(0,0\right)$ & $\mathbf{1}$  \\
			$\overline{\phi}$& $4$ & $\left(0,0\right)$ & $\mathbf{1}$ \\
			$T_{(\mu\nu)}$& $4$ & $\left(1,1\right)$ & $\mathbf{1}$  \\
			\bottomrule
		\end{tabular}
		\caption{Conformal primary operators in the $\mathcal{N}=4$ stress-tensor multiplet and their quantum numbers under the $\mathcal{N}=4$ superconformal algebra.}
		\label{tab:N4STOps}
	\end{table}
Let us denote an operator $\mathcal{O}$ with conformal dimension $\Delta$, R-symmetry representation $r$, and spin quantum numbers $\ell~\equiv~(\ell_{\text{left}},\ell_{\text{right}})$ by  $\tensor{\mathcal{O}}{^r_{\Delta, \ell}}$. In an SCFT, conformal symmetry and R-symmetry imply that any four-point correlation function can be expanded simultaneously in a basis of R-symmetry invariants as well as a basis of conformally-covariant functions of the coordinates of the four points, which we call the conformal basis:
	\es{eq:corrExpand} { &\braket{\tensor{\mathcal{O}}{_1^{\textit{r}_1}_{\Delta_1, \ell_1}}(x_1)\tensor{\mathcal{O}}{_2^{\textit{r}_2}_{\Delta_2, \ell_2}}(x_2)\tensor{\mathcal{O}}{_3^{\textit{r}_3}_{\Delta_3, \ell_3}}(x_3)\tensor{\mathcal{O}}{_4^{\textit{r}_4}}_{\Delta_4, \ell_4}(x_4)} \\
		=&\sum_{i,j} \tensor{g}{_{i j}^{\calallO}}(u,v)\,\tensor{\mathcal{R}}{^{i}_{\textit{r}_1,\textit{r}_2,\textit{r}_3,\textit{r}_4}}\tensor{\mathcal{S}}{^{j}_{(\Delta_1,\ell_1),(\Delta_2,\ell_2),(\Delta_3,\ell_3),(\Delta_4,\ell_4)}}(x_1,x_2,x_3,x_4)\, .}
	Here $\tensor{\mathcal{R}}{^{i}_{\{r_i\}}}$ are rank-four R-symmetry invariant tensors of dimension $\text{dim}(r_1) \times \text{dim}(r_2) \times \text{dim}(r_3) \times \text{dim}(r_4)$, the $\tensor{\mathcal{S}}{^{j}_{\{(\Delta_i,l_i)\}}}$ structures form the conformal basis, and $u$ and $v$ are the usual conformal cross-ratios 
	\es{eq:defuv}{ u=\frac{x^2_{12}x^2_{34}}{x^2_{13}x^2_{24}}\, , \qquad v=\frac{x^2_{14}x^2_{23}}{x^2_{13}x^2_{24}}\, ,}
	where $x_{ij} \equiv x_i - x_j$. The conformal basis can be taken as products of a finite number of conformally-invariant building blocks multiplied by an appropriate scalar prefactor determined by the conformal dimensions of the four operators, as explained in \cite{Cuomo:2017wme}. While the structures $\tensor{\mathcal{R}}{^{i}_{\{r_i\}}}$ and $\tensor{\mathcal{S}}{^{j}_{\{(\Delta_i,l_i)\}}}$ are fully specified by kinematics, the non-trivial data of the four-point function is contained in the conformally-invariant functions $\tensor{g}{_{i j}^{\calallO}}(u,v)$ in \eqref{eq:corrExpand}.
	
	Correlation functions of operators in the $\mathcal{N}=4$ stress-tensor multiplet are further constrained by supersymmetry, which implies that the correlators of any four such operators can be expressed in terms of the reduced correlator $\mathcal{T}(u,v)$ \cite{Dolan:2001tt}. For example, the four-point function of the stress-tensor superprimary $S$ can be written in the form \eqref{eq:corrExpand} as 
	\es{eq:gssssform}{ \braket{S(x_1)S(x_2)S(x_3)S(x_4)} &= \sum_{i=1}^6 \frac{1}{x^4_{12} x^4_{34}} \tensor{\mathcal{R}}{^{i}_{\mathbf{20}',\mathbf{20}', \mathbf{20}',\mathbf{20}'}}	g^{SSSS}_{i1}(u,v) \, , \\
		g^{SSSS}_{i1}(u,v) &=  \left(v \quad uv \quad u \quad u(u-v-1)\quad 1-u-v \quad v(v-u-1)\right)\mathcal{T}(u,v)\\
		& +  \left( 1 \quad u^2 \quad \frac{u^2}{v^2} \quad \frac{1}{c}\frac{u^2}{v} \quad \frac{1}{c}\frac{u}{v} \quad \frac{1}{c}u\right),}
	where $c$ is the conformal anomaly coefficient, and the six  $\tensor{\mathcal{R}}{^{i}_{\mathbf{20}',\mathbf{20}', \mathbf{20}',\mathbf{20}'}}$ tensors are defined as\footnote{The invariants listed here are equivalent to the R-symmetry invariants used in \cite{Binder:2019jwn,Chester:2020dja}, up to a reordering.}
	\es{eq:fourptinvs20p}{(\tensor{\mathcal{R}}{^{1}_{\mathbf{20}',\mathbf{20}', \mathbf{20}',\mathbf{20}'}})^{KLMN} &= \delta^{KL}_{\mathbf{20}',\mathbf{20}'} \delta^{MN}_{\mathbf{20}',\mathbf{20}'}\,,\\
		(\tensor{\mathcal{R}}{^{2}_{\mathbf{20}',\mathbf{20}', \mathbf{20}',\mathbf{20}'}})^{KLMN} &= (\tensor{\mathcal{R}}{^{1}_{\mathbf{20}',\mathbf{20}', \mathbf{20}',\mathbf{20}'}})^{KMNL} \,, \\
		 (\tensor{\mathcal{R}}{^{3}_{\mathbf{20}',\mathbf{20}', \mathbf{20}',\mathbf{20}'}})^{KLMN} &= (\tensor{\mathcal{R}}{^{1}_{\mathbf{20}',\mathbf{20}', \mathbf{20}',\mathbf{20}'}})^{KNLM} \,,\\
		(\tensor{\mathcal{R}}{^{4}_{\mathbf{20}',\mathbf{20}', \mathbf{20}',\mathbf{20}'}})^{KLMN} &= (C_{\mathbf{20}',\mathbf{6},\mathbf{6}})^{K}_{I_1 J_1}(C_{\mathbf{20}',\mathbf{6},\mathbf{6}})^{L}_{J_1 I_2} (C_{\mathbf{20}',\mathbf{6},\mathbf{6}})^{M}_{I_2 J_2} (C_{\mathbf{20}',\mathbf{6},\mathbf{6}})^{N}_{J_2 I_1}\,,\\
		(\tensor{\mathcal{R}}{^{5}_{\mathbf{20}',\mathbf{20}', \mathbf{20}',\mathbf{20}'}})^{KLMN} &= (\tensor{\mathcal{R}}{^{4}_{\mathbf{20}',\mathbf{20}', \mathbf{20}',\mathbf{20}'}})^{KMNL} \,,\\
		 (\tensor{\mathcal{R}}{^{6}_{\mathbf{20}',\mathbf{20}', \mathbf{20}',\mathbf{20}'}})^{KLMN} &= (\tensor{\mathcal{R}}{^{4}_{\mathbf{20}',\mathbf{20}', \mathbf{20}',\mathbf{20}'}})^{KNLM} \,,}
		where  $K,L,M,N$ are indices for the $\mathbf{20}'$ representation, $I_{i},J_{i}$ are indices for the $\mathbf{6}$ representation, and $C_{\mathbf{20}',\mathbf{6},\mathbf{6}}$ is the three-point $\mathfrak{su}(4)_R$-invariant between one copy of the $\mathbf{20}'$ rep and two copies of the $\mathbf{6}$ rep, normalized such that $(C_{\mathbf{20}',\mathbf{6},\mathbf{6}})^1_{1,1} = \frac{1}{\sqrt{2}}$.\footnote{Details regarding our $\mathfrak{su}(4)_R$  conventions, including the Clebsch-Gordan coefficients relating various representations, as well as an explanation on how three-point $\mathfrak{su}(4)_R$-invariants can be computed is provided in Appendix~\ref{sec:su4andN4conv}.} In the expression \eqref{eq:gssssform}, the second line is the four-point function in a free ${\cal N} = 4$ superconformal theory with central charge $c$, while the reduced correlator $\mathcal{T}(u,v)$ in the first line depends non-trivially on the dynamics of the theory.
	One should note that since we expand the correlation functions in a basis of conformally-covariant structures, as given in \eqref{eq:corrExpand}, only the Ward identities for Poincar\'e SUSY are needed to derive the expressions for all possible $\tensor{g}{_{i j}^{\calallO}}(u,v)$ in terms of $\mathcal{T}(u,v)$.\footnote{Using the superconformal algebra, one can show that invariance under special conformal transformations and Poincar\'e SUSY implies invariance under superconformal SUSY. Thus, the Ward identities for superconformal SUSY would not provide any additional information beyond what is contained in the Poincar\'e SUSY Ward identities.} We use these Poincar\'e SUSY Ward identities to compute the correlation functions of operators in the $\mathcal{N}=4$ stress-tensor multiplet that contribute to the squashed integral constraints in \eqref{eq:bbmmresult}. This computation is done in flat space, since correlation functions on $S^4$ can be obtained from flat space correlators via the conformal transformation 
	\es{eq:R4toS4corr}{\braket{\mathcal{O}_{1}(x_1)\mathcal{O}_{2}(x_2)\mathcal{O}_{3}(x_3)\mathcal{O}_{4}(x_4)}_{S^4} = \frac{\braket{\mathcal{O}_{1}(x_1)\mathcal{O}_{2}(x_2)\mathcal{O}_{3}(x_3)\mathcal{O}_{4}(x_4)}_{\mathbb{R}^4}}{\Omega(x_1)^{\Delta_1}\Omega(x_2)^{\Delta_2}\Omega(x_3)^{\Delta_3}\Omega(x_4)^{\Delta_4}},}
	where $\Delta_i$ is the conformal dimension of $\mathcal{O}_{i}(x_i)$ and $\Omega(x_i)$ is the Weyl factor in \eqref{eq:S4metric}.
	
	\begin{table}[!htb]
		\centering
		\renewcommand{\arraystretch}{1.5}
		\begin{tabular}{c|c|c}
			\Xhline{1.2pt}
			Variation & Correlators used & Correlators obtained  \\
			\hline
			$\delta \braket{SSS\chi} $ & $\braket{SSSS}$ & $\braket{SS\chi\bar{\chi}} \, \braket{SSSJ} $\\
			
			$\delta \braket{SS\bar{P}\chi} $ & $\braket{SS\chi\bar{\chi}}$ & $\braket{S\bar{P}\chi\chi} \, \braket{SSP\bar{P}} \, \braket{SSB^-\bar{P}}$\\
			
			$\delta \braket{SP\bar{P}\bar{\chi}} $ & $\braket{SP\bar{\chi}\,\bar{\chi}}\,\braket{SSP\bar{P}}$ & $\braket{P\bar{P}\chi\bar{\chi}} \, \braket{S\bar{P}\lambda\bar{\chi}} \, \braket{SP\bar{P}J}$\\
			
			$\delta \braket{\bar{P}\,\bar{P}P\chi} $ & $\braket{P\bar{P}\chi\bar{\chi}}$ & $\braket{\bar{P}\,\bar{P}\lambda\chi} \, \braket{PP\bar{P}\,\bar{P}} \, \braket{\bar{P}\,\bar{P}PB^-}$\\
			
			$\delta \braket{SSJ\bar{\chi}} $ & $\braket{SS\chi\bar{\chi}} \, \braket{SSSJ}$ & $\braket{SJ\chi\bar{\chi}} \, \braket{SS\psi\bar{\chi}} \, \braket{SSJJ}$\\
			
			$\delta \braket{SSB^+\chi} $ & $\braket{SS\chi\bar{\chi}}$ & $\braket{SB^+\chi\chi} \, \braket{SSB^-B^+} \, \braket{SSPB^+}\, \braket{SS\bar{\psi}\chi}$\\
			
			$\delta \braket{S\bar{P}J\chi} $ & $\braket{SJ\chi\bar{\chi}}\,\braket{S\bar{P}\chi\chi}$ & $\braket{\bar{P}J\chi\chi} \, \braket{S\bar{P}\psi\chi} \, \braket{SP\bar{P}J}\,\braket{S\bar{P}B^-J}$\\
			
			$\delta \braket{P\bar{P}J\bar{\chi}} $ & $\braket{PJ\bar{\chi}\,\bar{\chi}}\,\braket{P\bar{P}\chi\bar{\chi}}\,\braket{SP\bar{P}J}$ & $\braket{\lambda\bar{P}J\bar{\chi}} \, \braket{P\bar{P}\psi\bar{\chi}}\,\braket{P\bar{P}JJ}$\\
			
			$\delta \braket{SPB^+\bar{\chi}} $ & $\braket{SP\bar{\chi}\,\bar{\chi}}\,\braket{SSPB^+}$ & $\braket{PB^+\chi\bar{\chi}} \, \braket{SB^+\lambda\bar{\chi}}\,\braket{SP\bar{\psi}\bar{\chi}}\,\braket{SPB^+J}$\\
			
			$\delta \braket{P\bar{P}B^+\chi} $ & $\braket{PB^+\chi\bar{\chi}}\,\braket{P\bar{P}\chi\bar{\chi}}$ & $\braket{\bar{P}B^+\lambda\chi} \, \braket{P\bar{P}\,\bar{\psi}\chi}\,\braket{PP\bar{P}B^+}\,\braket{P\bar{P}B^-B^+}$\\
			
			$\delta \braket{\bar{P}\,\bar{P}B^-\chi} $ & $\braket{\bar{P}B^-\chi\bar{\chi}}$ & $\braket{\bar{P}\,\bar{P}\lambda\chi} \, \braket{\bar{P}\,\bar{P}B^-B^-}\, \braket{\bar{P}\,\bar{P}PB^-}$\\
			
			$\delta \braket{SJJ\bar{\chi}}$ & $\braket{SJ\chi\bar{\chi}}$\,$\braket{SSJJ}$ & $\braket{SJJJ}$ \, $\braket{JJ\chi\bar{\chi}}$ \, $\braket{SJ\psi\bar{\chi}}$\\
			
			$\delta \braket{JJJ\bar{\chi}}$ &  $\braket{JJ\chi\bar{\chi}}$\,$\braket{SJJJ}$ & $\braket{JJJJ}$ \, $\braket{JJ\psi\bar{\chi}}$ \\
			
			$\delta \braket{SB^-B^+\bar{\chi}} $ & $\braket{SB^-\bar{\chi}\,\bar{\chi}}\,\braket{SSB^-B^+}$ & $\braket{B^-B^+\chi\bar{\chi}} \, \braket{SB^+\lambda\bar{\chi}} \, \braket{SB^-B^+J}$\\[-.5em]
			& & $\braket{SB^-\bar{\psi}\bar{\chi}}$ \\
			
			$\delta \braket{SB^+J\chi} $ & $\braket{SJ\chi\bar{\chi}}\,\braket{SB^+\chi\chi}$ & $\braket{B^+J\chi\chi} \, \braket{SJ\bar{\psi}\chi}$\\[-.5em]
			& & $ \braket{SB^+\psi\chi} \, \braket{SPB^+J}\,\braket{SB^+B^-J}$\\
			
			$\delta \braket{B^-B^+J\bar{\chi}} $ & $\braket{B^-J\bar{\chi}\,\bar{\chi}}\,\braket{B^- B^+\chi\bar{\chi}}$ & $\braket{\lambda B^+J\bar{\chi}} \, \braket{ B^- J\bar{\psi}\bar{\chi}} $\\[-.5em]
			& $\braket{SB^- B^+J}$ & $\braket{B^-B^+\psi\bar{\chi}}\,\braket{B^-B^+JJ}$  \\
			
			$\delta \braket{SB^-B^+\bar{\chi}} $ & $\braket{SB^-\bar{\chi}\,\bar{\chi}}\,\braket{SSB^-B^+}$ & $\braket{B^-B^+\chi\bar{\chi}} \, \braket{SB^+\lambda\bar{\chi}} \, \braket{SB^-B^+J}$\\[-.5em]
			& & $ \braket{SB^-\bar{\psi}\bar{\chi}} $\\
			
			$\delta \braket{B^+B^+B^-\chi} $ & $\braket{B^-B^+\chi\bar{\chi}}$ & $\braket{B^+B^+\lambda \chi} \, \braket{B^-B^-B^+B^+} \, \braket{B^+B^+PB^-}$\\[-.5em]
			& & $\braket{B^+B^-\bar{\psi}\chi}$ \\
			\Xhline{1.2pt}
		\end{tabular}
		\caption{Ward identities solved to calculate the $\mathcal{N}=4$ four-point functions that contribute to fourth derivatives of the sphere free energy. Each Ward identity can be used to express the correlators in the third column in terms of derivatives of the correlator(s) in the second column. For clarity, we do not write Lorentz indices explicitly. For all Ward identities listed above, apart from the first one, the conjugate Ward identities (with all the operators switched with their conjugates) are also computed to get the conjugate correlators not listed above (such as $\braket{PPB^+B^+}$ and other correlators needed at intermediate steps).}
		\label{tab:N4STWard}
	\end{table}

	Details of the Ward identity computation can be found in Appendix B of \cite{Binder:2019jwn} using the component field method of \cite{Dolan:2001tt}. The SUSY transformations for the $\mathcal{N}=4$ stress-tensor multiplet are necessary for this Ward identity computation. We describe in Appendix~\ref{sec:su4andN4conv} how these SUSY transformations can be obtained. The particular four-point correlation functions of $\mathcal{N}=4$ stress-tensor multiplet operators that contribute to the fourth derivatives of the sphere free energy, listed in \eqref{eq:independent}, are: $\braket{SSSS}$, $\braket{SSP\bar{P}}$, $\braket{P\bar{P}P\bar{P}}$, $\braket{SSJJ}$, $\braket{SSB^-B^+}$, $\braket{P\bar{P}JJ}$, $\braket{P\bar{P}B^-B^+}$, $\braket{PPB^+B^+}$, $\braket{\bar{P}\,\bar{P}B^-B^-}$, $\braket{SPB^+ J}$, $\braket{S\bar{P}B^-J}$, $\braket{JJJJ}$, $\braket{JJB^+B^+}$,  and $\braket{B^-B^-B^+B^+}$,  where we have suppressed Lorentz indices for clarity. The chain of Ward identities solved to compute the $g$-functions for these correlators are given in Table \ref{tab:N4STWard}. 
	
	We focus on computing the $\mathcal{T}(u,v)$-dependent part of the $g$-functions of these correlators. Free theory contributions to squashing derivatives of the sphere free energy can be easily computed using the relations between various integral constraints, given in \eqref{eq:relation1}, and the previously-derived expressions for the derivatives of the sphere free energy that do not involve squashing, given in \eqref{eq:knownintcorr}. This is done by setting $\mathcal{T}(u,v)$ to zero when substituting \eqref{eq:knownintcorr} in \eqref{eq:relation1}. Note that even though the relations \eqref{eq:relation1} are obtained using the partition function of $\SU(N)$ $\mathcal{N}=4$ SYM \cite{Chester:2020vyz}, the free theory contributions to the squashed integral constraints computed using \eqref{eq:relation1} are valid for any $\mathcal{N}=4$ SCFT. Indeed, the only theory-specific data that enters into the free theory part of \eqref{eq:gssssform} is the parameter $c$. Hence, we do not need to explicitly compute the free theory part for any correlation functions.
	
	 $\mathcal{N}=4$ superconformal symmetry does not place any contraints on $\mathcal{T}(u,v)$. However, $\mathcal{T}(u,v)$ is constrained by the fact that the four-point function $\braket{S(x_1)S(x_2)S(x_3)S(x_4)}$ in \eqref{eq:gssssform} must remain invariant under crossing. This condition implies
	\es{eq:Tcrossing}{
		\mathcal{T}(v,u) &= \mathcal{T}\left(\frac{1}{v},\frac{u}{v}\right) = \frac{v^2}{u^2}\mathcal{T}(u,v)\, ,\\
		\mathcal{T}\left(\frac{u}{v},\frac{1}{v}\right) &=  \mathcal{T}\left(\frac{v}{u},\frac{1}{u}\right) = v^2  \mathcal{T}(u,v) \, ,\\
		\mathcal{T}\left(\frac{1}{u},\frac{v}{u}\right)  &=  \mathcal{T}(u,v) \, .}
	We will use these identities in Sections \ref{sec:nosquash} and \ref{sec:squashintcorr}.
	
\subsection{\texorpdfstring{$\mathcal{N}=2$}{N = 2} SCFT operators embedded in \texorpdfstring{$\mathcal{N}=4$}{N = 4} stress-tensor multiplet}
\label{N2asN4}

In addition to computing $\mathcal{N}= 4$ four-point functions, we need to identify the particular components of $\mathcal{N}= 4$ stress-tensor multiplet operators that correspond to the $\mathcal{N}=2$ operators coupling to the squashing, mass, and $\tau$ parameters, described in Section~\ref{sec:background}. Any $\mathcal{N}= 4$ SCFT can be described as an $\mathcal{N}= 2$ SCFT with $\mathfrak{su}(2)_F$ flavor symmetry. In this language, the R-symmetry algebra of the $\mathcal{N}=4$ SCFT splits as
	\begin{equation}\label{eq:su4split}
		\mathfrak{su}(4)_R \to \mathfrak{su}(2)_R \times \mathfrak{su}(2)_F\times \mathfrak{u}(1)_R \, .
	\end{equation}
	The $\mathcal{N}=4$ stress-tensor multiplet splits into several $\mathcal{N}=2$ SUSY multiplets under this decomposition. This includes the $\mathcal{N}=2$ stress-tensor multiplet, the flavor current multiplet, and the chiral (and anti-chiral) multiplets, which contain the squashing, mass, and $\tau$ (and $\bar{\tau}$) deformation operators discussed in \eqref{eq:squashcoupl}, \eqref{eq:masscoupl}, and \eqref{eq:taupole} respectively. While the $\mathcal{N}=2$ stress-tensor, chiral, and anti-chiral multiplets are in the singlet representation of $\mathfrak{su}(2)_F$ flavor symmetry, the flavor multiplet is in the triplet representation.

The decomposition of the $\mathcal{N}=4$ stress-tensor multiplet into these submultiplets can be done by first identifying the superprimaries of the $\mathcal{N}=2$ chiral, anti-chiral, flavor current, and stress-tensor multiplets as particular components of the $\mathcal{N}=4$ stress-tensor superprimary. Following this, we can compare the $\mathcal{N}=4$ SUSY transformations of the stress-tensor multiplet to the $\mathcal{N}=2$ SUSY transformations of the submultiplets that it decomposes into, and identify linear combinations of the $\mathcal{N}=4$ operators with operators in the $\mathcal{N}=2$ multiplets. We provide more details on this computation in Appendix~\ref{sec:N4toN2decomp}, where we also fix the normalization factors in this decomposition. In this section, we summarize the results by giving the components of $\mathcal{N}=4$ operators corresponding to the various operators discussed in Section~\ref{sec:background}.
	
	We choose the action of $\mathfrak{su}(2)_R$, $\mathfrak{su}(2)_F$ and $\mathfrak{u}(1)_R$  on the fundamental representation of $\mathfrak{su}(4)_R$ to be given by the following $4\times4$ matrices $\tensor{(T_{\mathbf{4}})}{^A_B}$:
	\es{eq:gensu2bas4}{
		(T_{\mathbf{4}})^{i}_{\mathfrak{su}(2)_R} = \begin{pmatrix} \sigma^i & 0 \\	0 & 0 \\\end{pmatrix}\,, \quad 	(T_{\mathbf{4}})^{i}_{\mathfrak{su}(2)_F} = \begin{pmatrix} 0 & 0 \\	0 & \sigma^i \\\end{pmatrix}\,, \quad (T_{\mathbf{4}})_{\mathfrak{u}(1)_R} = \frac{1}{2}\begin{pmatrix} -\mathbbm{1} & 0 \\	0 & \mathbbm{1} \\\end{pmatrix}\,,}
	where $i=1,2,3$ and $A,B$ are $\mathfrak{su}(4)_R$ fundamental indices. Choosing the basis for the $\mathbf{6}$ representation such that the Clebsch-Gordan coefficients for the anti-symmetric product $(\mathbf{4}\otimes \mathbf{4})_a = \mathbf{6}$ are
	\es{eq:clebsch644}{C^{I}_{AB} =\Biggl\{ \begin{pmatrix} 0 & \frac{\sigma^1}{2} \\	-\frac{\sigma^1}{2} & 0 \\\end{pmatrix} \, , \begin{pmatrix} 0 & -\frac{\sigma^3}{2} \\	\frac{\sigma^3}{2} & 0 \\\end{pmatrix}\, , \begin{pmatrix} 0 & \frac{i}{2} \mathbbm{1} \\	-\frac{i}{2}\mathbbm{1} & 0 \\\end{pmatrix} \, ,  \begin{pmatrix} 0 &  -\frac{\sigma^2}{2} \\	-\frac{\sigma^2}{2} & 0 \\\end{pmatrix} \, ,  \begin{pmatrix}  i \frac{\sigma^2}{2} & 0 \\	0 & i\frac{\sigma^2}{2}  \\ \end{pmatrix} \, , \begin{pmatrix}  \frac{\sigma^2}{2} & 0 \\	0 & -\frac{\sigma^2}{2}\\\end{pmatrix} \,  \Biggr\},}
	where $I$ is a $\mathbf{6}$ index, the action of the $\mathfrak{su}(2)_R\times\mathfrak{su}(2)_F\times \mathfrak{u}(1)_R$ generators in \eqref{eq:gensu2bas4} on the $\mathbf{6}$ representation of $\mathfrak{su}(4)_R$ is given by the following matrices\footnote{These matrices are obtained by using the condition \eqref{eq:gen6rep}.} $\tensor{(T_{\mathbf{6}})}{^I_J}$:
	\es{eq:gensu2bas6}{	(T_{\mathbf{6}})^i_{\mathfrak{su}(2)_R} &= \left\{ \begin{pmatrix} 0 & -i \mathbbm{1} & 0  \\	i \mathbbm{1} & 0 & 0\\ 0 & 0 & 0\\\end{pmatrix} ,\begin{pmatrix} \sigma^2 & 0 & 0  \\	0 & -\sigma^2 & 0\\ 0 & 0 & 0\\\end{pmatrix} , \begin{pmatrix} 0 &   \sigma^2 & 0  \\	 \sigma^2 & 0 & 0\\ 0 & 0 & 0\\\end{pmatrix} \right\}\, , \\
		(T_{\mathbf{6}})^i_{\mathfrak{su}(2)_F} &= \left\{ \begin{pmatrix} 0 & -i \sigma^3 & 0  \\	i \sigma^3 & 0 & 0\\ 0 & 0 & 0\\\end{pmatrix} ,\begin{pmatrix} \sigma^2 & 0 & 0  \\	0 & \sigma^2 & 0\\ 0 & 0 & 0\\\end{pmatrix} , \begin{pmatrix} 0 &  i \sigma^1 & 0  \\	-i \sigma^1 & 0 & 0\\ 0 & 0 & 0\\\end{pmatrix} \right\}\, , \\
		(T_{\mathbf{6}})_{\mathfrak{u}(1)_R} &=  \begin{pmatrix} 0 & 0 & 0  \\	0 & 0 & 0\\ 0 & 0 & \sigma^2 \\\end{pmatrix} \,.
	}
	
	The dimension-two superprimary of the $\mathcal{N}=4$ stress-tensor multiplet in the $\mathbf{20}'$ representation of $\mathfrak{su}(4)_R$ can be represented as the rank-two traceless symmetric tensor $S_{IJ}$, where $I,J$ are $\mathbf{6}$ indices. The superprimary $S_{IJ}$ contains several $\mathcal{N}=2$ operators which can be identified with the various subrepresentations that appear in the decomposition of the $\mathbf{20}'$ representation under the split \eqref{eq:su4split}: 
	\begin{equation}
		\begin{aligned}\label{eq:Ssplit}
			{\begin{array}{c@{ \to\ }c@{\ \oplus\ }c@{\ \oplus\ }c@{\ \oplus\ }c  }
					\mathbf{20}^{'}\,  & 
					(\mathbf{3},\mathbf{3})_0 \,
					& (\mathbf{1},\mathbf{1})_0 \, 
					& (\mathbf{2},\mathbf{2})_{\pm 1}\,
					& (\mathbf{1},\mathbf{1})_{\pm 2} \,,
					\\
					S_{IJ\,} \,
					& J^{(mn)}_{(ij)} \,
					&  \Phi \,
					&   \{W^m_i, \overline{W}^m_i\} \,
					&  \{\mathcal{A},\bar{\mathcal{A}}\}\,,
			\end{array}}
		\end{aligned}
	\end{equation}
	where $i, j$ are $\mathfrak{su}(2)_R$ labels, $m,n$ are $\mathfrak{su}(2)_F$ labels, and for the subrepresentation $(R_1,R_2)_{r}$, $R_1$ is the $\mathfrak{su}(2)_R$ representation, $R_2$ is the $\mathfrak{su}(2)_F$ representation, and $r$ is the $\mathfrak{u}(1)_R$ charge. The primaries of the $\mathcal{N}=2$ chiral and antichiral multiplets (which give the $\tau$ and $\bar{\tau}$ deformations described in \eqref{eq:taupole}), namely $\mathcal{A}$ and $\bar{\mathcal{A}}$ respectively, correspond to the $(\mathbf{1},\mathbf{1})_{2}$ and $(\mathbf{1},\mathbf{1})_{-2}$ subrepresentations in \eqref{eq:Ssplit} respectively. With $S_{IJ}$ normalized as 
	\es{eq:Snorm}{
		\braket{S_{IJ}(x_1)S_{KL}(x_2)} &= \frac{1}{2x^4_{12}} \left(\delta_{IK}\delta_{JL}+\delta_{IL}\delta_{JK} - \frac{1}{3}\delta_{IJ}\delta_{KL}\right)\, ,\\
	}
	these two operators correspond to the $S_{IJ}$ components 
	\es{eq:AAbarcomps}{
		\mathcal{A} &= \frac{\sqrt{c}}{2\sqrt{2}\tau_2}(S_{55} + 2 i S_{56} -S_{66})\, ,\\
		\bar{\mathcal{A}} &= \frac{\sqrt{c}}{2\sqrt{2}\tau_2}(S_{55} - 2i S_{56} -S_{66})\, ,\\}
	which gives the two-point function
	\es{eq:AAbartwoptfunc}{
		\braket{\mathcal{A}(x_1)\bar{\mathcal{A}}(x_2)} &= \frac{c}{2x^4_{12}\tau^2_2}\, ,\\}
	normalized to match the conventions in \cite{Chester:2020vyz}.
	
	The $\mathcal{N}=2$ flavor multiplet operator $J_{12}$, which is a part of the mass deformation in \eqref{eq:masscoupl}, lies in the $(\mathbf{3},\mathbf{3})_0$ subrepresentation in the decomposition of $S_{IJ}$ in \eqref{eq:Ssplit}. It is given by
	\es{eq:J12comps}{
		J_{12} &=  \frac{i\sqrt{c}}{4\sqrt{2}\pi^2}(S_{11}-  S_{22} - S_{33} +S_{44})\, ;
	}
	note that we have chosen this combination of components so that the mass deformation lies in the Cartan of $\mathfrak{su}(2)_F$ generated by $T^3_{\mathfrak{su}(2)_F}$.
	
	The dimension-three scalar $P$ in the $\mathcal{N}=4$ stress-tensor multiplet, which is in the $\overline{\mathbf{10}}$ representation of $\mathfrak{su}(4)_R$, can be written as a rank-two symmetric tensor $P_{(AB)}$, where $A,B$ are $\overline{\mathbf{4}}$ indices. This representation splits as
	\begin{equation}
		\begin{aligned}\label{eq:Psplit}
			{\begin{array}{c@{\ \to\ }c@{\ \oplus\ }c@{\ \oplus\ }c  }
					\overline{\mathbf{10}}\,  & 
					(\mathbf{1},\mathbf{3})_{-1}  \,
					& (\mathbf{3},\mathbf{1})_{1} \,
					& (\mathbf{2},\mathbf{2})_{0}\,
					\\
					P_{(AB)} \,
					& K^{(mn)} \,
					&  N_{(ij)} \, 
					& L^m_i  \,
			\end{array}}
		\end{aligned}
	\end{equation}
	under the decomposition in \eqref{eq:su4split}. Similarly, the $\bar{P}$ in the $\mathbf{10}$ representation, represented as rank-two symmetric tensor $\bar{P}^{(AB)}$, where $A,B$ are $\mathbf{4}$ indices, decomposes as
	\begin{equation}
		\begin{aligned}\label{eq:Pbarsplit}
			{\begin{array}{c@{\ \to\ }c@{\ \oplus\ }c@{\ \oplus\ }c  }
					\mathbf{10}\,  & 
					(\mathbf{1},\mathbf{3})_{1}  \,
					& (\mathbf{3},\mathbf{1})_{-1} \,
					& (\mathbf{2},\mathbf{2})_{0}\,
					\\
					\bar{P}^{(AB)} \,
					& \overline{K}^{(mn)} \,
					&  \overline{N}_{(ij)} \, 
					& \bar{L}^m_i  \, .
			\end{array}}
		\end{aligned}
	\end{equation}
	The dimension-three operators $K$ and $\overline{K}$ in the flavor multiplet, which are part of the mass deformation in \eqref{eq:masscoupl}, belong to the $(\mathbf{1},\mathbf{3})_{-1}$ and $(\mathbf{1},\mathbf{3})_{1}$ subrepresentations in \eqref{eq:Psplit} and \eqref{eq:Pbarsplit} respectively. Using the fact that the mass deformation lies in the Cartan generated by $T^3_{\mathfrak{su}{(2)}_F}$, we can identify $K$ and $\overline{K}$ as the components
	\es{eq:KKbarcomps}{K &= \frac{2\sqrt{c}}{\pi^2}P_{34}\, ,\\
		\overline{K} &= \frac{2\sqrt{c}}{\pi^2}\bar{P}^{34}\, ,\\ }
	where the normalization of $P_{AB}$ and $\bar{P}^{AB}$ is given by the two-point function 
	\es{eq:PP2ptfunc}{\braket{P_{AB}(x_1)\bar{P}^{CD}(x_2)} &= \frac{1}{2x^6_{12}}\left(\tensor{\delta}{_{A}^{C}}\tensor{\delta}{_{B}^{D}} +\tensor{\delta}{_{A}^{D}}\tensor{\delta}{_{B}^{C}} \right)\, .\\}
	
	The anti-self-dual operator  $B^{-}_{I\, \mu\nu}$ in the $\mathcal{N}=4$ stress-tensor multiplet lies in the $\mathbf{6}$ representation of $\mathfrak{su}(4)_R$. Under the decomposition \eqref{eq:su4split}, this representation splits as
	\begin{equation}\label{eq:Bsplit}
		\begin{aligned}
			{\begin{array}{c@{\ \to\ }c@{\ \oplus\ }c @{\ \oplus\ }c }
					\mathbf{6} \, 
					& (\mathbf{1},\mathbf{1})_{-1} \,
					& (\mathbf{1},\mathbf{1})_{1} \,
					&  (\mathbf{2},\mathbf{2})_{0}  \, ,
					\\
					B^{-}_{I\, \mu\nu} \,
					& Z^{-}_{\mu\nu} \,
					& G^{-}_{\mu\nu} \,
					&  H^{m\, -}_{i\,\mu\nu} \, .
			\end{array}}
		\end{aligned}
	\end{equation}
	The anti-self-dual operator $Z^{-}_{\mu\nu}$ in the $\mathcal{N}=2$ stress-tensor multiplet, which is a part of the squashing deformation \eqref{eq:squashcoupl}, lies in the $(\mathbf{1},\mathbf{1})_{-1}$ subrepresentation. Likewise, the self-dual operator  $B^{+}_{I \mu\nu}$ of the $\mathcal{N}=4$ stress-tensor multiplet, which is in the $\mathbf{6}$ representation of $\mathfrak{su}(4)_R$, splits as
	\begin{equation}\label{eq:bBsplit}
		\begin{aligned}
			{\begin{array}{c@{\ \to\ }c@{\ \oplus\ }c @{\ \oplus\ }c }
					\mathbf{6} \, 
					& (\mathbf{1},\mathbf{1})_{1} \,
					& (\mathbf{1},\mathbf{1})_{-1} \,
					&  (\mathbf{2},\mathbf{2})_{0}  \, ,
					\\
					B^{+}_{I\, \mu\nu} \,
					& Z^{+}_{\mu\nu} \,
					& G^{+}_{\mu\nu} \,
					&  H^{m\, +}_{i\,\mu\nu} \, .
			\end{array}}
		\end{aligned}
	\end{equation}
	The $Z^{+}_{\mu\nu}$ in the $\mathcal{N}=2$ stress-tensor multiplet, which is also a part of the squashing deformation, is identified as the $(\mathbf{1},\mathbf{1})_{1}$ subrepresentation. 
	
	With the normalization
	\es{eq:BB2ptfunc}{\braket{B^{-}_{I\,\mu\nu}(x_1)B^{+}_{J\,cd}(x_2)} = \frac{\delta_{IJ}}{8x^8_{12}}\mathcal{S}^{\{(2,0),(0,2)\}}_{[\mu\nu][cd]}(x_{12})  \, , }
	where the tensor $\mathcal{S}^{\{(2,0),(0,2)\}}_{[\mu\nu][cd]}(x)$ is
	\es{eq:defSDtensor}{\mathcal{S}^{\{(2,0),(0,2)\}}_{[\mu\nu][\rho \sigma]}(x) &= x^2(\delta_{\mu \rho}\delta_{\nu\sigma}-\delta_{\mu \sigma}\delta_{\nu\rho}) +2 (x_{\mu} \epsilon_{\nu\rho \sigma \tau}x^\tau-x_{\nu} \epsilon_{\mu \rho \sigma \tau}x^\tau)  +x^2\epsilon_{\mu \nu\rho \sigma}\\
		&-2(x_{\mu}x_{\rho}\delta_{\nu\sigma}-x_{\nu}x_{\rho}\delta_{\mu \sigma}-x_{\mu}x_{\sigma}\delta_{\nu\rho}+x_{\nu}x_{\sigma}\delta_{\mu\rho})\, ,}
	the $\mathcal{N}=2$ operators $Z^{-}_{\mu\nu}$ and $Z^{+}_{\mu\nu}$ are given by the $\mathcal{N}=4$ operators 
	\es{eq:ZZbarcomps}{Z^{+}_{\mu\nu} &= \frac{\sqrt{3c}}{2\sqrt{2}\pi^2}(B^{+}_{\mu\nu \, 6} - i B^{+}_{\mu\nu \, 5}) \, ,\\
	Z^{-}_{\mu\nu} &= \frac{\sqrt{3c}}{2\sqrt{2}\pi^2}(B^{-}_{\mu\nu \, 6} + i B^{-}_{\mu\nu \,5}) \, .}
	
	Finally, the $\mathfrak{su}(4)_R$ symmetry current operator in the $\mathbf{15}$ representation of $\mathfrak{su}(4)_R$, which can be written as a $6\times6$ antisymmetric tensor $J_{[IJ]\,\mu}$, splits as
	\begin{equation}\label{eq:Jsplit}
		\begin{aligned}
			{\begin{array}{c@{\ \to\ }c@{\ \oplus\ }c @{\ \oplus\ }c @{\ \oplus\ }c }
					\mathbf{15} \, 
					& (\mathbf{3},\mathbf{1})_{0} \,
					& (\mathbf{1},\mathbf{3})_{0} \,
					&  (\mathbf{1},\mathbf{1})_{0}  \, 
					&  (\mathbf{2},\mathbf{2})_{\pm 1}  \, ,
					\\
					J_{\mu} \,
					& \tensor{j}{_{\mu\,j}^i} \,
					& \tensor{j}{_{F\,\mu\,m}^n} \,
					&  j_{\mu} \, 
					& \{\tensor{j}{_{\mu\,i}^m},\tensor{\overline{j}}{_{\mu\,i}^m}\}.
			\end{array}}
		\end{aligned}
	\end{equation}
	The dimension-three $\mathfrak{su}(2)_R$ symmetry current operator $\tensor{j}{_{\mu\,j}^i}$ in the $\mathcal{N}=2$ stress-tensor multiplet, given in Table \ref{tab:STMultTable}, can be identified as the $(\mathbf{3},\mathbf{1})_{0}$ subrepresentation. The $\tensor{j}{_{\mu\,1}^1}$ operator in the squashing deformation \eqref{eq:squashcoupl} can be identified as 
	\es{eq:jdiagcomps}{\tensor{j}{_{\mu\,1}^1} = \frac{i\sqrt{3c}}{\pi^2}(J_{23\,\mu} -J_{14\,\mu})\,,}
	where $J_{[IJ]\mu}$ is normalized as
	\es{eq:Jnorm}{\braket{J_{[IJ]\mu}(x_1)J_{[KL]\nu}(x_2)} = (\delta_{IK}\delta_{JL}-\delta_{IL}\delta_{JK})\frac{(x^2_{12}\delta_{\mu\nu}-2x_{12\, \mu}x_{12\, \nu})}{4x^4_{12}}\, .}

	\section{Revisiting non-squashed integral constraints}
	\label{sec:nosquash}
	
	In this section, we revisit the $\mmttZ$ and $\mmmmZ$ integral constraints that were previously computed in \cite{Binder:2019jwn} and \cite{Chester:2020dja}. We rederive these integral constraints using a different method in order to illustrate the approach we will use for computing integral constraints with squashing.
	
	\subsection{\texorpdfstring{$\mmttZ$}{Mass² tau² derivative}}
	\label{sec:mmtt}

	For an $\mathcal{N}=2$ SCFT, the integrated correlator $\mmttZ$ is given by the integral
	\es{eq:mmtt1}{\mmttZ &= \int d^4 x_3 \, d^4 x_4  \, \sqrt{g(x_3)} \sqrt{g(x_4)} \,\braket{\mathcal{A}(N) \,\bar{\mathcal{A}}(S)\,\mathcal{O}_m(x_3)\,\mathcal{O}_m(x_4)}_{S^4} \, ,\\}
	where $\mathcal{O}_m \equiv i(K - \overline{K}) + 2J_{12}$ is the mass deformation operator defined in \eqref{eq:masscoupl}, and $\mathcal{A}$ and $\bar{\mathcal{A}}$ are the chiral and antichiral primary operators whose insertions at the poles correspond to the $\tau$ deformation, as mentioned in \eqref{eq:taupole}. Note that $\sqrt{g(x)} = \Omega(x)^4$, where $\Omega(x) = \frac{2}{1+x^2}$ is the conformal factor relating $S^4$ to $\mathbb{R}^4$ \eqref{eq:S4metric}, and correlators on $S^4$ are related to flat space correlators by a conformal transformation \eqref{eq:R4toS4corr}.
	
	In a generic $\mathcal{N}=2$ SCFT, the integral in \eqref{eq:mmtt1} receives contributions from two four-point correlators: $\braket{\mathcal{A}(N)\bar{\mathcal{A}}(S)J_{12}(x_3)J_{12}(x_4)}$ and $ \braket{\mathcal{A}(N)\bar{\mathcal{A}}(S)K(x_3)\overline{K}(x_4)}$. As discussed in the previous section, we are computing integral constraints for theories with $\mathcal{N}=4$ superconformal symmetry in this paper. In this case, the mass and $\tau$ deformation operators correspond to particular components of operators appearing in the stress-tensor multiplet of the $\mathcal{N}=4$ theory, as given in \eqref{eq:AAbarcomps}, \eqref{eq:J12comps} and \eqref{eq:KKbarcomps}, and we can use $\mathcal{N}=4$ SUSY Ward identities to relate these correlators to the four-point function of the superprimary of the $\mathcal{N}=4$ stress-tensor multiplet, given in \eqref{eq:gssssform}. This superprimary four-point correlator contains two types of terms: those that depend on the reduced correlator of the $\mathcal{N}=4$ theory $\mathcal{T}(u,v)$, and a free theory contribution. The integrated correlator given in \eqref{eq:mmtt1} can hence be separated into two parts,
	\es{eq:mmtt2}{\mmttZ = 	\left(\mmttZ\right)_{\mathcal{T}} + \left(\mmttZ\right)_{\text{free}} \, , }
	where $\left(\mmttZ\right)_{\mathcal{T}}$ contains only the $\mathcal{T}(u,v)$-dependent part of the four-point correlation function given in \eqref{eq:bbtautau1}, while $\left(\mmttZ\right)_{\text{free}}$ contains the free theory part of the four-point correlator. The term $\left(\mmttZ\right)_{\text{free}}$ vanishes \cite{Binder:2019jwn}, because in a free ${\cal N} = 4$ theory the vector multiplet (for which $\tau$ is the complexified gauge coupling) decouples from the hypermultiplet (for which $m$ is the mass parameter). We therefore focus on recomputing just $\left(\mmttZ\right)_{\mathcal{T}}$ in this subsection. This term is given by the integral
	\es{eq:mmttT}{ 
		&\left(\mmttZ\right)_{\mathcal{T}} \\
		=& \int d^4 x_3 \, d^4 x_4 \,\bigg[ 4\Omega(x_3)^2 \Omega(x_4)^2 \braket{\mathcal{A}(N)\bar{\mathcal{A}}(S)J_{12}(x_3)J_{12}(x_4)}_{\mathcal{T},\mathbb{R}^4}  \\
		&{}  +  \Omega(x_3) \Omega(x_4) \left(\braket{\mathcal{A}(N)\bar{\mathcal{A}}(S)K(x_3)\overline{K}(x_4)}_{\mathcal{T},\mathbb{R}^4}+ \braket{\mathcal{A}(N)\bar{\mathcal{A}}(S)\overline{K}(x_3)K(x_4)}_{\mathcal{T},\mathbb{R}^4}\right) \bigg] \, .}
	The $\mathcal{T}(u,v)$-dependent parts of the flat space correlators appearing in this integral can be written as differential operators acting on $\mathcal{T}(u,v)$:
	\es{eq:corrmmttsp}{
		\braket{\mathcal{A}(N)\bar{\mathcal{A}}(S)J_{12}(x_3)J_{12}(x_4)}_{\mathcal{T},\mathbb{R}^4}&=  -c^2 \frac{v\mathcal{T}(u,v)}{16\pi^4 \tau^2_2 x^4_{34}} \, , \\
		\braket{\mathcal{A}(N)\bar{\mathcal{A}}(S)K(x_3)\overline{K}(x_4)}_{\mathcal{T},\mathbb{R}^4} &= 
		   c^2\Box_4\left[ \frac{ v^2 \mathcal{T}(u,v)}{8\pi^4 \tau^2_2 x^4_{34}} \right]\, , \\ 
		\braket{\mathcal{A}(N)\bar{\mathcal{A}}(S)\overline{K}(x_4)K(x_3)}_{\mathcal{T},\mathbb{R}^4} &= 
		   c^2 \Box_3\left[ \frac{ \mathcal{T}(u,v)}{8\pi^4 \tau^2_2 x^4_{34}} \right]\, , \\
	}	
	where the conformal cross-ratios, defined in \eqref{eq:defuv}, are now given by $u=\frac{x^2_{34}}{x^2_3}$ and $v=\frac{x^2_4}{x^2_3}$, since $x_1$ and $x_2$ are set to the origin and infinity, respectively. The expressions in \eqref{eq:corrmmttsp} are obtained by taking the appropriate limits of the correlators in \eqref{eq:corrmmtt}. The second and third expressions in \eqref{eq:corrmmttsp}  are related to each other by $(x_3 \leftrightarrow x_4)$ crossing, as can be checked explicitly using the crossing relation $\mathcal{T}\left(\frac{u}{v},\frac{1}{v}\right) = v^2\mathcal{T}(u,v)$, given in \eqref{eq:Tcrossing}.
	
	Plugging the expressions in \eqref{eq:corrmmttsp} into the integral in \eqref{eq:mmttT} and integrating by parts, we get
	\es{eq:mmttT2}{& \left(\mmttZ\right)_{\mathcal{T}} \\
		=& {}- \int d^4 x_3 \, d^4 x_4 \left[\Omega(x_3)^2\Omega(x_4)^2 v  +  \Omega(x_3)\Omega(x_4)^3 v^2   +  \Omega(x_3)^3\Omega(x_4)   \right] \frac{c^2 \mathcal{T}(u,v)}{4\pi^4 \tau^2_2 x^4_{34}} \, .}
	We can then use $\SO(4)$ rotations to fix $\vecx_3 = z(1,0,0,0)$ and $ \vecx_4 = zr(\cos\theta, \sin\theta, 0, 0)$. In this configuration (along with $\vecx_1 = 0, \vecx_2 = \infty$), the conformal cross-ratios $u$ and $v$ are simple functions of the variables $r$ and $\theta$,
	\es{eq:uvinrtheta}{
		(u,v) = \left(\frac{x^2_{34}}{x^2_3 },\frac{x^2_{4}}{x^2_3 }\right) = \left(1+r^2-2r\cos\theta,r^2\right) \, .
	} 
The integrand in \eqref{eq:mmttT2} is a total $z$-derivative after this change of variables and can be integrated easily, leaving a boundary term contribution coming from the $z=0$ boundary. The final expression is 
	\es{eq:mmttfinal}{\left(\mmttZ\right)_{\mathcal{T}} =  - \frac{8c^2}{\pi \tau^2_2}\int   dr \, d\theta \, \frac{r^3 \sin^2\theta}{u^2} \, \mathcal{T}(u,v) \, . }
	This is the same expression that was found in  \cite{Binder:2019jwn, Chester:2020dja} (see also \eqref{eq:knownintcorr}). Compared to \cite{Binder:2019jwn}, our approach to computing the $\mmttZ$ integrated correlator is simpler since we use the compact expressions \eqref{eq:corrmmttsp} for the correlators contributing to $\mmttZ$. Additionally, we directly obtain a crossing-symmetric form for $\mmttZ$ since we start with a manifestly crossing-symmetric integrand in \eqref{eq:mmttT}. This was not the case for the computation in \cite{Binder:2019jwn}, and the result was later simplified  in \cite{Chester:2020dja} using crossing symmetry.

	\subsection{\texorpdfstring{$\mmmmZ$}{Mass⁴ derivative}}
	\label{sec:m4}
	
	The integrated correlator $\mmmmZ$ in an $\mathcal{N}=2$ SCFT is given by the integral
	\es{eq:m4p1}{\mmmmZ = & \int  \prod_{i=1}^4\, d^4 x_i\, \sqrt{g(x_i)}  \, \braket{\mathcal{O}_m(x_1) \mathcal{O}_m(x_2)\mathcal{O}_m(x_3)\mathcal{O}_m(x_4)}_{S^4} \, . \\ }
	where $\mathcal{O}_m$ is given in \eqref{eq:masscoupl}. Since we are computing integral constraints in an $\mathcal{N}=4$ SCFT, the four-point function in the integral above can be related to the stress-tensor superprimary four-point correlator, given in \eqref{eq:gssssform}, using $\mathcal{N}=4$ SUSY Ward identities. The particular components of $\mathcal{N}=4$ stress-tensor multiplet operators that correspond to $\mathcal{O}_m$ are given in \eqref{eq:J12comps} and \eqref{eq:KKbarcomps}. As was the case for the $\mmttZ$ integrated correlator in the previous subsection, we can separate out the integrated correlator in \eqref{eq:m4p1} into two parts 
	\es{eq:m4split}{\mmmmZ = \left(\mmmmZ\right)_{\mathcal{T}} + \left(\mmmmZ\right)_{\text{free}}\,,} 
	where $\left(\mmmmZ\right)_{\mathcal{T}}$ is the contribution of the purely $\mathcal{T}(u,v)$-dependent part of the four-point correlator in \eqref{eq:m4p1}, given by the integral
	\es{eq:m4p2}{\mmmmZT &= \int  \prod_{i=1}^4\, d^4 x_i\, \sqrt{g(x_i)}  \, \braket{\mathcal{O}_m(x_1) \mathcal{O}_m(x_2)\mathcal{O}_m(x_3)\mathcal{O}_m(x_4)}_{\mathcal{T},S^4}\, ,\\ }
	while $\left(\mmmmZ\right)_{\text{free}}$ is the contribution from the free theory part of the four-point correlator in \eqref{eq:m4p1}.  The term $\left(\mmmmZ\right)_{\text{free}}$ was computed in \cite{Chester:2020dja} to be 
	\es{eq:m4p3}{\left(\mmmmZ\right)_{\text{free}} = 48 c \,\zeta(3)\, . }
	We focus on rederiving $\left(\mmmmZ\right)_{\mathcal{T}}$ in this subsection.

	\subsubsection{Integration by parts}
	The $\mathcal{T}(u,v)$-dependent parts of the four-point correlators that contribute to the integral \eqref{eq:m4p2} are given in \eqref{eq:corrmmmm}. Substituting these correlators into \eqref{eq:m4p2} and integrating by parts, we have
	\es{eq:m4p4}{\mmmmZT &= \int \left( \prod_{i=1}^4\, d^4 \Omega^i_{S^4}\right) \, H_{m^4}(y_1,y_2,y_3,y_4) \frac{ \mathcal{T}(u,v)}{y^4_{12}y^4_{34}y^4_{13}y^4_{24} }   \, .\\ }
	where $d^4 \Omega^i_{S^4}\equiv d^5 y_i\, \delta (y^2_i-1) $ is the measure on $S^4$, $y^2_{ij} = \Omega(x_i)\Omega(x_j)x^2_{ij}$ is the square of the chordal distance between the points $y_i$ and $y_j$ on the unit sphere in $\mathbb{R}^5$,  the conformal cross-ratios $u$ and $v$, defined in \eqref{eq:defuv}, can be written in terms of the unit vectors on $\mathbb{R}^5$ as
	\es{eq:uvR5}{u = \frac{y^2_{12}y^2_{34}}{y^2_{13}y^2_{24}} \, , \qquad v = \frac{y^2_{14}y^2_{23}}{y^2_{13}y^2_{24}}\, ,} 
	and the factor $H_{m^4}(y_1,y_2,y_3,y_4)$ in \eqref{eq:m4p4} is given by the polynomial
	\es{eq:defHm4}{H_{m^4}&= \frac{ c^2 }{16\pi^8} (y^2_{12}y^2_{34}+y^2_{13}y^2_{24}+y^2_{14}y^2_{23 })  \\
		&\times  \left[y^2_{12}\left( \frac{1}{2}y^2_{12} +y^2_{13} +\frac{1}{8}y^2_{34} \right) + (\text{23 other permutations of\,} \{y_1,y_2,y_3,y_4\} )\right] \, .}
	Using the crossing relations given in \eqref{eq:Tcrossing}, one can check that the factor $\frac{\mathcal{T}(u,v)}{ y^4_{12}y^4_{34}y^4_{13}y^4_{24}}$ appearing in \eqref{eq:m4p4} is crossing-invariant. Since $H_{m^4}$ is invariant under any permutation of the four points $\{y_1,y_2,y_3,y_4\}$ on $S^4$, we observe that the expression \eqref{eq:m4p4} is invariant under crossing, as is expected from the integral \eqref{eq:m4p2}. 
	
	\subsubsection{Converting to $D$-functions in 6d}
	\label{sec:embed6d}
	
	The integral \eqref{eq:m4p4} can be further simplified into an integral over just the conformal cross-ratios $u$ and $v$, just like the $\mmttZ$ integrated correlator in \eqref{eq:mmttfinal}.  In this case, it is convenient to perform the simplification by embedding $S^4$ on the null lightcone in six dimensions. Since the conformal symmetry of $S^4$ is $\SO(5,1)$, this embedding makes conformal symmetry manifest.
	
	We will now write \eqref{eq:m4p4} in terms of coordinates in $\mathbb{R}^{5,1}$ with signature $( + + + + +\, -)$. A point $y$ on the unit sphere in $\mathbb{R}^5$ can be associated with the lightray $\mathbb{L}_{\vecy}$ defined as
	\es{eq:embed}{\mathbb{L}_{\vecy} \equiv \left\{\mathbf{X} = \left(\vec{X},X^{6}\right): \vec{X} = X^6 \vecy \right\}\, .}
	The $\SO(5)$-invariant $y^2_{ij}$ can be written in terms of $\mathbb{R}^{5,1}$ vectors as 
	\es{eq:invs6d}{y^2_{ij} = \frac{-2 \bold{X}_i\cdot \bold{X}_j}{(\bold{X}_i\cdot \bold{Y}_*)(\bold{X}_j\cdot \bold{Y}_*)}\, ,}
for any $\mathbf{X}_k \in \mathbb{L}_{\vecy_k}$, where $\mathbf{Y}_* = (0,0,0,0,0,1)$ is a future-directed time-like vector in $\mathbb{R}^{5,1}$. Note that \eqref{eq:invs6d} is the same no matter which $\mathbf{X}_j \in \mathbb{L}_{\vecy_j}$ is chosen to represent the lightray as it is invariant under rescalings along each of the rays: $\bold{X}_k \to \lambda_k\bold{X}_k$.  We can use \eqref{eq:invs6d} to write the integrand in \eqref{eq:m4p4} as a function in $\mathbb{R}^{5,1}$. Using \eqref{eq:invs6d}, the conformal cross-ratios, given in terms of the points $y_i$ on $S^4$  in \eqref{eq:uvR5}, can be written in terms of the 6d vectors $\bold{X}_i\in \mathbb{L}_{\vecy_j}$ as
	\es{eq:uv6d}{u = \frac{(\bold{X}_1\cdot \bold{X}_2) \, (\bold{X}_3\cdot \bold{X}_4)}{(\bold{X}_1\cdot \bold{X}_3) \,(\bold{X}_2\cdot \bold{X}_4)}\, ,  \qquad   v= \frac{(\bold{X}_1\cdot \bold{X}_4 )\, (\bold{X}_2\cdot \bold{X}_3)}{(\bold{X}_1\cdot \bold{X}_3 )\, (\bold{X}_2\cdot \bold{X}_4)} \, .}
	Note once again that these expressions are invariant under rescalings $\bold{X}_k \to \lambda_k\bold{X}_k$, and are thus independent of the choice of the representative point chosen for each lightray.
	
	Since our objective is to write the integrated correlator $\mmmmZT$ as an integral of a conformally-invariant function, it would be convenient to express \eqref{eq:m4p4} as an integral with a conformally-invariant measure. This is given by the expression
	\es{eq:mmmm5}{\left(\mmmmZ\right)_{\mathcal{T}} =  \int  \frac{\prod^4_{i=1} d^4\Omega^i_{S^4}}{(y^2_{12}y^2_{34})^4} \, F_{m^4}\left(\frac{\bold{X}_i\cdot \bold{X}_j}{(\bold{X}_i\cdot \bold{Y}_*)(\bold{X}_j\cdot \bold{Y}_*)}\right) \mathcal{T}(u,v)\, ,}
	whose measure $\frac{\prod^4_{i=1} d^4\Omega^i_{S^4}}{(y^2_{12}y^2_{34})^4}$ is invariant under conformal transformations.\footnote{In general, a conformal transformation on $S^4$ leaves the metric invariant up to a position-dependent Weyl rescaling factor $\lambda(y)$. Under such a conformal transformation, the numerator of the measure $\prod^4_{i=1} d^4\Omega^i_{S^4}$ goes to $ (\prod^4_{i=1} \lambda^4(y_i)) d^4\Omega^i_{S^4}$ and the factor $y^2_{12}y^2_{34}$ in the denominator goes to $ (\prod^4_{i=1}\lambda_i(y_i))y^2_{12}y^2_{34} $. The $\lambda(y_i)$ factors would cancel each other in $\frac{\prod^4_{i=1} d^4\Omega^i_{S^4}}{(y^2_{12}y^2_{34})^4}$ leaving it conformally invariant.} The function $F_{m^4}$ is related to $H_{m^4}$, defined in \eqref{eq:defHm4}, as
	\es{eq:defFmmmm}{F_{m^4}\left(\frac{\bold{X}_i\cdot \bold{X}_j}{(\bold{X}_i\cdot \bold{Y}_*)(\bold{X}_j\cdot \bold{Y}_*)}\right) &=  \left.\frac{y^4_{12}y^4_{34}}{y^4_{13}y^4_{24}}H_{m^4}(y^2_{ij})\right|_{y^2_{ij}\to \frac{-2 \bold{X}_i\cdot \bold{X}_j}{(\bold{X}_i\cdot\bold{Y}_*)\, (\bold{X}_j\cdot\bold{Y}_*)}} \, ,}
	for $\bold{X}_i \in \mathbb{L}_{\vec{y}_j}$.

	$F_{m^4}$ is not $\SO(5,1)$-invariant since it depends on $\bold{Y}_*$. However, it is invariant under $\SO(5,1)$ transformations acting simultaneously on the four lightrays $\mathbb{L}_{\vec{y}_i}$ as well as the point $\bold{Y}_*$. We can use $\SO(5,1)$ transformations to fix the points $\vec{y}_i$ to a specific configuration on $S^4$, while moving the point $\bold{Y}_*$ all over the locus of points satisfying $\bold{Y}_*\cdot \bold{Y}_* = -1$ and $Y^6_*>0$, also known as the AdS hyperboloid. This converts the integral over four points on $S^4$ in \eqref{eq:mmmm5} into an integral over the AdS hyperboloid as well as the conformal cross-ratios $u$ and $v$. Doing the integral over the AdS hyperboloid leaves us with a conformally-invariant form for \eqref{eq:mmmm5}. Details of this manipulation are provided in Appendix~\ref{sec:AdSavg}\@. Using the result given in \eqref{eq:bbmm10} and \eqref{eq:defFcal}, we are left with a conformally-invariant form for $\mmmmZT$,
	\es{eq:mmmm6}{\mmmmZT =  \frac{4\pi^5}{3}\int dr \, d\theta \, \frac{r^3 \sin^2\theta}{u^4} \mathcal{F}_{m^4}(u,v) \mathcal{T}(u,v)\, ,}
	where $r$ and $\theta$ are related to the conformal cross-ratios $u$ and $v$ via $u = 1+r^2-2r\cos\theta,\, v = r^2$, and the function $\mathcal{F}_{m^4}(u,v)$ is defined in terms of $F_{m^4}$ as
	\es{eq:defFcalmmmm}{ \mathcal{F}_{m^4}(u,v) = \mathcal{J}[F_{m^4}] \, , }
	for the functional $\mathcal{J}$ defined in \eqref{eq:defFcal}. The function $\mathcal{F}_{m^4}(u,v)$ can be expressed in terms of four-point $D_{r_1,r_2,r_3,r_4}$-functions, which are defined as integrals over AdS space in \eqref{eq:Ddefnalt},
	or equivalently, in terms of the conformally-invariant $\bar{D}$-functions, which are related to four-point $D$-functions via \eqref{eq:DtoDbar2}. We find that $\mathcal{F}_{m^4}(u,v)$ is given by the sum of $\bar{D}$-functions
	\es{eq:Fmmmmexpr}{\mathcal{F}_{m^4}(u,v) &= \frac{4c^2 u^2 (1+u+v)}{\pi^6} \big[u^2 \bar{D}_{3,3,1,1} +2 u \bar{D}_{3,2,2,1} +2 u \bar{D}_{3,2,1,2}  +\bar{D}_{3,1,3,1}  +2 \bar{D}_{3,1,2,2}   \\
		&   {}  +\bar{D}_{3,1,1,3} +2 u v \bar{D}_{2,3,2,1} +2 u \bar{D}_{2,3,1,2} +2 v \bar{D}_{2,2,3,1}  +2 (u+v+1) \bar{D}_{2,2,2,2}\\
		& {}  +2 \bar{D}_{2,2,1,3}  +2 \bar{D}_{2,1,3,2} +2 \bar{D}_{2,1,2,3} +v^2 \bar{D}_{1,3,3,1} +2 v \bar{D}_{1,3,2,2}  +\bar{D}_{1,3,1,3}  \\
		& {} +2 v \bar{D}_{1,2,3,2}+2 \bar{D}_{1,2,2,3}  +\bar{D}_{1,1,3,3}  \big] \, . \\ }
	As pointed out in Appendix~\ref{sec:Dbarbasis}, each $\bar{D}$-function appearing in this sum can be expressed in terms of a basis of four $\bar{D}$-functions, which we choose to be $\{\bar{D}_{1,1,1,1},\bar{D}_{1,1,2,2},\bar{D}_{1,2,2,1},\bar{D}_{2,2,2,2}\}$. Substituting in the expressions for each $\bar{D}$-function into \eqref{eq:Fmmmmexpr},  we obtain a simple expression for $\mathcal{F}_{m^4}(u,v)$,
	\es{eq:Fmmmmexpr2}{\mathcal{F}_{m^4}(u,v) & = \frac{24c^2}{\pi^6} u^2 (1+u+v) \bar{D}_{1,1,1,1}(u,v) \, .}

	Plugging this expression into \eqref{eq:mmmm5}, we obtain the final form for the integrated correlator  $\mmmmZT$ 
	\es{eq:mmmm7}{\mmmmZT =  \frac{32c^2}{\pi}\int dr \, d\theta \, \frac{r^3 \sin^2\theta}{u^2} (1+u+v) \bar{D}_{1,1,1,1}(u,v) \mathcal{T}(u,v)\, ,}
	which is the formula previously found in  \cite{Chester:2020dja} (see also \eqref{eq:knownintcorr}).
	
	Note that the computation of this integrated correlator in \cite{Chester:2020dja} had an apparently divergent contribution from the $\mathcal{N}=4$ stress-tensor correlator $\braket{P\bar{P}P\bar{P}}$. In \cite{Chester:2020dja}, this divergent contribution was regulated by deforming the conformal dimensions of one of the $P$ and $\bar{P}$ operators from $\Delta = 3$ to $\Delta = 3-\epsilon$ and changing just the kinematic prefactor of the four-point correlator. It was only after integrating by parts that this contribution was found to be finite in the $\epsilon \to 0$ limit. This deformation was not appropriately justified in \cite{Chester:2020dja}. In this section, we have avoided divergences  by using the compact expressions for the Ward identities in \eqref{eq:corrmmmm} and integrating by parts first. All contributions to $\mmmmZ$ are finite using this approach. 
	
	Additionally, note that in this approach we are using the mass deformation of the form \eqref{eq:masscoupl}, rather than the mass deformation used in \cite{Binder:2019jwn,Chester:2020dja} which is of the form \eqref{DefMass} and has subleading $O(m^2)$ terms. The mass deformation \eqref{eq:masscoupl} corresponds to a particular regularization scheme for which the SUSY Ward identities in Appendix~\ref{sec:corrs} hold even at coincident points. In this scheme, the $K\times \overline{K}$ OPE contains terms that contribute to the integral constraint in the same way that the $\mathcal{O}(m^2)$ terms of \eqref{eq:masscoupl} would via lower point correlators. Either scheme is valid and we get the same integral constraint as in \cite{Binder:2019jwn} and \cite{Chester:2020dja}.

	\section{Integral constraints with squashing}
	\label{sec:squashintcorr}

	In this section, we will derive the main results of this paper: the integral constraints corresponding to the derivatives $\bbttZ$, $\bbmmZ$, and $\bbbbZ$.
	
	\subsection{\texorpdfstring{$\bbttZ$}{b² tau² derivative}}
	\label{sec:bbtt}
	
	We begin with the $\bbttZ$ integral constraint. In a general $\mathcal{N}=2$ SCFT, this constraint is given by
	\es{eq:bbtautau1}{  
		\frac{\bbttZ}{\partial_\tau \partial_{\bar\tau}\log Z} &= \int d^4 x_3 \, d^4 x_4  \, \sqrt{g(x_3)} \sqrt{g(x_4)} \,\braket{\mathcal{A}(N) \,\bar{\mathcal{A}}(S)\,\mathcal{O}_b(x_3)\,\mathcal{O}_b(x_4)}_{S^4}\\
		& + (\text{lower-point contributions})\, , \\}
	where ${\cal O}_b$ is the squashing deformation operator, defined earlier in \eqref{eq:squashcoupl}. Note that we have used \eqref{eq:taupole} to relate the $\tau$ and $\bar{\tau}$ derivatives in \eqref{eq:bbtautau1} to insertions of $\mathcal{A}$ and $\bar{\mathcal{A}}$ at the north and south poles of $S^4$.
	
	In $\mathcal{N}=2$ language, the four-point correlation functions that contribute to this integrated correlator are $\braket{\mathcal{A}\bar{\mathcal{A}}Z^{-}_{\mu\nu}Z^{+}_{\rho\sigma}}$ and $\braket{\mathcal{A}\bar{\mathcal{A}}\tensor{j}{_{\mu\,1}^1} \tensor{j}{_{\nu\,1}^1}}$. For theories with $\mathcal{N}=4$ superconformal symmetry, the squashing deformation operators $\tensor{j}{_{\mu\,1}^1}$, $Z^{-}_{\mu\nu}$ and $Z^{+}_{\mu\nu}$ can be identified as specific components of operators in the $\mathcal{N}=4$ stress-tensor multiplet, given in \eqref{eq:ZZbarcomps} and \eqref{eq:jdiagcomps}. We can then use $\mathcal{N}=4$ SUSY Ward identities to relate the $\braket{\mathcal{A}\bar{\mathcal{A}}Z^{-}_{\mu\nu}Z^{+}_{\rho\sigma}}$ and $\braket{\mathcal{A}\bar{\mathcal{A}}\tensor{j}{_{\mu\,1}^1} \tensor{j}{_{\nu\,1}^1}}$ correlators to the four-point function of the superprimary of the $\mathcal{N}=4$ stress-tensor multiplet, given in \eqref{eq:gssssform}. The integrated correlator given in \eqref{eq:bbtautau1} can hence be separated into two parts,
	\es{eq:bbttpart1}{\bbttZ  = \left(\bbttZ\right)_{\mathcal{T}} + \left(\bbttZ\right)_{\text{free}}\, ,} 
	where $\left(\bbttZ\right)_{\mathcal{T}}$ contains only the $\mathcal{T}(u,v)$-dependent part of the four-point correlation function given in \eqref{eq:bbtautau1}, while $\left(\bbttZ\right)_{\text{free}}$ contains all the remaining terms: the free theory part of the four-point correlator as well as contributions from lower-point correlators. The term $\left(\bbttZ\right)_{\text{free}}$ can be easily computed using the relations given in \eqref{eq:relation1} and \eqref{eq:knownintcorr}, as was pointed out earlier in Section~\ref{WardSolve}. Therefore, we focus on computing just the $\left(\bbttZ\right)_{\mathcal{T}}$ piece in this section. The purely $\mathcal{T}(u,v)$-dependent part of the $\braket{\mathcal{A}\bar{\mathcal{A}}Z^{-}_{\mu\nu}Z^{+}_{\rho\sigma}}$ and $\braket{\mathcal{A}\bar{\mathcal{A}}\tensor{j}{_{\mu\, 1}^1} \tensor{j}{_{\nu\, 1}^1}}$ correlators in $\mathbb{R}^4$ are given by
	\es{eq:bbttcorrNS}{\braket{\mathcal{A}(N)\bar{\mathcal{A}}(S) \tensor{j}{_{\mu\,1}^1}(x_3) \tensor{j}{_{\nu\,1}^1}(x_4)}_{\mathcal{T},\mathbb{R}^4} &= \frac{c^2}{4\pi^4 \tau^2_2}(\delta_{\mu\nu}\partial_{3}^{ \lambda}\partial_{4 \lambda} - \partial_{3 \nu}\partial_{4 \mu} +  \epsilon_{\mu\nu\rho\sigma} \partial_{3}^{ \rho}\partial_{4}^{\sigma})\left[\frac{x^2_4 \mathcal{T}(u,v)}{x^2_3 x^4_{34}}\right] \, ,\\
		\braket{\mathcal{A}(N)\bar{\mathcal{A}}(S) Z^{-}_{\mu_1\nu_1} (x_3) Z^{+}_{\mu_2\nu_2}(x_4)}_{\mathcal{T},\mathbb{R}^4} &= -\frac{c^2}{32\pi^4 \tau^2_2} \tensor{{P^-}}{_{[\mu_1\nu_1]}^{[\mu_3\nu_3]}} \tensor{{P^+}}{_{[\mu_2\nu_2]}^{[\mu_4\nu_4]}} \delta_{\mu_3\mu_4}\partial_{4\nu_3}\partial_{4\nu_4} \left[\frac{x^4_4 \mathcal{T}(u,v)}{x^4_3 x^4_{34}}\right]\, , \\ 
		\braket{\mathcal{A}(N)\bar{\mathcal{A}}(S) Z^{+}_{\mu_1\nu_1} (x_3) Z^{-}_{\mu_2\nu_2}(x_4)}_{\mathcal{T},\mathbb{R}^4} &= -\frac{c^2}{32\pi^4 \tau^2_2} \tensor{{P^+}}{_{[\mu_1\nu_1]}^{[\mu_3\nu_3]}} \tensor{{P^-}}{_{[\mu_2\nu_2]}^{[\mu_4\nu_4]}} \delta_{\mu_3\mu_4}\partial_{3\nu_3}\partial_{3\nu_4} \left[\frac{ \mathcal{T}(u,v)}{x^4_{34}}\right],\\
	}
	where  $\tensor{{P^-}}{_{[\mu\nu]}^{[\rho\sigma]}}$ projects out the anti-self-dual part of a tensor, $\tensor{{P^+}}{_{[\mu\nu]}^{[\rho\sigma]}} $ projects out the self-dual part, and $u$ and $v$, defined in \eqref{eq:defuv}, are now given by $u=\frac{x^2_{34}}{x^2_3}$, $v=\frac{x^2_4}{x^2_3}$, since we have set $x_1$ to the origin and $x_2$ to infinity. These expressions come from taking appropriate limits of the expressions in \eqref{eq:corrbbtt}. The second and third correlators above are equal to each other after the permutation of points $x_3\leftrightarrow x_4$, as can be checked using the crossing relation $\mathcal{T}(\frac{u}{v},\frac{1}{v}) = v^2 \mathcal{T}(u,v)$.  Note that we are once again working in a regularization scheme in which the Ward identities in \eqref{eq:bbttcorrNS} are valid at coincident points, as was the case in Section~\ref{sec:nosquash}. We employ a similar regularization scheme choice for the computation of the remaining squashed integral constraints that appear later on in Sections \ref{sec:bbmm} and \ref{sec:bbbb}.
	
 Substituting \eqref{eq:bbttcorrNS} into \eqref{eq:bbtautau1} and integrating by parts, we obtain
	\es{eq:bbtautau2}{\left(\bbttZ\right)_{\mathcal{T}} &= \int d^4 x_3 \, d^4 x_4 \,   H^{bb\tau\bar{\tau}}(x_3,x_4)\mathcal{T}(u,v)\, , }
	where
	\es{eq:Hbbttdef}{H^{bb\tau\bar{\tau}}&= -\frac{c^2 \, x^2_{4}}{8\pi^4\tau^2_2 x^2_{3} x^4_{34}}\Omega(x_3)^3\Omega(x_4)^3\left(2 + \frac{3}{2} \frac{x^2_{4}\Omega(x_4)}{x^2_{3}\Omega(x_3)}+\frac{3}{2} \frac{x^2_{3}\Omega(x_3)}{x^2_{4}\Omega(x_4)}\right) \, .\\}
	
	This integral is manifestly $\SO(4)$-invariant. Using rotational symmetry, we can do 5 of the 8 remaining integrals easily and fix $x_3$ and $x_4$ to the configuration: $x_3 = z(1,0,0,0),$ and $x_4 = r z (\cos\theta,\sin\theta,0,0)$. 
	Then the integral in \eqref{eq:bbtautau2} is reduced to an integral over $r$, $\theta$, and $z$. We further find that the integrand is a total  $z$-derivative, so the $z$-integral can be carried out easily, leaving behind a boundary term contribution from $z=0$. After this step, \eqref{eq:bbtautau2} reduces to the simple form
	\es{eq:bbtautau3}{\left(\bbttZ\right)_{\mathcal{T}} &=   - \frac{8c^2}{\pi^2\tau^2_2} \int dr \, d\theta \, \frac{r^3 \sin^2\theta}{u^2} \mathcal{T}(u,v)\, ,\\}
	where $u=\frac{x^2_{34}}{x^2_3}= 1+r^2-2r\cos\theta$ and $v=\frac{x^2_{4}}{x^2_3} = r^2$. This is exactly what is expected from the first relation in \eqref{eq:relation1} after substituting in the $\partial^2_\tau \partial^2_{\bar{\tau}} \partial^2_m F$ integral constraint given in \eqref{eq:knownintcorr}. As we noted in the introduction, this means that the integral constraint from $\partial_\tau \partial_{\bar\tau} \partial_b^2 F$ is equivalent to the one from  $\partial_\tau \partial_{\bar\tau} \partial_m^2 F$.

	\subsection{\texorpdfstring{$\bbmmZ$}{b² m² derivative}}
	\label{sec:bbmm} 
	
	The $\bbmmZ$ integral constraint for an $\mathcal{N}=2$ SCFT is given by
	\es{eq:bbmm1}{\bbmmZ &= \int  \prod_{i=1}^4\, d^4 x_i\, \sqrt{g(x_i)}  \, \braket{\mathcal{O}_m(x_1) \mathcal{O}_m(x_2)\mathcal{O}_b(x_3)\mathcal{O}_b(x_4)}_{S^4}\\
		& + \text{(lower-point contributions)} \, , \\ }
	where $\mathcal{O}_m$ is the mass deformation, defined in \eqref{eq:masscoupl}, and $\mathcal{O}_b$ is the squashing deformation, defined in \eqref{eq:squashcoupl}.  For an $\mathcal{N}=4$ SCFT, the stress-tensor multiplet operators corresponding to the mass deformation and squashing deformation are given in \eqref{eq:J12comps}, \eqref{eq:KKbarcomps}, \eqref{eq:ZZbarcomps}, and \eqref{eq:jdiagcomps}. The four-point correlators of these $\mathcal{N}=4$ stress-tensor multiplet operators contributing to $\bbmmZ$ can be related to $\mathcal{T}(u,v)$ using $\mathcal{N}=4$ SUSY Ward identities. The purely $\mathcal{T}(u,v)$-dependent parts of these correlators are given in \eqref{eq:corrbbmm}. Once again, we can separate $\bbmmZ$ into two terms,
	\es{eq:bbmmpart1}{\bbmmZ = \bbmmZT  + \left(\bbmmZ\right)_{\text{free}}\, ,}
	where $\bbmmZT $ is the $\mathcal{T}(u,v)$-dependent part from the integral of the four-point correlator in \eqref{eq:bbmm1}, 
	\es{eq:bbmm2}{\bbmmZT  &= \int  \prod_{i=1}^4\, d^4 x_i\, \sqrt{g(x_i)}  \, \braket{\mathcal{O}_m(x_1) \mathcal{O}_m(x_2)\mathcal{O}_b(x_3)\mathcal{O}_b(x_4)}_{\mathcal{T}, S^4}\, ,}
	while $\left(\bbmmZ\right)_{\text{free}}$ consists of the integral of the free theory part of the four-point correlator as well as the integral of the lower-point correlator contributions, and does not have any $\mathcal{T}(u,v)$ dependence. One can proceed to simplify the integral \eqref{eq:bbmm2} by substituting in the correlators given in \eqref{eq:corrbbmm} and integrating by parts to get
	\es{eq:bbmm3_old}{\bbmmZT = \int \left(\prod^4_{i=1}d^4x_i \, \Omega(x_i)^4\right) \,  \mathcal{K}^{(2)}(x_i)H_{b^2m^2}(x_i) \mathcal{T}(u,v) \, ,}
	where the scalar field $\mathcal{K}^{(2)}$ is defined as 
	\es{eq:defKSc}{\mathcal{K}^{(2)}(x_1,x_2,x_3,x_4) &= 16 (\nabla^{[\mu_1}\xi^{\nu_1]})(x_3) (\nabla^{[\mu_2}\xi^{\nu_2]})(x_4) \mathcal{K}_{[\mu_1 \nu_1][\mu_2 \nu_2]}(x_1,x_2,x_3,x_4) \, ,\\}
	using the $\mathcal{K}$-tensor defined in \eqref{eq:Kcaltensdefn}, and the factor $H_{b^2m^2}(x_i)$ is given by
	\es{eq:Hbbmm}{
		H_{b^2m^2}{}&= \frac{c^2}{128\pi^8 y^4_{12}y^4_{34}y^4_{13}y^4_{24}}\bigl(3(y^4_{13}+y^4_{14}+y^4_{23}+y^4_{24})+9y^4_{34} + y^4_{12}+2y^2_{12}y^2_{34} \\
		{} &  +2(3y^2_{34}+y^2_{12})(y^2_{13}+y^2_{14}+y^2_{23}+y^2_{24})+4(y^2_{13}y^2_{14}+y^2_{23}y^2_{24})\\
		{}& +2(y^2_{14}y^2_{23}+y^2_{13}y^2_{24})+3(y^2_{13}y^2_{23}+y^2_{14}y^2_{24})\bigr),\\}
	where $y^2_{ij} = \Omega(x_i)\Omega(x_j)x^2_{ij}$ is the square of the chordal distance between the points $y_i$ and $y_j$ on the unit sphere in $\mathbb{R}^5$.
	
	Since we wish to express $\bbmmZT$ as an integral over the conformal cross-ratios $u$ and $v$, we would need to manipulate the integral \eqref{eq:bbmm2} into a conformally-invariant form. The first step in this process is to manipulate the integrand of \eqref{eq:bbmm2} into a rotationally-invariant function on $S^4$. We now describe how this can be done conveniently by averaging over background fields.
	
	\subsubsection{$\SO(5)$-averaging}
	\label{sec:avgso5}
	
	Under a rotation by an $\SO(5)$ group element $\mathcal{R}$, represented by an orthogonal matrix $\tensor{\mathcal{R}}{^p_q}$ when acting on an $\SO(5)$ vector, the $\mathbb{R}^5$ embedding space coordinates change linearly as $y^p \to y'^p = \tensor{\mathcal{R}}{^p_q} y^q$.\footnote{Raising and lowering of $\SO(5)$ indices are done with Kronecker deltas $\delta^p_q$.} The stereographic coordinates $x$ inherit their $\SO(5)$ transformations from their definitions \eqref{eq:varsR4toS4} in terms of the $y$ variables. The integrand of \eqref{eq:bbmm2} can be converted to an $\SO(5)$-invariant form by changing the integration variables: $x \to x' = \mathcal{R}(x)$, and averaging $\mathcal{R}$ over $\SO(5)$. For a generic function $f(x_i)$, its $\SO(5)$ average $\avg{f(x_i)}$ is given by the integral
	\es{eq:genso5avg}{\avg{f(x_i)} = \frac{1}{\text{vol}(\SO(5))} \int d\mathcal{R}& \, f(\mathcal{R}(x_i))\, , }
	where $d\mathcal{R}$ is the Haar measure on $\SO(5)$. Since $\mathcal{K}^{(2)}$ is the only factor in \eqref{eq:bbmm2} that is not $\SO(5)$-invariant, it suffices to replace $\mathcal{K}^{(2)}$ with its $\SO(5)$ average $\avg{\mathcal{K}^{(2)}}$. 
	
	Computing $\avg{\mathcal{K}^{(2)}}$ is simplified by the fact that the background field $d\xi \equiv  \partial_\mu \xi_\nu \, dx^\mu~\wedge~dx^\nu$ is a particular combination of the tensor harmonics $\mathcal{Y}^{[p q]}$ in the $\bold{10}$ of $\SO(5)$, 
	\es{eq:Yharmonic}{
	   d\xi =  M_{pq} \mathcal{Y}^{[p q]} \,, \qquad 
	    \mathcal{Y}^{[p q]}\equiv \frac{1}{2}\,dy^p\wedge dy^q = \frac 12 \mathcal{Y}^{[p q]}_{\mu \nu}(x)\, dx^{\mu} \wedge dx^{\nu} \, ,}
	where $p,q$ are $\SO(5)$ vector indices and the $M_{[p q]}$ matrix is 
	\es{eq:bkgrndvec}{M = \begin{pmatrix}
			0 & 1 & 0 & 0 & 0 \\
			-1 & 0 & 0 & 0 & 0 \\
			0 & 0 & 0 & -1 & 0 \\
			0 & 0 & 1 & 0 & 0 \\
			0 & 0 & 0 & 0 & 0 \\
		\end{pmatrix} \, .} 
	 We therefore have
	\es{eq:defKSc2}{\avg{\mathcal{K}^{(2)}} &= \avg{{M_{p_1q_1}M_{p_2q_2}\mathcal{L}_{{b^2m^2}}^{[p_1q_1]\,[p_2q_2]}}}  \, ,\\}
	where
	\es{eq:Ldefbbmm}{ \mathcal{L}_{{b^2m^2}}^{[p_1q_1]\,[p_2q_2]} = 4 \mathcal{Y}^{[p_1 q_1][\mu_1\nu_1]}(x_3)  \mathcal{Y}^{[p_2 q_2][\mu_2\nu_2]}(x_4) \mathcal{K}_{[\mu_1 \nu_1][\mu_2 \nu_2]}(x_1,x_2,x_3,x_4)\, .}
	 Note that the tensor $\mathcal{K}(x_1,x_2,x_3,x_4) \equiv \frac{1}{4}\mathcal{K}_{[\mu_1 \nu_1][\mu_2 \nu_2]}(x_1,x_2,x_3,x_4)\, (dx^{\mu_1}_3 \wedge dx^{\nu_1}_3) \, (dx^{\mu_2}_4 \wedge dx^{\nu_2}_4)$, defined in \eqref{eq:Kcaltensdefn}, is invariant under $\SO(5)$ transformations.\footnote{This follows from the $\SO(5)$ transformation properties of the $\bold{K}$-tensor used for defining the $\mathcal{K}$-tensor using \eqref{eq:Kdefn}, \eqref{eq:Ktensdefn}, and \eqref{eq:Kcaltensdefn}. The $\bold{K}$-tensor is defined using 6d embedding space in \cite{Cuomo:2017wme}. In this embedding space, the $\bold{K}$-tensor is manifestly covariant under the $\SO(1,5)$ symmetry group of the 6d embedding space, which contains the $\SO(5)$ as a subgroup. The invariance of the $\mathcal{K}$-tensor under $\SO(5)$ transformations thus follows.} Using \eqref{eq:genso5avg}, the invariance of $\mathcal{K}$, and the fact that $\mathcal{Y}^{[pq]}$ lies in the $\mathbf{10}$ of $\SO(5)$, we can express the $\SO(5)$-average $\avg{{\mathcal{K}^{(2)}}} $ as
	\es{eq:defKSc3}{\avg{\mathcal{K}^{(2)}} &= P_{\mathbf{1}}[M_{p_1q_1} M_{p_2 q_2}] \, \mathcal{L}_{{b^2m^2}}^{[p_1q_1]\,[p_2q_2]}(x_1,x_2,x_3,x_4)\, ,}
	where 
	\es{eq:defMMproj}{P_{\mathbf{1}}[M_{p_1q_1} M_{p_2 q_2}]  = \frac{1}{\text{vol}(\SO(5))}\int d\mathcal{R} \, (\mathcal{R}M\mathcal{R}^{-1})_{p_1q_1} (\mathcal{R}M\mathcal{R}^{-1})_{p_2q_2}}
	is the projection of the $\mathbf{10}\times\mathbf{10}$ tensor product onto the singlet representation. This projection is given by
	\es{eq:MMproj}{P_{\mathbf{1}}[M_{p_1q_1} M_{p_2 q_2}] = (\mathbf{I}^{[p_1q_1][p_2q_2]} M_{p_1q_1} M_{p_2 q_2}) \bold{I}_{[p_1q_1][p_2q_2]}\, ,}
	where 
	\es{eq:inv2ptso5}{\mathbf{I}_{[p_1q_1][p_2q_2]} = \frac{1}{\sqrt{40}}(\delta_{p_1 p_2}\delta_{q_1 q_2} - \delta_{p_1 q_2}\delta_{q_1 p_2}) 
	}
	is the unique $\SO(5)$-invariant tensor that can be constructed out of two $\mathbf{10}$ representations, normalized as $\mathbf{I}_{[p_1q_1][p_2q_2]} \mathbf{I}^{[p_1q_1][p_2q_2]} = 1$.
	
	Plugging \eqref{eq:MMproj} into \eqref{eq:defKSc3} and using $M_{pq}$ given in \eqref{eq:bkgrndvec}, we have
	\es{eq:Krotavg3}{\avg{\mathcal{K}^{(2)}} &= \frac{2}{5} \tensor{\mathcal{L}}{_{b^2m^2}^{[p_1 q_1]}_{[p_1 q_1]} } \, , \\ }
	with $\mathcal{L}_{{b^2m^2}}$ given in \eqref{eq:Ldefbbmm}.  Further using \eqref{eq:Ktenscorrindex}, \eqref{eq:Kcaltensdefn}, and \eqref{eq:Yharmonic}, $\avg{\mathcal{K}^{(2)}}$ can be computed explicitly and expressed in a manifestly $\SO(5)$-invariant way using $\mathbb{R}^5$ coordinates as
	\es{eq:Krotavg4}{\avg{\mathcal{K}^{(2)}} &= {} \frac{2}{5}\bigl((8y^2_{13}y^2_{23}+8y^2_{14}y^2_{24}+4y^2_{13}y^2_{24}+4y^2_{14}y^2_{23})+(8-y^2_{12})y^4_{34}\\
		{}&+(4y^2_{12}-8y^2_{13}-8y^2_{14}-8y^2_{23}-8y^2_{24}+y^2_{14}y^2_{23}+y^2_{13}y^2_{24})y^2_{34}\bigr)\,. }
	
	We have now converted \eqref{eq:bbmm2} into a $\SO(5)$-invariant integral of the form
	\es{eq:bbmm3}{\left(\bbmmZ\right)_{\mathcal{T}} = \int \left(\prod^4_{i=1}d\Omega^i_{S^4}\right) \,  \avg{\mathcal{K}^{(2)}} H_{b^2m^2} \mathcal{T}(u,v)\, ,}
	where $d\Omega^i_{S^4}= d^5y_i\, \delta(y_i^2-1)$ is the measure on $S^4$,  and $\avg{\mathcal{K}^{(2)}}$ and $H_{b^2m^2}$ are given in \eqref{eq:Krotavg4} and \eqref{eq:Hbbmm}, respectively. The conformal cross-ratios $u$ and $v$ are defined in terms of the $\mathbb{R}^5$ unit vectors in \eqref{eq:uvR5}.
	
	\subsubsection{Crossing symmetry}
	
	The integrand in \eqref{eq:bbmm3} is invariant only under $(y_1 \leftrightarrow y_2)$ and $(y_3 \leftrightarrow y_4)$  permutations, which are inherited from the permutation symmetries of the correlator in \eqref{eq:bbmm2}. It was observed in \cite{Chester:2020dja}  that writing the non-squashed integral constraints in a crossing-invariant form resulted in a simpler expression. We anticipate that expressing \eqref{eq:bbmm3} in a crossing-invariant form would make later manipulations simpler as well. This crossing-invariant form can be obtained by replacing the integrand in \eqref{eq:bbmm3} with the average of the crossed versions of the integrand under the 24 permutations of the 4 points on $S^4$. Using the crossing properties of $\mathcal{T}(u,v)$, given in \eqref{eq:Tcrossing}, one can observe that the factor of $\frac{\mathcal{T}(u,v)}{y^4_{12}y^4_{34}y^4_{13}y^4_{24}}$ is crossing-invariant. Therefore, we can rewrite \eqref{eq:bbmm3} as
	\es{eq:bbmm3cross}{ \left(\bbmmZ\right)_{\mathcal{T}} &= \int \left(\prod^4_{i=1}d\Omega^i_{S^4}\right)  N_{b^2m^2}(y^2_{ij}) \frac{\mathcal{T}(u,v)}{y^4_{12}y^4_{34}y^4_{13}y^4_{24}} \, ,  \\}
	where
	\es{eq:defNbbmm}{N_{b^2m^2}(y^2_{ij})&=\frac{1}{24}\left(y^4_{12}y^4_{34}y^4_{13}y^4_{24}\avg{\mathcal{K}^{(2)}} H_{b^2m^2} +  (\text{23 other permutations of } \{y_1,y_2,y_3,y_4\}) \right) \\
		&= \frac{c^2}{1920 \pi^8}\bigg[\biggl( y^8_{12} \biggl( 18 - \frac{9}{4}y^2_{34}\biggr) + y^2_{12}y^2_{13} \left(14 y_{23}^2 y_{24}^2 +8 y_{24}^2 y_{34}^2 + 3y_{14}^2 y_{24}^2  y_{34}^2  \right)  \\
		&+  y^6_{12} \left(y^2_{13}\left(- 24 +\frac{9}{2}y^2_{24}  -3 y^2_{34}\right)  - \frac{3}{4}y^4_{34} + 26 y^2_{34}\right) \\
		&+   y^4_{12}\biggl(y^4_{13}(3y^2_{34}-12) + 8 y_{34}^4+y^2_{13}\biggl(-y_{34}^4+\frac{3}{2} y_{14}^2 y_{34}^2+2 y_{23}^2 y_{34}^2 + \frac{5}{2} y_{24}^2 y_{34}^2  \\
		& +    84 y_{34}^2 -68 y_{23}^2+9 y_{23}^2 y_{24}^2 +20 y_{24}^2 \biggr)\biggr)  \biggr) \\
		&+ (\text{23 other permutations of } \{y_1,y_2,y_3,y_4\})\bigg]\, ,}
	where $\avg{\mathcal{K}^{(2)}}$ and $H_{b^2m^2}$ are given in \eqref{eq:Krotavg4} and \eqref{eq:Hbbmm} respectively.
	
	\subsubsection{Simplification in 6d}
	\label{sec:simp6dbbmm}
	
	The next step in our simplification is to convert the integral \eqref{eq:bbmm3cross} into a conformally-invariant form such that the measure could be expressed in terms of $\bar{D}$-functions as was done for the $\mmmmZ$ integrated correlator in Section~\ref{sec:m4}. Following the approach described in Section~\ref{sec:embed6d}, we begin this manipulation by embedding a point $y_i$ on $S^4$ in $\mathbb{R}^{5,1}$ as the light-ray $\mathbb{L}_{\vec{y}_j}$ defined in \eqref{eq:embed}. The integral \eqref{eq:bbmm3cross} then becomes
	\es{eq:bbmm4}{\left(\bbmmZ\right)_{\mathcal{T}} =  \int  \frac{\prod^4_{i=1} d^4\Omega^i_{S^4}}{(y^2_{12}y^2_{34})^4} \, F_{b^2m^2}\left(\frac{\bold{X}_i\cdot \bold{X}_j}{(\bold{X}_i\cdot \bold{Y}_*)(\bold{X}_j\cdot \bold{Y}_*)}\right) \mathcal{T}(u,v)\, ,}
	where  $F_{b^2m^2}$ is related to $N_{b^2m^2}$, defined in \eqref{eq:defNbbmm}, as
	\es{eq:defFbbmm}{F_{b^2m^2}\left(\frac{\bold{X}_i\cdot \bold{X}_j}{(\bold{X}_i\cdot \bold{Y}_*)(\bold{X}_j\cdot \bold{Y}_*)}\right) &=  \left.\frac{y^4_{12}y^4_{34}}{y^4_{13}y^4_{24}}N_{b^2m^2}(y^2_{ij})\right|_{y^2_{ij}\to \frac{-2 \bold{X}_i\cdot \bold{X}_j}{\bold{X}_i\cdot\bold{Y}_*\, \bold{X}_j\cdot\bold{Y}_*}} \, ,}
	for $\bold{X}_i \in \mathbb{L}_{\vec{y}_j}$. 
	
	Following the procedure outlined in Appendix~\ref{sec:AdSavg}, we can use $\SO(5,1)$ transformations to fix the four $S^4$ points $y_i$ to a particular kinematic configuration leaving \eqref{eq:bbmm4} as an integral over the conformal cross-ratios $u$ and $v$. Using the results given in \eqref{eq:bbmm10} and \eqref{eq:defFcal}, we are left with a conformally-invariant form for $\bbmmZT$,
	\es{eq:bbmm5}{\bbmmZT =  \frac{4\pi^5}{3}\int dr \, d\theta \, \frac{r^3 \sin^2\theta}{u^4} \mathcal{F}_{b^2m^2}(u,v) \mathcal{T}(u,v),}
	where the conformal cross-ratios $u$ and $v$ are related to $r$ and $\theta$ by $u = 1+r^2-2r\cos\theta,\, v = r^2$, and the function $\mathcal{F}_{b^2m^2}(u,v)$ is defined in terms of $F_{b^2m^2}$ using the functional $\mathcal{J}$, defined in \eqref{eq:defFcal}, 
	\es{eq:defFcalbbmm}{ \mathcal{F}_{b^2m^2}(u,v) = \mathcal{J}[F_{b^2m^2}] \, . }
	The $\mathcal{F}_{b^2m^2}(u,v)$ function can be expressed in terms of $D_{r_1,\ldots,r_n}$-functions, defined as integrals in \eqref{eq:Ddefnalt}. A naive application of \eqref{eq:Ddefnalt} to the integral \eqref{eq:defFcalbbmm} would result in an apparently divergent $\mathcal{F}_{b^2m^2}(u,v)$ function. Carefully regulating the integral in \eqref{eq:defFcalbbmm} and using various identities of $D$-functions, one can show that $\mathcal{F}_{b^2m^2}(u,v)$ is finite. We go over the details of this process in Appendix~\ref{sec:FAdSsimp}. The final form for the integrated correlator $\bbmmZT$ after this simplification is
	\es{eq:bbmmintcorr}{
		\bbmmZT = \frac{32c^2}{3\pi}\int dr d\theta \, \frac{r^3 \sin^2 \theta}{u^2}((1+u+v)\bar{D}_{1,1,1,1}(u,v)-4)\mathcal{T}(u,v) \, ,}	
	using the simplified expression for $\mathcal{F}_{b^2m^2}(u,v)$ given in \eqref{eq:Fbbmmfinal}. Using \eqref{eq:knownintcorr}, we can see that this result trivially satisfies the $\mathcal{T}(u,v)$ part of the second relation in \eqref{eq:relation1}.

	\subsection{\texorpdfstring{$\bbbbZ$}{b⁴ derivative}}
	\label{sec:bbbb}
	
	We now compute the final integrated correlator with squashing $\bbbbZ$. For an $\mathcal{N}=2$ SCFT, it is given by the integral
	\es{eq:bbbb1}{\bbbbZ &= \int  \prod_{i=1}^4\, d^4 x_i\, \sqrt{g(x_i)}  \, \braket{\mathcal{O}_b(x_1) \mathcal{O}_b(x_2)\mathcal{O}_b(x_3)\mathcal{O}_b(x_4)}_{S^4}\\
		& + \text{(lower-point contributions)} \, , \\ }
	where, once again, $\mathcal{O}_b$ is the squashing deformation defined in \eqref{eq:squashcoupl}. For an $\mathcal{N}=4$ SCFT, the purely $\mathcal{T}(u,v)$-dependent parts of the four-point correlation functions appearing in \eqref{eq:bbbb1} are given in \eqref{eq:corrbbbb}. These relations arise from using \eqref{eq:ZZbarcomps} and \eqref{eq:jdiagcomps} to identify the $\mathcal{N}=4$ operators that correspond to the squashing deformation and solving $\mathcal{N}=4$ SUSY Ward identities. Once again, we focus on computing the $\mathcal{T}(u,v)$-dependent part of $\bbbbZ$ which is the integral
	\es{eq:bbbbT1}{\bbbbZT &= \int  \prod_{i=1}^4\, d^4 x_i\, \sqrt{g(x_i)}  \, \braket{\mathcal{O}_b(x_1) \mathcal{O}_b(x_2)\mathcal{O}_b(x_3)\mathcal{O}_b(x_4)}_{\mathcal{T},S^4} \, .\\ }
	
	We begin our computation by substituting the correlators in \eqref{eq:corrbbbb} into the integral \eqref{eq:bbbbT1} and integrating by parts to get the expression
	\es{eq:bbbb2}{\bbbbZT = \int \left(\prod^4_{i=1}d^4 x_i \, \Omega(x_i)^4\right) \,  \mathcal{M}^{(2)}(x_i)H_{b^4}(x_i) \mathcal{T}(u,v) \, .}
	Here, the scalar field $\mathcal{M}^{(2)}$ is defined as
	\es{eq:defMSc}{\mathcal{M}^{(2)}(x_1,x_2,x_3,x_4) &= 256 (\nabla^{[\mu_1}\xi^{\nu_1]})(x_1) (\nabla^{[\mu_2}\xi^{\nu_2]})(x_2)  (\nabla^{[\mu_3}\xi^{\nu_3]})(x_3) (\nabla^{[\mu_4}\xi^{\nu_4]})(x_4) \\
		& \times \mathcal{M}_{[\mu_1 \nu_1][\mu_2 \nu_2][\mu_3 \nu_3][\mu_4 \nu_4]}(x_1,x_2,x_3,x_4) \, ,\\}
	where the $\mathcal{M}_{[\mu_1 \nu_1][\mu_2 \nu_2][\mu_3 \nu_3][\mu_4 \nu_4]}$ tensor is defined in \eqref{eq:Mtenscorrindex}, and the factor $H_{b^4}(x_i)$ is given by
	\es{eq:Hbbbb}{
		H_{b^4}&= \frac{c^2}{4096\pi^8 y^4_{12}y^4_{34}y^4_{13}y^4_{24}y^2_{14}y^2_{23}}\bigg[\left(-\frac{9}{4} y^4_{12}-6 y^2_{12} y^2_{13}-y^2_{12} y^2_{34}\right) \\
		&{} + (\text{23 other permutations of } \{y_1,y_2,y_3,y_4\}\bigg]\,,\\ 
	}
	where $y^2_{ij} = \Omega(x_i)\Omega(x_j)x^2_{ij}$ is the chordal distance on $S^4$.
	
	\subsubsection{$\SO(5)$-averaging}
	\label{sec:avgso5bbbb}
	
	Just like in the previous section, we can continue simplifying \eqref{eq:bbbb2} by converting its integrand into an $\SO(5)$-invariant function on $S^4$. Following the discussion in Section~\ref{sec:avgso5}, this is done by replacing $\mathcal{M}^{(2)}$ in \eqref{eq:bbbb2} with its $\SO(5)$-average $\avg{\mathcal{M}^{(2)}}$, where the averaging operation $\avg{...}$ is defined by \eqref{eq:genso5avg}. Substituting in the expression for $\nabla^{[\mu} \xi^{\nu]}$  as a tensor harmonic in \eqref{eq:Yharmonic}, we have
	\es{eq:Mrotavg}{\avg{\mathcal{M}^{(2)}}= \avg{M_{p_1q_1}M_{p_2q_2}M_{p_3q_3}M_{p_4q_4} \mathcal{L}^{[p_1q_1][p_2q_2][p_3q_3][p_4q_4]}_{b^4}}\,, }
	where
	\es{eq:Ldefbbbb}{\mathcal{L}^{[p_1q_1][p_2q_2][p_3q_3][p_4q_4]}_{b^4} &= \,  16 \mathcal{Y}^{[p_1q_1][\mu_1\nu_1]}(x_1)\mathcal{Y}^{[p_1q_1][\mu_1\nu_1]}(x_1)\mathcal{Y}^{[p_2q_2][\mu_3\nu_3]}(x_3)\mathcal{Y}^{[p_4q_4][\mu_4\nu_4]}(x_4) \\
		& \times	\mathcal{M}_{[\mu_1 \nu_1][\mu_2 \nu_2][\mu_3 \nu_3][\mu_4 \nu_4]}(x_1,x_2,x_3,x_4) \, , \\ }
	$\mathcal{M}_{[\mu_1 \nu_1][\mu_2 \nu_2][\mu_3 \nu_3][\mu_4 \nu_4]}(x_1,x_2,x_3,x_4)$ is the tensor given in \eqref{eq:Mtenscorrindex}, and the tensor harmonics $\mathcal{Y}^{[pq]}$  and the matrix $M_{pq}$ are given in \eqref{eq:Yharmonic} and \eqref{eq:bkgrndvec} respectively. The tensor $\mathcal{M}\equiv \frac{1}{16}\mathcal{M}_{[\mu_1 \nu_1][\mu_2 \nu_2][\mu_3 \nu_3][\mu_4 \nu_4]} dx_1^{\mu_1}\wedge dx_1^{\nu_1} \, dx_2^{\mu_2}\wedge dx_2^{\nu_2} \, dx_3^{\mu_3}\wedge dx_3^{\nu_3} \, dx_4^{\mu_4}\wedge dx_4^{\nu_4}$ is $\SO(5)$-invariant. This is made manifest by the expression \eqref{eq:MtensHarmonic} in which it is written as an $\SO(5)$-invariant sum over the $\mathbf{10}$ tensor harmonics. Using the invariance of $\mathcal{M}$ and the fact that the $\mathcal{Y}^{[pq]}$ tensor harmonics lie in the $\mathbf{10}$ of $\SO(5)$, \eqref{eq:Mrotavg} becomes 
	\es{eq:Mrotavg2}{ \avg{\mathcal{M}^{(2)}}= P_{\mathbf{1}}[M_{p_1q_1} M_{p_2q_2} M_{p_3q_3} M_{p_4q_4}] \mathcal{L}_{b^4}^{[p_1 q_1][p_2 q_2][p_3 q_3][p_4 q_4]}(x_1,x_2,x_3,x_4) \, .}
	Here, $P_{\mathbf{1}}$ is the projector onto the singlet subspace of the tensor product of the four $M_{pq}$ matrices. It is given explicitly by the integral
	\es{eq:ProjMbbbb}{&P_{\mathbf{1}}[M_{p_1q_1} M_{p_2q_2} M_{p_3q_3} M_{p_4q_4}] \\
		&= \frac{1}{\vol(\SO(5))}\int d\mathcal{R}\, (\mathcal{R}M\mathcal{R}^{-1})_{p_1q_1} (\mathcal{R}M\mathcal{R}^{-1})_{p_2q_2}(\mathcal{R}M\mathcal{R}^{-1})_{p_3q_3}(\mathcal{R}M\mathcal{R}^{-1})_{p_4q_4} \, ,}
	where $d\mathcal{R}$ is the Haar measure on $\SO(5)$.
	
	The space of $\SO(5)$-invariant tensors in the tensor product of four copies of $\mathbf{10}$ is six-dimensional. We define our choice for basis tensors $T^i$ for this space  in terms of the contracted objects $ \Theta^i = T^i_{[p_1q_1][p_2q_2][p_3q_3][p_4q_4]} M^{p_1q_1}_1 M^{p_2q_2}_2 M^{p_3q_3}_3 M^{p_4q_4}_4$,
	\es{eq:Tso5basis}{ \Theta  =& \, \frac{1}{4}\left\{ \text{tr}(M_1M_2)\text{tr}(M_3M_4),\text{tr}(M_1M_3)\text{tr}(M_2M_4),\text{tr}(M_1M_4)\text{tr}(M_2M_3),\right.\\
		&\left. \text{tr}(M_1M_2M_3M_4),\text{tr}(M_1M_2M_4M_3),\text{tr}(M_1M_3M_2M_4)\right\}\, ,  \\}
	for generic antisymmetric rank-two matrices $M_i$. The integral \eqref{eq:ProjMbbbb} can be written in this basis as
	\es{eq:Mproddecom}{P_{\mathbf{1}}[M_{p_1q_1} M_{p_2q_2} M_{p_3q_3} M_{p_4q_4}] = \sum_{i=1}^6 a_i T^i_{[p_1q_1][p_2q_2][p_3q_3][p_4q_4]}\,,}
	where $\vec{a} = \frac{1}{21}\left(1,1,1,-\frac{4}{5},-\frac{4}{5},-\frac{4}{5}\right)$. This specific vector $\vec{a}$ is obtained by using \eqref{eq:Mproddecom} as an ansatz and requiring that both sides of \eqref{eq:Mproddecom} give the same number when contracted with the six $T^i$ tensors.
	
	Using this result along with the definition of $\mathcal{L}_{b^4}$ in \eqref{eq:Ldefbbbb} and the expression for $\mathcal{M}$ an $\SO(5)$-invariant sum in \eqref{eq:MtensHarmonic}, \eqref{eq:Mrotavg2} can be written as
	\es{eq:MScSO5avg}{\avg{\mathcal{M}^{(2)}} 
		&=  -\frac{8}{105} y^2_{23}y^2_{14}  \bigg[ \bigg(y^4_{12} \left(\frac{11}{8} y^4_{34}-44 y^2_{34}+160\right)\\
		&{} +  y^2_{12} \left(16 y^2_{13} y^2_{23}+40 y^2_{13} y^2_{24}-\frac{5}{4} y^2_{13} y^2_{24} y^2_{34}-320 y^2_{13}+120 y^2_{34}\right)\bigg)\\
		&{}+ (\text{23 other permutations of} \{y_1,y_2,y_3,y_4\}) \bigg]\, .\\}
	
	We now have a $\SO(5)$-invariant form for $\bbbbZT$,
	\es{eq:bbbb3}{\bbbbZT = \int \left(\prod^4_{i=1}d^4 \Omega^i_{S^4}\right) \,  \avg{\mathcal{M}^{(2)}} H_{b^4} \mathcal{T}(u,v) \, ,}
	where $H_{b^4}$ and $\avg{\mathcal{M}^{(2)}}$ are given explicitly in \eqref{eq:Hbbbb} and \eqref{eq:MScSO5avg} respectively. The expression for $\bbbbZT$ can be simplified further into a conformally-invariant form following the same procedure that we used for $\bbmmZT$ in Section~\ref{sec:simp6dbbmm}. This is done by first embedding the integrand in \eqref{eq:bbbb3} on the 6d lightcone, as discussed in Section~\ref{sec:embed6d}. The $\bbbbZT$ integrated correlator is then given by the integral
	\es{eq:bbbb4_old}{\bbbbZT = \int  \left(\frac{\prod^4_{i=1}d^4 \Omega^i_{S^4}}{(y^2_{12} y^2_{34})^4}\right) \mathcal{T}(u,v) \left.F_{b^4}\left(\frac{\bold{X}_i\cdot \bold{X}_j}{(\bold{X}_i\cdot \bold{Y}_*)(\bold{X}_j\cdot \bold{Y}_*)}\right)\right|_{\bold{X}_i \in \mathbb{L}_{\vec{y}_i}}\, ,}
	where $\bold{Y}_* = (0,0,0,0,0,1)$, the lightray $\mathbb{L}_{\vec{y}_i}$ is defined in \eqref{eq:embed}, and the function $F_{b^4}$ is related to  $H_{b^4}(y^2_{ij})$, given in \eqref{eq:Hbbbb}, and $\avg{\mathcal{M}^{(2)}} $, given in \eqref{eq:MScSO5avg}, by the substitution 
	\es{eq:defFbbbb}{F_{b^4}\left(\frac{\bold{X}_i\cdot \bold{X}_j}{(\bold{X}_i\cdot \bold{Y}_*)(\bold{X}_j\cdot \bold{Y}_*)}\right) &= [(y^2_{12}y^2_{34})^4\avg{\mathcal{M}^{(2)}(y^2_{ij})}H_{b^4}(y^2_{ij})]_{y^2_{ij}\to \frac{-2 \bold{X}_i\cdot \bold{X}_j}{\bold{X}_i\cdot\bold{Y}_*\, \bold{X}_j\cdot\bold{Y}_*}} \, .\\}
	
	Following the simplification in Appendix~\ref{sec:AdSavg}, we can write the integral \eqref{eq:bbbb4_old} in the conformally-invariant form
	\es{eq:bbbb4}{\bbbbZT =  \frac{4\pi^5}{3}\int dr \, d\theta \, \frac{r^3 \sin^2\theta}{u^4} \mathcal{F}_{b^4}(u,v) \mathcal{T}(u,v),}
	where $u = 1+r^2 -2 r \cos \theta$, $v = r^2$, and 
	\es{eq:defFcalbbbb}{\mathcal{F}_{b^4}(u,v) = \mathcal{J}[F_{b^4}]\, ,}
	with the functional $\mathcal{J}[\ldots]$ defined as in \eqref{eq:defFcal}. $\mathcal{F}_{b^4}(u,v)$ can be simplified using the same techniques that were used to simplify $\mathcal{F}_{b^2m^2}(u,v)$ in Appendix~\ref{sec:FAdSsimp}\@. This procedure involves using the $\epsilon$-deformation discussed in Appendix~\ref{sec:extrmsimp} to regulate apparent divergences coming from extremal two-point, three-point and four-point $D$-functions.
	Following this simplification, we obtain our final expression for $\bbbbZT$ :
	\es{eq:bbbbZfinal}{\bbbbZ  = \frac{32c^2}{\pi}\int dr d\theta \, \frac{r^3 \sin^2\theta}{u^2} ((1+u+v)\bar{D}_{1,1,1,1}(u,v)-8) \mathcal{T}(u,v) \, .}
	This expression satisfies the $\mathcal{T}(u,v)$-dependent part of the third relation in \eqref{eq:relation1}, once \eqref{eq:bbmmintcorr} and the second expression in \eqref{eq:knownintcorr} are substituted in.
	
	\section*{Acknowledgments} 
	
	We thank  Clay Cordova and David Simmons-Duffin for useful discussions.   SMC is supported by the UK Engineering and Physical Sciences Research council grant number EP/Z000106/1, and Royal Society under the grant URF\textbackslash R1\textbackslash 221310. The work of DP and SSP is supported in part by the U.S.~Department~of~Energy under Award No.~DE-SC0007968. The work of RD is supported by a Pappalardo Fellowship in Physics at MIT.

	\appendix
	
	\section{SUSY in  \texorpdfstring{$\mathcal{N}=2$}{N = 2} SCFTs}\label{sec:SUSYN2def}

	In this appendix, we give some details regarding the $\mathcal{N}=2$ superconformal algebra. In particular, we describe the field content of the $\mathcal{N}=2$ Weyl multiplet and give the SUSY transformations of these fields. We then explain how these transformations can be used to identify Killing spinors in flat space as well as round $S^4$. Before beginning this, we will specify the details of the spinor conventions used in this work. 
	
	While we primarily use four-component spinor notation in this paper, we shall occasionally switch to two-component spinor notation for convenience. A generic four-component spinor $\psi$ consists of a two-component left-handed spinor $\psi_L$ and a two-component right-handed spinor $\tilde{\psi}_R$,
	\es{eq:deffourcompspinor}{\psi_{\hat{\alpha}} = \begin{pmatrix}
			\psi_{L\alpha} \\ \tilde{\psi}^{\dot{\beta}}_{R}
		\end{pmatrix} \, .}
	Note that we use undotted indices for left-handed two-component spinors, dotted indices for right-handed two-component spinors, and hatted indices for four-component spinors. The components of the Majorana conjugate $\bar{\psi}\equiv \psi^T C$ are given by the relation
	\es{eq:MajConj}{\bar{\psi}^{\hat{\alpha}} = \psi_{\hat{\beta}}C^{\hat{\beta}\hat{\alpha}} \, \qquad C = \begin{pmatrix}
			0 & i & 0 & 0 \\
			-i & 0 & 0 & 0 \\
			0 & 0 & 0 & i \\
			0 & 0 & -i & 0 \\
		\end{pmatrix}\,. }
	The raising and lowering conventions for the two-component spinors are
	\es{eq:RaiseLowerConv}{\tilde{\psi}_{R\dot{\alpha}}=\tilde{\psi}_{R}^{\dot{\beta}}\epsilon_{\dot{\beta}\dot{\alpha}}\,,\qquad \psi^{\alpha}_{L}=\epsilon^{\alpha\beta}\psi_{L\beta} \,,}
	where the rank-two Levi-Civita tensors are chosen such that 
	\es{eq:defepsRaiseLower}{\epsilon_{12} = \epsilon^{12} = \epsilon_{\dot{1}\dot{2}}= \epsilon_{\dot{1}\dot{2}} =1\,.}
	We take the Euclidean $\sigma$ and $\bar{\sigma}$ matrices to be
	\es{eq:sigmamatrices}{\sigma^{a}_{\alpha\dot{\alpha}} = \{ \vec{\sigma},i \mathbbm{1} \}\, ,& \qquad  \bar{\sigma}^{a\, \dot{\alpha}\alpha} = \{ \vec{\sigma},-i \mathbbm{1} \} \, . \\}
	where $\sigma^i$ with $i=1,2,3$ are the conventional Pauli matrices
	\es{eq:paulimat}{\vec{\sigma} = \left\{\begin{pmatrix}
			0& 1\\
			1& 0 \\
		\end{pmatrix} \,, \begin{pmatrix}
			0& -i\\
			i& 0 \\
		\end{pmatrix}\,,\begin{pmatrix}
			1& 0\\
			0& -1 \\
		\end{pmatrix}\right\} \,.} 
	The Euclidean gamma matrices are given by 
	\es{eq:gammaconv}{
		\gamma^a = \begin{pmatrix}
			0& \sigma^a\\
			\bar{\sigma}^a & 0 \\
		\end{pmatrix} \,, \qquad \gamma_* = \gamma^1\gamma^2\gamma^3\gamma^4 = \begin{pmatrix}
			\mathbbm{1}& 0\\
			0& -\mathbbm{1}\\
		\end{pmatrix} \, .}
		Spinning operators can be written either using coordinate indices or using spinor indices. These two forms are related by
		\es{eq:frametospinor}{\tensor{j}{_{\alpha\dot{\alpha}}} \equiv \tensor{j}{_{\mu}} \sigma^\mu_{\alpha\dot{\alpha}}\, , \qquad  Z^{-}_{\alpha\beta} \equiv Z^{-}_{\mu\nu} \sigma^{\mu\nu}_{\alpha\beta} \,  , \qquad  Z^{+}_{\dot{\alpha}\dot{\beta}} \equiv Z^{+}_{\mu\nu} \bar{\sigma}^{\mu\nu}_{\dot{\alpha}\dot{\beta}} \,  ,}
		where $j_\mu$ is any spin-one field, $Z^+_{\mu\nu}$ is any self-dual field, $Z^-_{\mu\nu}$ is any anti-self-dual field, and 
		\es{eq:sigmunudef}{\sigma^{\mu\nu}_{\alpha\beta}&= \frac{1}{4}(\sigma^\mu_{\alpha\dot{\alpha}}\bar{\sigma}^{\nu\dot{\alpha}\gamma}-\sigma^\nu_{\alpha\dot{\alpha}}\bar{\sigma}^{\mu\dot{\alpha}\gamma})\epsilon_{\beta\gamma}\,, \\
			\bar{\sigma}^{\mu\nu}_{\dot{\alpha}\dot{\beta}}&= \frac{1}{4}\epsilon_{\dot{\gamma}\dot{\alpha}}(\bar{\sigma}^{\mu\dot{\gamma}\alpha} \sigma^\nu_{\alpha\dot{\beta}}-\bar{\sigma}^{\nu\dot{\gamma}\alpha} \sigma^\mu_{\alpha\dot{\beta}})\,.}
		Note that self-dual and anti-self-dual fields in Euclidean signature satisfy the relations
		\es{eq:defSD}{Z^{\pm}_{ab} = \pm\frac{1}{2}\epsilon_{abcd}Z^{\pm cd}\,.}
	
	\subsection{Superconformal algebra and the Weyl multiplet}
	
	We follow the conventions for $\mathcal{N}=2$ superconformal symmetry used in \cite{Freedman:2012zz} and  \cite{Lauria:2020rhc}. A general $\mathcal{N}=2$ superconformal transformation acts on a field $\mathcal{O}$ via the Poisson bracket
	\begin{equation}\label{eq:SUSYgen}
		\delta(\epsilon, \eta) \mathcal{O} = \{\bareps^i Q_i + \bareps_i Q^i + \bar{\eta}^i S_i + \bar{\eta}_i S^i, \mathcal{O}\}\, ,
	\end{equation}
	where $Q_i$ and $Q^i$ generate Poincar\'e supersymmetry (SUSY) transformations, $S_i$ and $S^i$ generate conformal SUSY transformations, and  $\bar{\psi}\equiv \psi^T C$ is the Majorana conjugate for a four-component spinor as defined in \eqref{eq:MajConj}. We pass from a Poisson bracket to the corresponding commutator in the quantum algebra via the relation $\{O,\ldots\} = -i [\mathbf{O},\ldots]$ (or an analogous relation involving anti-commutators for fermionic operators), where $\mathcal{O}$ is any operator in the classical field theory and $\mathbf{O}$ is its quantum version.

	\begin{table}[!tb]
		\centering
		\renewcommand{\arraystretch}{1.2}
		\begin{tabular}{ccccc}
			\toprule
			Field & Weyl weight & Lorentz rep & $\mathfrak{su}(2)_R$ rep & $\mathfrak{u}(1)_R$ charge \\
			\midrule
			$Q_i$& $\frac{1}{2}$ & $\left(\frac{1}{2},0\right)$ & $\mathbf{2}$ & $-\frac{1}{2}$ \\
			$Q^i$& $\frac{1}{2}$ & $\left(0,\frac{1}{2}\right)$ & $\mathbf{2}$ & $\frac{1}{2}$ \\
			$S_i$& $-\frac{1}{2}$ & $\left(0,\frac{1}{2}\right)$ & $\mathbf{2}$ & $-\frac{1}{2}$ \\
			$S^i$& $-\frac{1}{2}$ & $\left(\frac{1}{2},0\right)$ & $\mathbf{2}$ & $\frac{1}{2}$ \\
			\bottomrule
		\end{tabular}
		\caption{Fermionic operators for the $\mathcal{N}=2$ superconformal algebra and their quantum numbers.}
		\label{tab:ScftOpTable}
	\end{table}  
	
	The quantum numbers of the supersymmetry generators are given in Table \ref{tab:ScftOpTable}. The quantum numbers for the $\epsilon^i$, $\epsilon_i$, $\eta^i$ and $\eta_i$ spinors parametrizing SUSY transformations can be inferred from Table \ref{tab:ScftOpTable} and \eqref{eq:SUSYgen}. The SUSY generators satisfy the anticommutator relations
	\es{eq:SUSYAlg}{
		\{\tensor{Q}{^i}{_{\hat{\alpha}}},\tensor{Q}{_j}{^{\hat{\beta}}}\} &= -\frac{1}{2}\tensor{\delta}{_j^i}\tensor{(\gamma^a)}{_{\hat{\alpha}}^{\hat{\beta}}}P_a\,,\\
		\{\tensor{S}{^i}{_{\hat{\alpha}}},\tensor{S}{_j}{^{\hat{\beta}}}\} &= -\frac{1}{2}\tensor{\delta}{_j^i}\tensor{(\gamma^a)}{_{\hat{\alpha}}^{\hat{\beta}}}K_a\,,\\
		\{\tensor{Q}{_{i\hat{\alpha}}},\tensor{S}{^{j\hat{\beta}}}\} &= -\frac{1}{2}\tensor{\delta}{_i^j}\tensor{\delta}{_{\hat{\alpha}}^{\hat{\beta}}}D-\frac{1}{4}\tensor{\delta}{_i^j}\tensor{(\gamma^{ab})}{_{\hat{\alpha}}^{\hat{\beta}}}M_{ab}+\frac{i}{2}\tensor{\delta}{_i^j}\tensor{\delta}{_{\hat{\alpha}}^{\hat{\beta}}}T -\tensor{\delta}{_{\hat{\alpha}}^{\hat{\beta}}}\tensor{U}{_i^j}\,,\\
		\{\tensor{Q}{^i}{_{\hat{\alpha}}},\tensor{S}{_j}{^{\hat{\beta}}}\} &= -\frac{1}{2}\tensor{\delta}{_j^i}\tensor{\delta}{_{\hat{\alpha}}^{\hat{\beta}}}D-\frac{1}{4}\tensor{\delta}{_j^i}\tensor{(\gamma^{ab})}{_{\hat{\alpha}}^{\hat{\beta}}}M_{ab}-\frac{i}{2}\tensor{\delta}{_j^i}\tensor{\delta}{_{\hat{\alpha}}^{\hat{\beta}}}T +\tensor{\delta}{_{\hat{\alpha}}^{\hat{\beta}}}\tensor{U}{_j^i}\,,
	}
	where $P_a$ is the generator for translations, $K_a$ is the generator for special conformal transformations, $M_{ab}$ is the generator for rotations, $D$ is the generator for dilations, $T$ is the generator for $\mathfrak{u}(1)_R$ transformations, and $\tensor{U}{_j^i}$ is the generator for $\mathfrak{su}(2)_R$ transformations. All other anticommutators of the $Q$ and $S$ operators vanish. Additional details of the $\mathcal{N}=2$ superconformal algebra in the conventions we follow, including all other (anti-)commutators, can be found in  \cite{Freedman:2012zz} and \cite{Lauria:2020rhc}.
	
	\begin{table}[!tb]
		\centering
		\renewcommand{\arraystretch}{1.2}
		\begin{tabular}{ccccc}
			\toprule
			Field & Weyl weight & Lorentz rep & $\mathfrak{su}(2)_R$ rep & $\mathfrak{u}(1)_R$ charge \\
			\midrule
			$\tensor{e}{_\mu^a}$ & $-1$& $\left(\frac{1}{2},\frac{1}{2}\right)$ & $\mathbf{1}$ & $0$ \\
			$\tensor{\psi}{_a^i}$ & $\frac{1}{2}$& $\left(1,\frac{1}{2}\right)$ & $\mathbf{2}$ & $\frac{1}{2}$ \\
			$\psi_{a i}$ & $\frac{1}{2}$& $\left(\frac{1}{2},1\right)$ & $\mathbf{2}$ & $-\frac{1}{2}$ \\
			$b_{a}$ & $1$ & $\left(\frac{1}{2},\frac{1}{2}\right)$ & $\mathbf{1}$ & $0$ \\
			$A_{a}$ & $1$ & $\left(\frac{1}{2},\frac{1}{2}\right)$ & $\mathbf{1}$ & $0$ \\
			$\tensor{V}{_{a\,j}^i}$ & $1$ & $\left(\frac{1}{2},\frac{1}{2}\right)$ & $\mathbf{3}$ & $0$ \\
			$T^{-}_{ab}$ & $1$ & $\left(1,0\right)$ & $\mathbf{1}$ & $1$ \\
			$T^{+}_{ab}$ & $1$ & $\left(0,1\right)$ & $\mathbf{1}$ & $-1$ \\
			$\chi^i$ & $\frac{3}{2}$ & $\left(\frac{1}{2},0\right)$ & $\mathbf{2}$ & $\frac{1}{2}$ \\
			$\chi_i$ & $\frac{3}{2}$ & $\left(0,\frac{1}{2}\right)$ & $\mathbf{2}$ & $-\frac{1}{2}$ \\
			$D$ & 2 & $(0,0)$ & $\mathbf{1}$ & 0 \\
			\bottomrule
		\end{tabular}
		\caption{Unconstrained fields in the Weyl multiplet and their transformation properties under the $\mathcal{N}=2$ superconformal algebra.  }
		\label{tab:WeylMultTable}
	\end{table}
	
	An $\mathcal{N}=2$ SCFT can be placed on a curved manifold by coupling the theory to a background $\mathcal{N}=2$ Weyl multiplet \cite{Festuccia:2011ws}. This will generically break supersymmetry; in order to preserve supersymmetry, the background must be chosen carefully. The Weyl multiplet contains the gauge fields for $\mathcal{N}=2$ superconformal symmetry, as well as auxiliary fields that are required for the closure of the superconformal algebra. Some of the fields in the Weyl multiplet, such as the spin connection $\omega_\mu{}^{ab}$, are determined in terms of the others by suitably chosen curvature constraints \cite{DEWIT1980186}. The quantum numbers for the unconstrained Weyl multiplet fields are given in Table \ref{tab:WeylMultTable}. The Poincar\'e and conformal SUSY transformations for these unconstrained fields are (see  (20.69) in~\cite{Freedman:2012zz}):
	\es{eq:WeylSUSY}{\delta \tensor{e}{_\mu^a}&= \frac{1}{2}\bareps^{i}\gamma^a \tensor{\psi}{_\mu_i} + \frac{1}{2}\bareps_{i}\gamma^a \tensor{\psi}{_\mu^i}\, , \\
		\delta b_{\mu} &= \frac{1}{2}\bareps^i\tensor{\phi}{_\mu_i}- \frac{1}{2}\bar{\eta}^i\tensor{\psi}{_\mu_i}-\frac{3}{8}\bareps^{i}\gamma_{\mu}\chi_i + \frac{1}{2}\bareps_i\tensor{\phi}{_\mu^i}- \frac{1}{2}\bar{\eta}_i\tensor{\psi}{_\mu^i}-\frac{3}{8}\bareps_{i}\gamma_{\mu}\chi^i\, , \\
		\delta A_{\mu} &= -\frac{i}{2}\bareps^i\tensor{\phi}{_\mu_i}- \frac{i}{2}\bar{\eta}^i\tensor{\psi}{_\mu_i}-\frac{3}{8}i\bareps^{i}\gamma_{\mu}\chi_i + \frac{i}{2}\bareps_i\tensor{\phi}{_\mu^i}+ \frac{i}{2}\bar{\eta}_i\tensor{\psi}{_\mu^i}+\frac{3}{8}i\bareps_{i}\gamma_{\mu}\chi^i\, , \\
		\delta \tensor{V}{_{\mu i}^j}&=  -\bareps_i \phi^j_{\mu} - \bar{\eta}_i \psi^j_\mu + \frac{3}{4}\bareps_i\gamma_\mu\chi^j + \bareps^j\tensor{\phi}{_\mu_i} + \bar{\eta}^j\tensor{\psi}{_\mu_i} -\frac{3}{4}\bareps^j\gamma_\mu\chi_i \\
		&\, -\frac{1}{2}\tensor{\delta}{_i^j}\left(-\bareps_k\tensor{\phi}{_\mu^k}-\bar{\eta}_k \tensor{\psi}{_\mu^k}+ \frac{3}{4}\bareps_k\gamma_\mu\chi^k+\bareps^k\tensor{\phi}{_\mu_k}+\bar{\eta}^k\tensor{\psi}{_\mu_k}-\frac{3}{4}\bareps^k\gamma_\mu\chi_k\right)\, , \\
		\delta \tensor{\psi}{_\mu^i}&=\left(\partial_\mu+ \frac{1}{2}b_\mu+\frac{1}{4}\gamma^{ab}\omega_{\mu ab}-\frac{1}{2}iA_\mu\right)\epsilon^i-\tensor{V}{_{\mu\,j}^i}\epsilon^j - \frac{1}{16}\gamma^{ab}T^{-}_{ab}\varepsilon^{ij}\gamma_\mu \epsilon_j - \gamma_\mu \eta^i\, , \\
		\delta \tensor{\psi}{_\mu_i}&=\left(\partial_\mu+ \frac{1}{2}b_\mu+\frac{1}{4}\gamma^{ab}\omega_{\mu ab}+\frac{1}{2}iA_\mu\right)\epsilon_i+\tensor{V}{_{\mu\,i}^j}\epsilon_j - \frac{1}{16}\gamma^{ab}T^{+}_{ab}\varepsilon_{ij}\gamma_\mu \epsilon^j - \gamma_\mu \eta_i\, , \\
		\delta T^{-}_{ab} &= 2\bareps^i \hat{R}_{ab}(Q^j) \varepsilon_{ij}\, , \\
		\delta T^{+}_{ab} &= 2\bareps_i \hat{R}_{ab}(Q_j) \varepsilon^{ij}\, , \\
		\delta \chi^i &= \frac{1}{2}D\epsilon^i + \frac{1}{6}\gamma^{ab}\left[-\frac{1}{4}\slashed{\mathcal{D}}T^{-}_{ab}\varepsilon^{ij}\epsilon_j - \hat{R}_{ab}(\tensor{U}{_j^i})\epsilon^j+ i\hat{R}_{ab}(T)\epsilon^i + \frac{1}{2}T^{-}_{ab}\varepsilon^{ij}\eta_j\right]\, ,\\
		\delta \chi_i &= \frac{1}{2}D\epsilon_i + \frac{1}{6}\gamma^{ab}\left[-\frac{1}{4}\slashed{\mathcal{D}}T^{+}_{ab}\varepsilon_{ij}\epsilon^j + \hat{R}_{ab}(\tensor{U}{_i^j})\epsilon_j- i\hat{R}_{ab}(T)\epsilon_i + \frac{1}{2}T^{+}_{ab}\varepsilon_{ij}\eta^j\right]\, ,\\
		\delta D &= \frac{1}{2}\bareps^i\slashed{\mathcal{D}}\chi_i+\frac{1}{2}\bareps_i\slashed{\mathcal{D}}\chi^i \, ,
	}
	where the derivative $\slashed{\mathcal{D}}\equiv \gamma^\mu \mathcal{D}_\mu$ contains all the gauge fields in the Weyl multiplet except for the frame field.\footnote{This also includes the gauge fields of Poincar\'e SUSY ($\psi_{\mu i}$, $\tensor{\psi}{_{\mu}^i}$) and conformal SUSY ($\phi_{\mu i}$, $\tensor{\phi}{_{\mu}^i}$), along with the gauge fields of all the bosonic symmetries of the $\mathcal{N}=2$ superconformal group.} $\hat{R}_{ab}$ is the covariant curvature for the corresponding gauge symmetry and  $\phi^i_{\mu}$ is a constrained field that is determined in terms of the unconstrained fermionic fields in the Weyl multiplet. 
	
	The spin connection $\tensor{\omega}{_\mu^{ab}}$ and the $\mathfrak{su}(2)_R$ gauge field $\tensor{V}{_{\mu\,i}^j}$ are the only gauge fields whose contributions to the covariant derivative are relevant for our discussion of Weyl multiplet backgrounds. The expressions for the curvatures appearing in the above SUSY transformations as well as the constrained fields $\phi^i_{\mu}$ and $\omega^{ab}_\mu$ are 
	\es{eq:Curvs}{\hat{R}_{\mu\nu}(T) &= \partial_\mu A_\nu - \partial_\nu A_\mu + (\text{higher order terms})\, ,\\
		\hat{R}_{\mu\nu}(\tensor{U}{_i^j}) &= \partial_\mu \tensor{V}{_{\nu\, i}^j} - \partial_\nu \tensor{V}{_{\mu\, i}^j} + (\text{higher order terms})\, ,\\
		\hat{R}_{\mu\nu}(Q^i) &= \partial_\mu \psi^i_\nu - \partial_\nu \psi^i_\mu -\gamma_\mu \phi^i_\nu + \gamma_\nu \phi^i_\mu + (\text{higher order terms})\, ,\\
		\phi^i_\mu&= -\frac{1}{2}\gamma^\nu(\partial_\mu \psi^i_\nu-\partial_\nu \psi^i_\mu) + \frac{1}{6}\gamma_\mu \gamma^{ab}\partial_a \psi^i_b + \frac{1}{4}\gamma_\mu\chi^i+ (\text{higher order terms})\, ,\\
		\omega^{ab}_\mu &= 2 e^{\nu\,[a}\partial_{[\mu}e_{\nu]}^{b]}  - e^{\nu[a}e^{b]\sigma}e_{\mu c}\partial_\nu e^c_\sigma +2e^{[a}_\mu e^{b]\,\nu} b_\nu + (\text{higher order terms}) \,, }
	where the higher order terms are quadratic in Weyl multiplet fields (like $T_{ab}^{-}\gamma_\mu\psi_{\nu\,j}$). These higher-order terms are irrelevant for our discussion; the complete expressions for the curvature and all the constrained fields can be found in \cite{Freedman:2012zz} and \cite{Lauria:2020rhc}. We will now describe how the SUSY transformations in \eqref{eq:WeylSUSY} can be used to identify Killing spinors on $\mathbb{R}^4$ and on round $S^4$. 
	
	\subsection{Killing spinors in \texorpdfstring{$\mathbb{R}^4$}{R⁴} and round \texorpdfstring{$S^4$}{S⁴}}
	
	When placing an ${\cal N} = 2$ SCFT on any manifold, the background values of all the fermionic operators in the Weyl multiplet are set to zero. The bosonic operators thus remain invariant under any SUSY transformation. The background values of the bosonic operators then have to be chosen such that the background values of the fermionic operators remain zero under the preserved SUSY transformation. That is, they must obey
	\es{eq:confkilling}{
	  \delta \chi^i=\delta \chi_i=\delta \tensor{\psi}{_\mu^i}=\delta \tensor{\psi}{_\mu_i}=0\,.
	}
	
	For an $\mathcal{N}=2$ SCFT in flat space, the simplest supersymmetry-preserving background has all the fields in the Weyl multiplet except for the frame field set to zero. The resulting Killing equations are
	\begin{equation}\label{eq:killingspin}
		\partial_{\mu}\epsilon^i = \gamma_\mu \eta^i\, , \qquad 
		\partial_{\mu}\epsilon_i = \gamma_\mu \eta_i\, . 
	\end{equation}
	These equations have the general solution
	\begin{equation}
		\label{eq:epsflat}
		\epsilon_{\mathbb{R}^4}^i= \alpha^i +  \slashed{x} \beta^i\, , \qquad\epsilon_{\mathbb{R}^4\,i}= \alpha_i + \slashed{x} \beta_i\, , \qquad \eta_{\mathbb{R}^4}^i=\beta^i\, , \qquad\eta_{\mathbb{R}^4\,i}=\beta_i\, ,
	\end{equation}
	where $\slashed{x}\equiv x_{\mu}\gamma^\mu$, and $\alpha^i, \,\beta^i,\, \alpha_i,\, \text{and\,} \beta_i$ are arbitrary chiral spinors.  These spinors contain $16$ independent parameters, corresponding to the $16$ supersymmetries of $\mathcal{N}=2$ SCFTs on $\mathbb{R}^4$. The quantum numbers of these spinors are listed in Table \ref{tab:killingSolnTable}.
	
	\begin{table}[!tb]
		\centering
		\renewcommand{\arraystretch}{1.2}
		\begin{tabular}{ccc}
			\toprule
			Spinor & Lorentz rep  & $\mathfrak{u}(1)_R$ charge \\
			\midrule
			$\alpha^i$& $\left(\frac{1}{2},0\right)$ & $\frac{1}{2}$ \\
			$\alpha_i$&  $\left(0,\frac{1}{2}\right)$ &  $-\frac{1}{2}$ \\
			$\beta^i$& $\left(0,\frac{1}{2}\right)$ &  $\frac{1}{2}$ \\
			$\beta_i$& $\left(\frac{1}{2},0\right)$ &  $-\frac{1}{2}$ \\
			\bottomrule
		\end{tabular}
		\caption{Quantum numbers of spinors parametrizing superconformal transformations in \eqref{eq:epsflat} and \eqref{eq:epsS4}.}
		\label{tab:killingSolnTable}
	\end{table}

	Similarly, to place the theory on round $S^4$, we can take the frame field to be given by \eqref{frame} and set all other independent fields in the Weyl multiplet to zero. The resulting Killing equations are
	\begin{equation}\label{eq:killingspinround}
		D_{\mu}\epsilon^i=\left(\partial_{\mu} + \frac{1}{4}\gamma^{ab}\omega_{\mu ab}\right)\epsilon^i = \gamma_\mu \eta^i\, , \qquad 
		D_{\mu}\epsilon_i=\left(\partial_{\mu} + \frac{1}{4}\gamma^{ab}\omega_{\mu ab}\right)\epsilon_i = \gamma_\mu \eta_i\, .
	\end{equation}
	The general solution to the above equations are related to \eqref{eq:epsflat} by the Weyl transformation given in \eqref{eq:S4metric}. The explicit solution is
	\es{eq:epsS4}{
		\epsilon_{S^4}^i= \frac{\sqrt{2}}{\sqrt{1+x^2}}(\alpha^i +  \slashed{x} \beta^i)\, , & \qquad\epsilon_{S^4\,i}= \frac{\sqrt{2}}{\sqrt{1+x^2}}(\alpha_i +  \slashed{x} \beta_i)\, ,\\
		\eta_{S^4}^i= \frac{1}{\sqrt{2}\sqrt{1+x^2}}(\beta^i -  \slashed{x} \alpha^i)\, , & \qquad\eta_{S^4\,i}= \frac{1}{\sqrt{2}\sqrt{1+x^2}}(\beta_i -  \slashed{x} \alpha_i)\, ,\\
	}
	where again there are $16$ parameters and therefore $16$ supersymmetries.  The $\alpha^i, \alpha_i, \beta^i$ and $\beta_i$ spinors have the same chirality as the flat space case.  
	The $\tau$ deformation described in Section~\ref{sec:taudef} is a deformation by an exactly marginal operator which does not break any of these 16 supersymmetries. In contrast, the other two deformations that we consider in this work, mass and squashing, break several of the supersymmetries of the $\mathcal{N}=2$ superconformal group, leaving only a proper subgroup unbroken. The derivation of two deformations, given by \eqref{eq:masscoupl} and \eqref{eq:squashcoupl}, is provided in Appendices \ref{sec:massdeformderiv} and \ref{sec:squashedSUSYconstr} respectively. We also discuss the supersymmetries that are preserved by these deformations.
	
	\section{Mass deformation on \texorpdfstring{$S^4$}{S⁴}}
	\label{sec:massdeformderiv}
	
	To introduce a mass deformation for an $\mathcal{N}=2$ SCFT on $S^4$ while preserving supersymmetry, we can couple the $\mathcal{N}=2$ flavor current multiplet to a background vector multiplet. Supersymmetry is preserved by choosing appropriate background values for the scalar fields in the vector multiplet.
	
	\subsection{Background vector multiplet}
	\label{sec:vecmultN2}
	The field content of an $\mathcal{N}=2$ vector multiplet is summarized in Table \ref{tab:vecMultTable}. The Poincar\'e and conformal SUSY transformations for the vector multiplet fields are \cite{Lauria:2020rhc}
	\es{eq:vecmult}{
		\delta X &=\frac{1}{2}\bareps^{i}\Omega_i \, , \\
		\delta \overline{X} &= \frac{1}{2} \bareps_{i}\Omega^i \, , \\
		\delta A_{\mu} &= \frac{1}{2} \varepsilon^{ij}\bareps_i \gamma_{\mu}\Omega_j+\frac{1}{2}\varepsilon_{ij}\bareps^i \gamma_{\mu}\Omega^j \, ,\\
		\delta \Omega_i &= \slashed{\partial}X \epsilon_i +\frac{1}{4}\gamma^{ab}F^{-}_{ab}\varepsilon_{ij}\epsilon^{j}+Y_{ij}\epsilon^{j} + 2 X \eta_i \, ,\\
		\delta \Omega^i &= \slashed{\partial}\overline{X} \epsilon^i +\frac{1}{4}\gamma^{ab}F^{+}_{ab}\varepsilon^{ij}\epsilon_{j}+Y^{ij}\epsilon_{j} + 2 \overline{X} \eta^i\, ,\\
		\delta Y_{ij} &= \frac{1}{2}\bareps_{(i} \slashed{D}\Omega_{j)}+\frac{1}{2}\varepsilon_{ik}\varepsilon_{jl}\bareps^{(k} \slashed{D}\Omega^{l)}\, .
	}
	
	\begin{table}[!tb]
		\centering
		\renewcommand{\arraystretch}{1.2}
		\begin{tabular}{ccccc}
			\toprule
			Field & Conformal dimension & Lorentz rep & $\mathfrak{su}(2)_R$ rep & $\mathfrak{u}(1)_R$ charge \\
			\midrule
			$X$ & 1 & $(0,0)$ & $\mathbf{1}$ & $1$ \\	
			$\overline{X}$ & 1 & $(0,0)$ & $\mathbf{1}$ & $-1$ \\
			$A_{\mu}$ & 1 & $\left(\frac{1}{2},\frac{1}{2}\right)$ & $\mathbf{1}$ & $0$ \\
			$\Omega_i$ & $\frac{3}{2}$& $\left(\frac{1}{2},0\right)$ & $\mathbf{2}$ & $\frac{1}{2}$ \\
			$\Omega^i$ & $\frac{3}{2}$ & $\left(0,\frac{1}{2}\right)$ & $\mathbf{2}$ & $-\frac{1}{2}$ \\
			$Y_{ij}$ & 2 & $\left(0,0\right)$ & $\mathbf{3}$ & $0$ \\
			\bottomrule
		\end{tabular}
		\caption{Fields in the vector multiplet and their quantum numbers under the $\mathcal{N}=2$ superconformal algebra.}
		\label{tab:vecMultTable}
	\end{table}
	
	The background values of the fermionic fields as well as the vector field are set to zero,
	\begin{equation}
		A_{\mu} = \Omega_i =\Omega^i = 0 \, ,
	\end{equation}
	in order to preserve the $\mathfrak{so}(5)$ isometry of $S^4$. In order to have $\delta \Omega_i = \delta \Omega^i = 0$, the values for the $X$, $\overline{X}$ and $Y_{ij}$ fields must satisfy the constraints
	\begin{equation}\label{eqn:massinv}
		Y_{ij}\epsilon^j + 2 X \eta_i = 0\, , \qquad Y^{ij}\epsilon_j + 2\overline{X} \eta_i = 0 \, .
	\end{equation}
	Substituting in the expressions in \eqref{eq:epsS4} for the Killing spinors on $S^4$ and setting $X=\mathbf{m}$ and  $ \overline{X}=\overline{\mathbf{m}}$,  we find the relations
	\es{eq:massbkgrnd}{Y_{ij}\alpha^j = -\mathbf{m}\beta_i\, ,\qquad& Y_{ij}\beta^j = \mathbf{m}\alpha_i \, , \\
		Y^{ij}\alpha_j = -\overline{\mathbf{m}}\beta^i\, ,\qquad& Y^{ij}\beta_j = \overline{\mathbf{m}}\alpha^i\, . \\
	}
	This system of equations has non-trivial solutions if
	\begin{equation}
		Y^{ij}Y_{jk} = -\mathbf{m}\overline{\mathbf{m}}\tensor{\delta}{^i_k}\, .
	\end{equation}
	Without loss of generality, we can fix this background value of $Y_{ij}$ to lie along a Cartan direction and hence set $Y_{11} = Y_{22} = 0$. Our chosen background for the mass deformation is thus given by
	\es{eq:massvecbkgrnd}{
		X = \mathbf{m} \, , \qquad \overline{X} = \overline{\mathbf{m}}\, ,   \qquad
		& -Y_{12} = -Y_{21} = Y^{12} = Y^{21}  = -\sqrt{\mathbf{m}\overline{\mathbf{m}}}\, , \\
		\hspace{2mm} Y_{11} = Y_{22} &= Y^{11} = Y^{22}  =0\, .}
			  
	This mass background breaks conformal symmetry. It also breaks $\mathfrak{u}(1)_R$ symmetry since the $\mathfrak{u}(1)_R$\nobreakdash-charged fields $X$ and $\overline{X}$ are given non-zero background values. Additionally, the non-zero background value of $Y_{ij}$ on $S^4$ breaks $\mathfrak{su}(2)_R$ down to the $\mathfrak{so}(2)_R$ Cartan generated by $\sigma^3$.  Equations \eqref{eq:massbkgrnd} and \eqref{eq:massvecbkgrnd} can be used to expressed the $\beta^i$ spinors in terms of $\alpha_i$ spinors as well as the $\beta_i$ spinors in terms of $\alpha^i$ spinors. Thus, a massive $\mathcal{N}=2$ theory on $S^4$ has only 8 supersymmetries. The supersymmetry algebra of this theory is $\mathfrak{osp}(2\,|\,4)$, which contains the bosonic subalgebra $\mathfrak{so}(5)\times \mathfrak{so}(2)_R$, where $\mathfrak{so}(5)$ corresponds to the isometries of $S^4$.

	\subsection{Flavor current multiplet}
	\label{sec:flavmultN2}
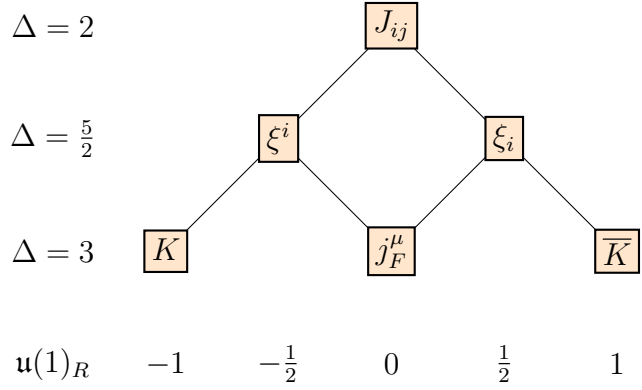
\begin{figure}
		\centering
		\begin{tikzpicture}[box/.style = {draw,black,thick,rectangle,fill=orange!20,inner sep=.1cm},scale=1.5]
			\draw (0, -4) -- (-1, -5);
			\draw (0, -4) -- (1, -5);
			\draw (-1, -5) -- (0, -6);
			\draw (-1, -5) -- (-2, -6);
			\draw (1, -5) -- (0, -6);
			\draw (1, -5) -- (2, -6);
			\node[box] at (0, -4) {$J _{ij}$};
			\node[box] at (-1, -5) {${\xi^i}$};
			\node[box] at (1, -5) {${\xi_i}$};
			\node[box] at (0, -6) {${j_F^{\mu}}$};
			\node[box] at (-2, -6) {$ K _{\text{}}$};
			\node[box] at (2, -6) {$\overline{K}$};
			\node[] at (-3, -4) {$\Delta = 2$};
			\node[] at (-3, -5) {$\Delta = \frac{5}{2}$};
			\node[] at (-3, -6) {$\Delta = 3$};
			\node[] at (0, -7) {$0$};
			\node[] at (1, -7) {$\frac{1}{2}$};
			\node[] at (2, -7) {$1$};
			\node[] at (-1, -7) {$-\frac{1}{2}$};
			\node[] at (-2, -7) {$-1$};
			\node[] at (-3, -7) {$\mathfrak{u}(1)_R $};
		\end{tikzpicture}
		\caption{Operators in the flavor multiplet of $\mathcal{N}=2$ SCFT. The $\mathfrak{u}(1)_R$ charge of the operators are listed below.}
		\label{fig:flavormult}
	\end{figure}

	The vector multiplet  background in \eqref{eq:massvecbkgrnd} couples to a flavor current multiplet to give the mass deformation. The flavor current multiplet contains the flavor current $j_\mu$, two complex scalars $K$ and $\overline{K}$, an $\mathfrak{su}(2)_R$ doublet of left-handed fermionic operators $\xi^i$, an $\mathfrak{su}(2)_R$ doublet of two right-handed fermionic operators $\xi_i$, and an $\mathfrak{su}(2)_R$ triplet of real scalar operators $J_{ij}=J_{ji}=\epsilon_{ik}\epsilon_{jl}J^{kl}= (J^{ij})^*$, as depicted in Figure \ref{fig:flavormult}. The SUSY transformations for the flavor multiplet are \cite{Dempsey:2024vkf}
	\es{eq:SUSYflavor}{\delta J_{ij} &= \bareps_{(i}\xi_{j)}+\varepsilon_{ik}\varepsilon_{jl}\epsilon^{(k}\xi^{l)}\, ,\\
		\delta \xi^{i} &= \frac{1}{2}(\slashed{\partial}J^{ij})\epsilon_j+ \frac{1}{2}\varepsilon^{ij}\slashed{j}_F\epsilon_j-\frac{1}{2} K\epsilon^{i}+2J^{ij}\eta_j\, ,\\
		\delta \xi_{i} &= \frac{1}{2}(\slashed{\partial}J_{ij})\epsilon^j+ \frac{1}{2}\varepsilon_{ij}\slashed{j}_F\epsilon^j-\frac{1}{2}\overline{K}\epsilon_{i}+2J_{ij}\eta^j \, ,\\
		\delta j_F^\mu &= \frac{1}{2}\bareps^{i}\gamma^{\mu\nu}(D_\nu \xi^{j})\varepsilon_{ij}+\frac{1}{2}\bareps_{i}\gamma^{\mu\nu}(D_\nu \xi_{j})\varepsilon^{ij}+\frac{3}{2}\bar{\eta}^i\gamma^a\xi^j\varepsilon_{ij}+\frac{3}{2}\bar{\eta}_i\gamma^a\xi_j\varepsilon^{ij}\, ,\\
		\delta K &= -\bareps_i \slashed{D}\xi^{i}+2\bar{\eta}_i\xi^i\, ,\\
		\delta \overline{K} &= -\bareps^i \slashed{D}\xi_{i}+2\bar{\eta}^i\xi_i \, .
	}

	The coupling of the flavor current multiplet and the vector multiplet is given by
	\begin{equation}\label{eq:flavveccoupl}
		S_{m} = \int d^4 x \sqrt{g(x)} \,(A_{\mu}j_F^{\mu}- X K - \overline{X}\,\overline{K}+J_{ij}Y^{ij} - \overline{\xi}^{i}\Omega_i- \overline{\xi}_{i}\Omega^i)\, .
	\end{equation}
	This coupling term is invariant under the superconformal transformations listed in \eqref{eq:vecmult} and \eqref{eq:SUSYflavor}. Substituting in the vector background in \eqref{eq:massvecbkgrnd} gives the mass deformation
		\begin{equation}\label{eq:massdefp1}
		S = - \int d^4 x \sqrt{g(x)} \,( \mathbf{m} K + \overline{\mathbf{m}}\,\overline{K}+ 2\sqrt{\mathbf{m}\overline{\mathbf{m}}}J_{12})\, .
	\end{equation}
	We point out later on, below \eqref{eq:masssquashmix}, that consistency with the squashing deformation requires setting $\mathbf{m} = im$ and $\overline{\mathbf{m}}=-im$. Fixing these values in \eqref{eq:massdefp1} gives the expression for the mass deformation identified in \eqref{eq:masscoupl}.
	
\section{SUSY on squashed $S^4$}
\label{sec:squashedSUSYconstr}

We now move on to discussing $\mathcal{N}=2$ supersymmetric theories on a squashed $S^4$ that can also accomodate mass and $\tau$ deformations. The squashed $S^4$ metric is given in \eqref{eq:squashembed}. When the $S^4$ is squashed, some of the bosonic fields of the Weyl multiplet other than the frame field must also have non-trivial background values in order for some SUSY to be preserved. In this section, we first demonstrate that such theories can preserve at most two supersymmetries of the $\mathcal{N}=2$ superconformal algebra. We then review the squashed $S^4$ Weyl multiplet background that was identified in \cite{Pestun:2014mja}, correcting some typographical errors. Finally, we compute the SUSY transformations of the $\mathcal{N}=2$ stress-tensor multiplet which must be coupled to the Weyl multiplet background to obtain the squashing deformation in \eqref{eq:squashcoupl}. In the following discussion we put tildes on all background fields and supersymmetry parameters corresponding to the squashed sphere in order to distinguish them from those on the round sphere.

\subsection{Preserved supersymmetries on squashed \texorpdfstring{$S^4$}{S⁴}}

We will start by expanding both the background fields and the SUSY parameters $(\tilde \epsilon^i, \tilde \epsilon_i)$ up to first order in $(b-1)$, for instance 
\es{eq:exsq1}{
	\tensor{\widetilde{V}}{_{\mu\,i}^j} &= (b-1) \Delta \tensor{{V}}{_{\mu\,i}^j}+O((b-1)^2) \,,\\
	\widetilde{\epsilon}^i &= \epsilon^{i}+(b-1) \Delta \epsilon^{i}+O((b-1)^2) \,,\\
	\widetilde{\epsilon}_i &= \epsilon_i+(b-1) \Delta \epsilon_i+O((b-1)^2) \,.  
}
Note that 
an $\mathfrak{su}(2)_R$ gauge transformation with parameter $\tensor{\widetilde{U}}{_i^j} = \delta_i^j + (b-1)\,\Delta \tensor{{U}}{_i^j} + O((b-1)^2)$ changes the first order coefficients as
\es{eq:exsq2}{
	\Delta \tensor{{V}}{_{\mu\,i}^j}&\xrightarrow{\mathfrak{su}(2)_R \,\text{gauge}} \Delta \tensor{{V}}{_{\mu\,i}^j}-\partial_{\mu} \Delta \tensor{{U}}{_i^j}\,,\\
	\epsilon^{i} &\xrightarrow{\mathfrak{su}(2)_R\,\text{gauge}}\epsilon^{i}\,,\\
	\Delta \epsilon^{i} &\xrightarrow{\mathfrak{su}(2)_R \,\text{gauge}} \Delta \epsilon^{i} + \Delta \tensor{{U}}{^{i}_j}\epsilon^{j}\,,\\
	\epsilon_i &\xrightarrow{\mathfrak{su}(2)_R\,\text{gauge}}\epsilon_i\,,\\
	\Delta \epsilon_i &\xrightarrow{\mathfrak{su}(2)_R \,\text{gauge}} \Delta \epsilon_i + \Delta \tensor{{U}}{_i^j}\epsilon_j\,,  }
where $\Delta \tensor{{U}}{_i^j} =- \Delta \tensor{{U}}{^j_i} $, while leaving the zeroth order terms unchanged. Working at first order in perturbation theory in $(b-1)$, we can ask which of the $16$ Killing spinors \eqref{eq:epsS4} would be preserved by the squashing up to correction terms that are subleading in $(b-1)$.

 Since the squashing deformation in \eqref{eq:squashembed} breaks the isometry of $S^4$ from $\mathfrak{so}(5)$ to ${\mathfrak{so}(2)\times\mathfrak{so}(2)}$, we would expect the supersymmetry algebra on the squashed $S^4$ to be a proper subalgebra of the $\mathcal{N}=2$ superconformal algebra. In particular, the spacetime part of the bosonic generators of the supersymmetry algebra on squashed $S^4$ must be a subset of the generators of the $\mathfrak{so}(2)\times\mathfrak{so}(2)$ isometries of the squashed sphere. We can use our knowledge of these isometries of squashed $S^4$ to identify the Killing spinors of round $S^4$ that might be preserved by the chosen squashing deformation. The Killing spinors of any supersymmetry algebra on the squashed $S^4$ at zeroth order in $(b-1)$ must be some linear combination of the subset of the round $S^4$ Killing spinors that are identified by this procedure. The subleading terms in $(b-1)$ of the squashed $S^4$ Killing spinors cannot be constrained by this procedure since they can be changed by an $\mathfrak{su}(2)_R$ gauge transformation, as shown in \eqref{eq:exsq2}, or similarly, by an $\mathfrak{u}(1)_R$ gauge transformation.

We elaborate in detail how this set of preserved Killing spinors can be identified. The Killing vectors for the two unbroken isometries on squashed $S^4$ are dual to the $\xi^{12}$ and $\xi^{34}$ one-forms, where $\xi^{pq}$ is defined in \eqref{KVGen}. In terms of the embedding space variables used in \eqref{eq:squashembed}, these isometry transformations are just rotations in the $(y_1,y_2)$ plane and $(y_3,y_4)$ plane respectively. At a generic point on the sphere except the poles, the action of these isometry generators is equivalent to a translation on the sphere along with a local rotation. However, at the poles of the sphere, an isometry transformation is equivalent to just a local rotation. Hence, it is convenient to focus on studying the action of the supersymmetry generators of squashed $S^4$ at the north pole ($\vec{x} = 0$) to identify the preserved Killing spinors.

Any supersymmetry generator $\mathcal{Q}$ on squashed $S^4$ can be written in the form \eqref{eq:SUSYgen},
\es{eq:squashedSUSYgen}{\mathcal{Q} = \bareps^i Q_i + \bareps_i Q^i + \bar{\eta}^i S_i + \bar{\eta}_i S^i\, ,}
where the $\epsilon^i$, $\epsilon_i$, $\eta^i$ and $\eta_i$ spinors at zeroth order in $(b-1)$ are of the form \eqref{eq:epsS4}. When evaluated at $x^\mu = 0$, the anti-commutator $\{\mathcal{Q},\mathcal{Q}\}$ must be some linear combination of the two isometry generators---which at the north pole are equivalent to the rotation generators $M_{12}$ and $M_{34}$ of the $\mathcal{N}=2$ superconformal algebra given in \eqref{eq:SUSYAlg}---and possibly the $\SU(2)_R$ and $\U(1)_R$ generators. This condition constrains the $\alpha_i$, $\alpha^i$, $\beta^i$, and $\beta_i$ spinors that define the supersymmetry algebra via \eqref{eq:epsS4}. Since we eventually want to describe a mass-deformed theory on squashed $S^4$ we additionally require that the $\alpha_i$, $\alpha^i$, $\beta^i$, and $\beta_i$ spinors satisfy \eqref{eq:massbkgrnd}, where the vector multiplet background fields $Y_{ij}$, $X$, and $\overline{X}$ characterizing the mass deformation were identified in  \eqref{eq:massvecbkgrnd}.   Using the $\{Q_i, Q^j\}$ and $\{S_i, S^j\}$ anticommutators in \eqref{eq:SUSYAlg}, and requiring that $P_\mu$ and $K_\mu$ generators do not appear in the $\{\mathcal{Q},\mathcal{Q}\}$ anticommutator evaluated at the north pole gives the constraints
\es{eq:restrict1}{\bar{\alpha}^i\gamma^a\alpha_i=\bar{\beta}^i\gamma^a\beta_i=0\, .}
Similarly, using the $\{Q_i, S^i\}$ and $\{ Q^i, S_j\}$ anticommutators in \eqref{eq:SUSYAlg} and requiring that the rotation generators $M_{13}, M_{14}, M_{23}$ and $M_{24}$ do not appear in  $\{\mathcal{Q},\mathcal{Q}\}$ anticommutator evaluated at the north pole, we find the constraints
\es{eq:restrict2}{\bar{\alpha}^i\gamma^{ab}\beta_i=\bar{\alpha}_i\gamma^{ab}\beta^i=0\, ,}
for $\gamma^{ab} \in \{\gamma^{13},\gamma^{14},\gamma^{23},\gamma^{24}\}$.

In the set of squashed-$S^4$ supersymmetry algebras permitted by the constraints \eqref{eq:restrict1} and \eqref{eq:restrict2}, there are algebras with only nilpotent supercharges $\mathcal{Q}$ (satisfying $\mathcal{Q}^2 = 0$). We disregard these solutions. This is motivated by the fact that the localization calculations in  \cite{Pestun:2007rz} and \cite{Pestun:2014mja} used a SUSY generator that squared to an isometry generator along with a $\mathfrak{su}(2)_R$ generator. If an algebra satisfies the two constraints \eqref{eq:restrict1} and \eqref{eq:restrict2}, and contains non-nilpotent supercharges, we find that it must additionally have either $\alpha_i = 0$ or $\alpha^i=0$ for $i=1,2$. Since we wish to study the $\tau$ deformation along with the squashing deformation, and we will be using \eqref{eq:taupole} to relate this deformation to an insertion of the chiral primary $\mathcal{A}$ at the north pole, we  choose the solution $\alpha^i = 0$. This choice ensures that this insertion is invariant under SUSY transformations, in light of \eqref{eq:DefChiral}.\footnote{Equivalently, the choice $\alpha^i = 0$ follows from demanding that the SUSY generator $\mathbb{Q}$ used to obtain the result  \eqref{eq:taupole} is a part of the SUSY algebra on squashed $S^4$. This SUSY generator $\mathbb{Q}$ satisfies the condition $\mathbb{Q}\mathcal{A}(N)= 0$ \cite{Gomis:2014woa}.}

With this choice, and after using all the constraints in \eqref{eq:restrict1} and \eqref{eq:restrict2}, we find that the isometries of squashed sphere constrain SUSY algebras of mass-deformed $\mathcal{N}=2$ SCFTs on squashed $S^4$ to have at most two SUSY generators, with the generic SUSY generator $\mathcal{Q}$, defined in \eqref{eq:squashedSUSYgen} using $\eqref{eq:epsS4}$, parametrized by the spinors  
\es{eq:sqS4susysym}{\alpha_1 = \begin{pmatrix}
		0 \\ 0\\ 0 \\ \mathbf{a}_1 \\
	\end{pmatrix} \,, \qquad
	\alpha_2 = \begin{pmatrix}
		0 \\ 0\\ \mathbf{a}_2 \\ 0 \\
	\end{pmatrix} \,, \qquad \alpha^i = 0\, ,
}
where $\mathbf{a}_1$ and $\mathbf{a}_2$ are arbitrary parameters, and $\beta_i$ and $\beta^i$ are related to $\alpha_i$ and $\alpha^i$ by \eqref{eq:massbkgrnd} and \eqref{eq:massvecbkgrnd}.

	\subsection{Details on the Pestun squashing background}
	\label{sec:bkgrndsq}

	A family of SUSY-preserving Weyl multiplet backgrounds on a squashed $S^4$ was obtained in \cite{Pestun:2014mja} by picking a particular SUSY generator that is left unbroken by the squashing deformation, and then requiring that the background fields satisfy the conditions \eqref{eq:confkilling} for this generator using the SUSY transformations given in \eqref{eq:WeylSUSY}.\footnote{The same family of backgrounds was also obtained in \cite{Hama:2012bg}.} We restate this background below, fixing some typographical errors in \cite{Pestun:2014mja}. 
	
	The background was written in \cite{Pestun:2014mja} in terms of the angular variables $(\rho, \theta, \phi_1, \phi_2)$ that parametrize the squashed $S^4$ via
	\es{eq:angvardef}{
	  \begin{pmatrix} 
	   y_1 \\ 
	   y_2 \\ 
	   y_3 \\
	   y_4 \\
	   y_5
	  \end{pmatrix} = 
	   \begin{pmatrix}
	     b\sin\rho \cos\theta \cos\phi_1 \\
	     b\sin\rho \cos\theta \sin\phi_1 \\
	     b^{-1} \sin\rho \sin\theta \cos\phi_2 \\
	     b^{-1} \sin\rho \sin\theta \sin\phi_2 \\
	      \cos\rho
	   \end{pmatrix} \,,
		}
		where $y_i$ are the coordinates appearing in \eqref{eq:squashembed}.
	The frame was chosen to be 
	\es{eq:framePestun}{
	     \widetilde{e}^1 &= b\sin\rho \cos\theta\, d\phi_1 \, , \quad \widetilde{e}^2 = \frac{\sin\rho \sin\theta}{b}\, d\phi_2 \, , \quad 
		\widetilde{e}^3 = \sin\rho f_1\, d\theta + f_3 \,d\rho\, , \quad \widetilde{e}^4 = f_2 \,d\rho \, , }
	where the functions $f_i$ are defined as 
	\es{eq:defnfs}{
	         f_1(\theta, b) &= \sqrt{b^2\sin^2\theta + \frac{\cos^2\theta}{b^2}} \,,  \qquad
		f_2(\theta, \rho,  b) = \sqrt{\sin^2\rho + \frac{\cos^2\rho}{f_1(\theta, b)^2}} \,,\\ 
		f_3(\theta, \rho,  b) &= \frac{b^{-2}-b^2}{f_1(\theta,b)}\cos\rho \sin\theta \cos\theta \,.
	}

	The background values of the Weyl multiplet fields are as follows. The $\mathfrak{u}(1)_R$ background vector field $\tilde{A}_\mu$ is set to zero, and the $\mathfrak{su}(2)_R$ background vector field $ \tensor{\widetilde{V}}{^j_i} \equiv i \widetilde{V}_I \tensor{(\sigma^I)}{^j_i} = -\tensor{\widetilde{V}}{_i^j}$ is given by
	\es{eq:VhattoV}{ \begin{pmatrix}
			\widetilde{V}_1 \\ \widetilde{V}_2 \end{pmatrix} &= \begin{pmatrix}
			\cos(\phi_1 + \phi_2) & \sin(\phi_1 + \phi_2) \\
			-\sin(\phi_1 + \phi_2) & \cos(\phi_1 + \phi_2) \\
		\end{pmatrix}  \begin{pmatrix}
			\hat{V}_1 \\ \hat{V}_2 \end{pmatrix} \, , \qquad
		\widetilde{V}_3 = \hat{V}_3 \,, }
where the one-forms  $\hat{V}_I$ are defined by
	\es{eq:Vhatbkgrnd}{ \hat{V}_1 & = \frac{(f_1 -f_2)\cos\rho + \sin\rho\, \partial_\rho f_1 -\partial_\theta f_3}{2f_2} \, d\theta + \frac{f_3(\sin\rho \, \partial_\rho f_1 + f_1 \cos\rho - \partial_\theta f_3) - f_2\partial_\theta f_2}{2f_1 f_2 \sin\rho}\,d\rho  \\
		& -\frac{1}{4}c_1 \sin^2\rho \, f_1 d\theta - \frac{1}{4}(\sin\rho \, c_1 f_3  + c_2 \sin \rho \, f_2 ) \,d\rho \,, \\[1em]
		\hat{V}_2 & =  \left(\frac{b \sin ^2 \theta f_3}{2 f_1 f_2}-\frac{1}{4} b \sin(2\theta) \cos \rho \left(\frac{1}{f_1}-\frac{1}{f_2}\right)\right)\, d\phi_1 \\
		& + \left(\frac{\cos^2\theta f_3}{2 b f_1 f_2}+\frac{\sin (2\theta) \cos\rho }{4 b} \left(\frac{1}{f_1}-\frac{1}{f_2}\right) \right) \,d\phi_2\\
		& - \frac{b}{8} c_1 \sin(2\theta) \sin^2\rho \, d\phi_1 + \frac{1}{8b} c_1 \sin(2\theta)\sin^2\rho \, d\phi_2 - \frac{1}{4}c_3 \sin\rho f_2 \,d\rho \, ,  \\[1em]
		\hat{V}_3 & = \left(\frac{b\sin(2\theta)\cos\rho f_3}{4f_1 f_2} + \frac{b\sin^2\theta}{2f_1} + \frac{b\cos^2\theta}{2f_2} -\frac{1}{2}\right)\, d\phi_1 \\
		& +\left(-\frac{\sin(2\theta)\cos\rho f_3}{4 bf_1f_2}+\frac{\cos^2\theta}{2bf_1}+\frac{\sin^2\theta }{2bf_2}-\frac{1}{2}\right) \,d\phi_2 \\ 
		& - \frac{b}{8} c_2 \sin(2\theta) \sin^2\rho \, d\phi_1 + \frac{1}{8b} c_2 \sin(2\theta)\sin^2\rho \, d\phi_2 + \frac{1}{4}c_3 \sin^2\rho f_1 \,d\theta + \frac{1}{4} c_3 \sin\rho f_3 \,d\rho  \, ,  }
	where the three functions $c_i$, with $i = 1,2,3$, are arbitrary functions on the squashed $S^4$.
	Meanwhile, the background tensor $\widetilde{T}_{ab}$ is given by 
	\es{eq:tilTbkgrnd}{
	         \widetilde{T}_{34} &= c_3\, ,\\
		\widetilde{T}_{24} &= -2\sin\theta\left(\frac{1}{f_1} 
		    - \frac{1}{f_2}\right) + c_1 \cos\rho \sin\theta + c_2 \cos\theta  \, , \\
		\widetilde{T}_{23} &=- 2\sin\theta \frac{f_3}{f_1f_2} 
		    + c_1  \cos\theta - c_2 \cos\rho \cos\theta  \, , \\
		\widetilde{T}_{14} &=- 2\cos\theta\left(\frac{1}{f_1} 
		     - \frac{1}{f_2}\right) + c_1  \cos\rho\cos\theta - c_2 \sin\theta  \, , \\
		\widetilde{T}_{13} &= -2\cos\theta \frac{f_3}{f_1f_2}  
		     - c_1 \sin\theta - c_2 \cos\rho\cos\theta  \, , \\
		\widetilde{T}_{12} &= c_3 \cos\rho \, .
	}
	The self-dual and anti-self-dual tensor backgrounds are given by the linear combinations
	\es{eq:Tpmdefn}{
	  \widetilde{T}^{\pm}_{ab} 
	     = \frac{1}{2} \left( \widetilde{T}_{ab} \pm \frac 12 \epsilon_{abcd} \widetilde{T}^{cd} \right) \,.
	 }
In addition, the scalar field $\widetilde{D}$ is given by
	\es{eq:DRbkgrnd}{
	-\frac{1}{2}\left(\widetilde{D}+\frac{\widetilde{R}}{6}\right)
		&= \frac{1}{4f_1^2} -\frac{1}{4f_2^2} -\frac{1}{f_1 f_2} + \frac{f_3^2}{4f_1^2f_2^2}  - \frac{1}{16}\sin^2 \rho \, (c_1^2 + c_2^2 + c^2_3) \\
		& + c_1 \left( - \frac{\cos\rho}{4f_1}+ \frac{3\cos\rho}{4f_2} - \frac{f_3\cot (2\theta)}{2f_1f_2} + \frac{\sin\rho \, \partial_\rho f_1}{4f_1 f_2} - \frac{\partial_\theta f_3}{4f_1 f_2}\right)   \\
		& + c_2 \left(-\frac{\cot 2\theta}{2f_1} + \frac{f_3 \cos \rho }{4f_1 f_2} - \frac{\partial_\theta f_2}{4f_1 f_2}\right) + \frac{\sin \rho}{4f_2} \partial_\rho c_1 - \frac{f_3}{4f_1 f_2} \partial_\theta c_1  - \frac{1}{4f_1}\partial_\theta c_2  \,,
	}
where $\widetilde{R}$ is the Ricci scalar.

The functions $c_1$, $c_2$, and $c_3$ should be chosen appropriately to ensure that the Weyl multiplet backgrounds are well-defined on $S^4$. One such suitable choice  for these functions, at linear order in $(b-1)$, is 
	\es{eq:cfuncs}{ c_1 &= 2(b-1) \cos(2\theta) \cos\rho + O((b-1)^2)\, ,\\
		c_2 &= 2(b-1) \sin(2\theta)  + O((b-1)^2) \, ,\\
		c_3 &= 0\, .}

With this choice of the $c_i$ functions, we can expand the background fields $\widetilde{T}_{ab}$, $\tensor{\widetilde{V}}{_{\mu\, i }^j}$, and $\widetilde{D}$ in $(b-1)$ to identify a SUSY-preserving Weyl multiplet background at linear order in $(b-1)$, given by the fields
	\es{eq}{
	        \widetilde{T}_{ab} & = (b-1) \Delta T_{ab} + O((b-1)^2) \, ,\\
		\tensor{\widetilde{V}}{_{\mu\, i }^j } & = (b-1)\Delta \tensor{V}{_{\mu\, i }^j } + O((b-1)^2) \, ,   \\ 
		\widetilde{D} & = (b-1) \Delta D + O((b-1)^2) \, ,} 
	where 
		\begin{equation}\label{eq:tilTlin}
		\begin{aligned}
			&\Delta T_{34} = 0\, , \qquad             
			&\Delta T_{24} &= 2\sin\theta\, , \qquad           
			&\Delta T_{23} &= 2 \cos\rho \cos\theta  \, , \\
			&\Delta T_{14} =- 2\cos\theta \, , \qquad 
			&\Delta T_{13} &= 2 \cos\rho\sin\theta  \, , \qquad	
			&\Delta T_{12} &= 0 \, ,
		\end{aligned}
	\end{equation}
	and
	\es{eq:Vlin}{
	  \Delta \tensor{V}{_1^1} &= 
	     - \Delta \tensor{V}{_2^2} = -\frac{i }{4}  \sin^2\rho \,
	     ((d\phi_1+d\phi_2) \cos (2\theta)+d\phi_1-d\phi_2) \,, \\
	  \Delta \tensor{V}{_1^2} & = 
	     - d \left[ \frac{i}{2} e^{i(\phi_1+\phi_2)} \sin (2 \theta ) \cos \rho \right]  \,, \\
	  \Delta \tensor{V}{_2^1} &= 
	     - d \left[ \frac{i}{2}  e^{-i(\phi_1+\phi_2)} \sin (2 \theta ) \cos \rho \right] \,.
	}
The expressions for $\Delta \tensor{V}{_1^2}$ and $\Delta \tensor{V}{_2^1}$ are total derivatives and can therefore be set to zero by a gauge transformation, leaving $\Delta \tensor{V}{_1^1} = - \Delta \tensor{V}{_2^2}$ as the only non-zero component of $\Delta \tensor{V}{_i^j}$.  
	Additionally, the scalar $\Delta D$ is also set to zero due to the suitable choice of the $c_i$ functions made in \eqref{eq:cfuncs}. 
	
	After the gauge transformation mentioned above and the choices for the $c_i$, the non-zero linearized background fields can be written as
	\es{bkgrndintermsofxi}{\Delta \tensor{V}{_1^1} &= - \Delta\tensor{V}{_2^2} = -i \xi\,, \qquad \Delta T = -2\, *\!d\xi\,, }
	where 
	\es{xidef}{\xi = \frac{1}{2}\sin^2 \rho \,(\cos^2\theta \, d\phi_1 + \sin^2\theta\, d\phi_2) \,,}
	 is a one-form dual to a particular Killing vector on $S^4$, as described in \eqref{KVExplicit} using \eqref{KVGen}.
	Alternatively, the non-zero background fields can be expressed in terms of the modified stereographic coordinates \eqref{eq:varsR4toSquashedS4} as 
	\es{eq:squashbkgrnd}{ 
	    \Delta \tensor{V}{_1^1} &= - \Delta \tensor{V}{_2^2} 
	       = -\frac{2i \bold{M}_{\mu\nu}x^\mu\, dx^\nu}{(1+x^2)^2} \,, \\
	    \Delta \tensor{T}{^{-}} &= \left(\frac{2}{1+x^2}\right)^3 
	       \bold{M}_{\mu\nu} \,dx^\mu \wedge dx^\nu  \,, \\ 
	    \Delta \tensor{T}{^{+}} &= -\left(\frac{2}{1+x^2}\right)^3 
	       \left(x^2 \bold{M}_{\mu\nu} + 4 x^\rho \delta_{\rho\left[\mu\right.} 
	       \bold{M}_{\left. \nu \right]\lambda}x^\lambda \right)  
	       \,dx^\mu \wedge dx^\nu\,,
	}
where the matrix $\bold{M}$ is defined by
	\es{eq:matfields}{
	   \bold{M} = \begin{pmatrix}
			0&1&0&0\\
			-1&0&0&0\\
			0&0&0&-1\\
			0&0&1&0\\
		\end{pmatrix}\,.
	}
	
In the frame 
	\begin{equation}
		\begin{split}
			\widetilde{e}_i &= \frac{2 dx_i}{1+x^2} 
			   + (b-1)\left[\frac{2(M_* \cdot d\vec{x})_i}{1+x^2} 
			   +  \frac{8x_i (\vec{x}\cdot M_* \cdot d\vec{x})\, 
			   \vec{x}\cdot d\vec{x}}{(1+x^2)^3} 
			   -\frac{4x_i \vec{x}\cdot (s_i \mathbbm{1}_4 + M_*)
			   \cdot d\vec{x}}{(1+x^2)^2} \right]\\
			&{}+ \mathcal{O}\left((b-1)^2\right) \,,
		\end{split}
	\end{equation}
where $s_i = 1$ when $i = 1,2$, and $s_i = -1$ when $i=3,4$, and the matrix $M_* \equiv  \text{diag}(1,1,-1,-1)$, the SUSY generator that \cite{Pestun:2014mja} chose to leave unbroken by the squashing deformation is given by the spinors
	\es{eq:squasheps}{
	  \widetilde{\epsilon}_1
	     &= \frac{\sqrt{2}}{\sqrt{1+x^2}}
	        \begin{pmatrix}	0 \\ 0 \\ 0 \\ i \end{pmatrix} + O(b-1) \,, \qquad \quad \ 
	  \widetilde{\epsilon}_2
	     = \frac{\sqrt{2}}{\sqrt{1+x^2}}
	       \begin{pmatrix}	0 \\ 0 \\ -1 \\ 0 \end{pmatrix} + O(b-1) \,,
	       \\
	  \widetilde{\epsilon}^1
	     &= \frac{\sqrt{2}}{\sqrt{1+x^2}}
	       \slashed{x}\begin{pmatrix}	0 \\ 0 \\ -i \\ 0 \end{pmatrix} + O(b-1) \,, \qquad
	  \widetilde{\epsilon}^2
	     = \frac{\sqrt{2}}{\sqrt{1+x^2}} 
	         \slashed{x}\begin{pmatrix}	0 \\ 0 \\  0 \\ -1 \end{pmatrix}  + O(b-1) \,.
	 }
Comparing to \eqref{eq:epsS4}, we observe that the $\beta^i$ and $\alpha_i$ spinors for the squashing background satisfy \es{eq:masssquashmix}{\beta^i = i(\sigma_1)^{ij}\alpha_j\, .}
	Comparing \eqref{eq:masssquashmix} to the Killing spinors of the massive theory in \eqref{eq:massbkgrnd} and \eqref{eq:massvecbkgrnd}, we see that for the squashing and mass deformations to be compatible, we must have $\bold{m} = im$ and $\overline{\bold{m}} = -im$ in \eqref{eq:massvecbkgrnd}.

	Using  the SUSY transformations in \eqref{eq:WeylSUSY}, one can check that the background in \eqref{eq:squashbkgrnd} satisfies the conditions $\delta \tensor{\widetilde{\psi}}{_\mu^i} = \delta \tensor{\widetilde{\psi}}{_{\mu i}} = \delta \widetilde{\chi}^i = \delta \widetilde{\chi}_i = 0$  for the spinors in \eqref{eq:squasheps}, with no additional corrections to \eqref{eq:squasheps} at linear order in $(b-1)$. In fact, at linear order in $(b-1)$, we find that the background in \eqref{eq:squashbkgrnd} accomodates two possible supersymmetries, parametrized at zeroth order by the $\alpha_i$ spinors that we had identified in \eqref{eq:sqS4susysym}, with the corresponding $\beta^i$ spinors given by \eqref{eq:masssquashmix}, and with no additional corrections at linear order in $(b-1)$. It is unclear whether it is possible to choose the $c_i$ functions such that we obtain a Weyl multiplet background that preserves two supersymmetries on squashed $S^4$ at all orders in $(b-1)$, but this question is irrelevant for the analysis in this paper.
	
	\subsection{\texorpdfstring{$\mathcal{N}=2$}{N = 2} stress-tensor multiplet}
	\label{sec:stmultN2}
	
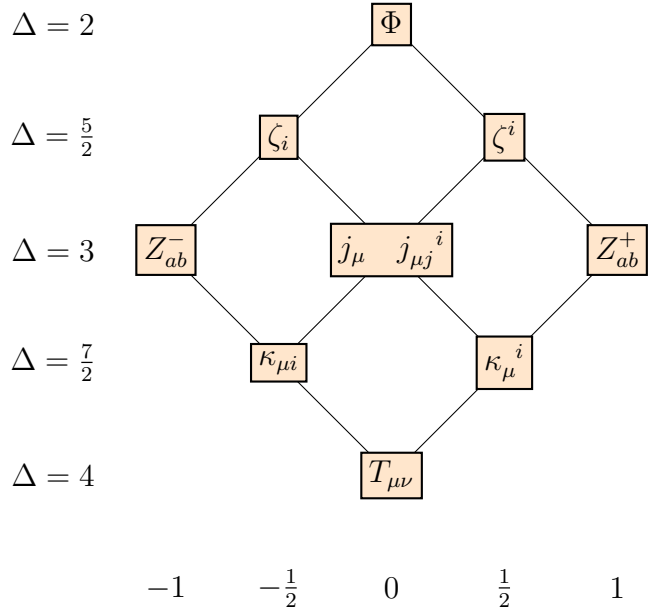
\begin{figure}[htb]
		\centering
		\usetikzlibrary{arrows.meta}
		\begin{tikzpicture}[box/.style = {draw,black,thick,rectangle,fill=orange!20,inner sep=.1cm},scale=1.5]
			\draw  (0, -4) -- (-1, -5);
			\draw (0, -4) -- (1, -5);
			\draw (-1, -5) -- (-2, -6);
			\draw (-1, -5) -- (-0, -6);
			\draw (1, -5) -- (2, -6);
			\draw (1, -5) -- (0, -6);
			\draw (-2, -6) -- (-1, -7);
			\draw (2, -6) -- (1, -7);
			\draw (0, -6) -- (-1, -7);
			\draw (0, -6) -- (1, -7);
			\draw (-1, -7) -- (-0, -8);
			\draw (1, -7) -- (0, -8);
			\node[box] at (0, -4) {$\Phi_{\text{}}$};
			\node[box] at (-1, -5) {$\zeta_i$};
			\node[box] at (1, -5) {${\zeta^{i}}$};
			\node[box] at (-2, -6) {${Z^{-}_{ab}}$};
			\node[box] at (2, -6) {${Z^{+}_{ab}}$};
			\node[box] at (0, -6) {${j}_{\mu} \quad {\tensor{j}{_{\mu j}^i}}$};
			\node[box] at (-1, -7) {$\kappa _{\mu i}$};
			\node[box] at (1, -7) {$\tensor{\kappa}{_{\mu}^i}$};
			\node[box] at (0, -8) {$T_{\mu\nu}$};
			\node[] at (-3, -4) {$\Delta = 2$};
			\node[] at (-3, -5) {$\Delta = \frac{5}{2}$};
			\node[] at (-3, -6) {$\Delta = 3$};
			\node[] at (-3, -7) {$\Delta = \frac{7}{2}$};
			\node[] at (-3, -8) {$\Delta = 4$};
			\node[] at (0, -9) {$0$};
			\node[] at (1, -9) {$\frac{1}{2}$};
			\node[] at (2, -9) {$1$};
			\node[] at (-1, -9) {$-\frac{1}{2}$};
			\node[] at (-2, -9) {$-1$};
		\end{tikzpicture}
		\caption{Operators appearing in the stress-tensor multiplet of $\mathcal{N}=2$ SCFTs. The $\mathfrak{u}(1)_R$ charges are listed for operators in each column are listed at the bottom.}
		\label{fig:stmultn2}
	\end{figure}

	Any $\mathcal{N}=2$ superconformal theory has several conserved currents, including a traceless stress-tensor $T_{\mu\nu}$, supercurrents for $\mathcal{N}=2$ SUSY $\kappa_{\mu i}$ and $\tensor{\kappa}{_\mu^i}$, the $\mathfrak{su}(2)_R$ current $\tensor{j}{_{\mu\,j}^i}$, and the $\mathfrak{u}(1)_R$ current $j_\mu$. These conserved currents along with several other fields together form a multiplet, called the stress-tensor multiplet. The complete field content of the $\mathcal{N}=2$ stress-tensor multiplet is depicted in Figure~\ref{fig:stmultn2}. The action of global Poincar\`e and conformal SUSY transformations on the operators in the $\mathcal{N}=2$ stress-tensor multiplet is given by\footnote{These transformations can be obtained by requiring that the SUSY algebra \eqref{eq:SUSYAlg} be satisfied. The normalization of the fields of this multiplet are chosen such that the coupling of the stress-tensor multiplet to the Weyl multiplet, which we discuss later on in \eqref{eq:weylst}, has simple numerical coefficients. The derivative $D_\mu$ in these transformations contains the gauge fields of the Weyl multiplet background that is used to define the $\mathcal{N}=2$ SCFT.}
	\es{eq:STmultSUSY}{\delta \Phi &= -\frac{1}{2} \bareps^i \zeta_i-\frac{1}{2} \bareps_i \zeta^i\,,\\
		\delta \zeta_i &= \,\epsilon_{ij}Z^{-}_{ab}\gamma^{ab}
		\epsilon^j +\frac{i}{2}\slashed{j}\epsilon_i +\frac{1}{2}\tensor{\slashed{j}}{_i^j}\epsilon_j -\frac{1}{2}\slashed{D}\Phi\epsilon_i- 2 \Phi \eta_i \,,\\
		\delta \zeta^i &= \epsilon^{ij}Z^{+}_{ab}\gamma^{ab}
		\epsilon_j-\frac{i}{2}\slashed{j}\epsilon^i  -\frac{1}{2}\tensor{\slashed{j}}{_j^i}\epsilon^j  -\frac{1}{2}\slashed{D}\Phi \epsilon^i - 2 \Phi \eta^i \,,\\
		\delta Z^{-}_{ab}&= \epsilon^{ij}\left(\frac{1}{24}\bareps_i\gamma^{\mu}\gamma_{ab}D_{\mu}\zeta_j -\frac{1}{16}\bareps_i\gamma^{\mu}\gamma_{ab}\tensor{\kappa}{_{\mu\,j}}-\frac{1}{16} \bar{\eta}_i \gamma_{ab} \zeta_j\right), \\
		\delta Z^{+}_{ab}&=\,\epsilon_{ij}\left(\frac{1}{24}\bareps^i\gamma^{\mu}\gamma_{ab}D_{\mu}\zeta^j - \frac{1}{16}\bareps^i\gamma^{\mu}\gamma_{ab}\tensor{\kappa}{_\mu^j}-\frac{1}{16} \bar{\eta}^i \gamma_{ab} \zeta^j\right), \\
		\delta j_\mu &= \frac{i}{3}\bareps^i \gamma_{\mu\nu}D^{\nu}\zeta_i-\frac{i}{3}\bareps_i \gamma_{\mu\nu}D^{\nu}\zeta^i +\frac{i}{2}\bareps^i \tensor{\kappa}{_{\mu i}}- \frac{i}{2}\bareps_i \tensor{\kappa}{_\mu^i} + i \bar{\eta}^i \gamma_\mu \zeta_i - i \bar{\eta}_i \gamma_\mu \zeta^i \,,\\
		\delta \tensor{j}{_{\mu\,j}^i} &= \left(-\frac{1}{3}\bareps^i \gamma_{\mu\nu}D^{\nu}\zeta_j+\frac{1}{3}\bareps_j \gamma_{\mu\nu}D^{\nu}\zeta^i +\bareps^i \tensor{\kappa}{_{\mu j}} -\bareps_j \tensor{\kappa}{_{\mu}^i} - \bar{\eta}^i\gamma_\mu\zeta_j +  \bar{\eta}_j\gamma_\mu\zeta^i\right)\\
		&-\frac{1}{2}\tensor{\delta}{_j^i}\left(-\frac{1}{3}\bareps^k \gamma_{\mu\nu}D^{\nu}\zeta_k+\frac{1}{3}\bareps_k \gamma_{\mu\nu}D^{\nu}\zeta^k +\bareps^k \tensor{\kappa}{_{\mu k}} -\bareps_k \tensor{\kappa}{_{\mu}^k}- \bar{\eta}^k\gamma_\mu\zeta_k +  \bar{\eta}_k\gamma_\mu\zeta^k\right)\,,\\
		\delta \tensor{\kappa}{_\mu_i} &= -2\varepsilon_{ij}\left(\frac{1}{3}\gamma_{\mu\nu}\gamma^{cd}\epsilon^j D^\nu Z_{cd}^{-}+ \gamma^{\nu \lambda}\epsilon^j D_\nu Z^{-}_{\mu \lambda} - \gamma_{\mu \lambda}\epsilon^j D_\nu Z^{-\,\nu \lambda}+\frac{2}{3} \gamma^{cd}\gamma_\mu \eta^j Z^-_{cd} \right)\\ 
		&-\frac{1}{2}T_{\mu\nu}\gamma^\nu\epsilon_i -\frac{i}{12}(\gamma^\nu\epsilon_i(D_\mu j_\nu+D_\nu j_\mu)+\gamma^{\lambda}\gamma^\nu\gamma_\mu\epsilon_i(D_\lambda j_\nu))-i \left(\eta_i j_\mu -\frac{1}{3}\gamma_{\mu\nu}\eta_i j^\nu \right)\\
		&+\frac{1}{6}(\gamma^\nu\epsilon_k(D_\mu \tensor{j}{_{\nu\,i}^k}+ D_\nu \tensor{j}{_{\mu\,i}^k})+\gamma^{\lambda}\gamma^\nu\gamma_\mu\epsilon_k(D_\lambda \tensor{j}{_{\nu\,i}^k})) + 2\left(\eta_k \tensor{j}{_{\mu \, i}^k} - \frac{1}{3} \gamma_{\mu\nu} \eta_k \tensor{j}{^{\nu \,}_{i}^k}\right)\,,\\
		\delta \tensor{\kappa}{_\mu^i} &= -2\varepsilon^{ij}\left(\frac{1}{3}\gamma_{\mu\nu}\gamma^{cd}\epsilon_j D^\nu Z_{cd}^{+}+ \gamma^{\nu \lambda}\epsilon_j D_\nu Z^{+}_{\mu \lambda} - \gamma_{\mu \lambda}\epsilon_j D_\nu Z^{+\,\nu \lambda}+\frac{2}{3} \gamma^{cd}\gamma_\mu \eta_j Z^+_{cd}\right)\\
		&-\frac{1}{2}T_{\mu\nu}\gamma^\nu\epsilon^i +\frac{i}{12}(\gamma^\nu\epsilon^i(D_\mu j_\nu+D_\nu j_\mu)+\gamma^{\lambda}\gamma^\nu\gamma_\mu\epsilon^i(D_\lambda j_\nu))+i \left(\eta^i j_\mu -\frac{1}{3}\gamma_{\mu\nu}\eta^i j^\nu \right)\\
		&-\frac{1}{6}(\gamma^\nu\epsilon^k(D_\mu \tensor{j}{_{\nu\,k}^i}+D_\nu \tensor{j}{_{\mu\,k}^i})+\gamma^{\lambda}\gamma^\nu\gamma_\mu\epsilon^k(D_\lambda \tensor{j}{_{\nu\,k}^i})) - 2\left(\eta^j \tensor{j}{_{\mu \, j}^i} - \frac{1}{3} \gamma_{\mu\nu} \eta^j \tensor{j}{^{\nu \,}_{j}^i}\right)\,,\\
		\delta T_{\mu\nu}&= -\frac{1}{2}\bareps^i\gamma_{ \tau(\mu}D^\tau \kappa_{\nu)\,i}-\frac{1}{2}\bareps_i\gamma_{\tau(\mu }D^\tau \tensor{\kappa}{_{\nu)}^i} + \frac{3}{8}\bar{\eta}^i \gamma_{(\mu} \kappa_{\nu) i} + \frac{3}{8}\bar{\eta}_i \gamma_{(\mu} \tensor{\kappa}{_{\nu)}^i}\,. 
	}

	When all the symmetries of the $\mathcal{N}=2$ superconformal group are elevated from global to local symmetries, the action of any $\mathcal{N}=2$ SCFT can be modified to include couplings to a background $\mathcal{N}=2$ Weyl multiplet in order to maintain local $\mathcal{N}=2$ superconformal symmetry. We express this action as $S[\mathcal{V}_{\text{Weyl}}]$, where $\mathcal{V}_{\text{Weyl}}$ represents the background Weyl multiplet. The Weyl multiplet contains the frame field $\tensor{e}{_\mu^a}$ which determines the spacetime metric $g_{\mu\nu}$. If the Weyl multiplet background has its frame field set to be $\tensor{e}{_\mu^a} = \tensor{\delta}{_\mu^a}$ and all other independent fields set to zero, we will reobtain the action of this $\mathcal{N}=2$ SCFT around flat space. If instead the Weyl multiplet background has its frame field set to the form \eqref{frame} and all other independent fields set to zero again, we will obtain the action of this $\mathcal{N}=2$ SCFT around a $S^4$ background. Let us refer to this background as $\mathcal{V}_{\text{Weyl, $S^4$ bk}}$. Under a general transformation parametrized by a small parameter $\lambda$ that changes the background Weyl multiplet as $\tilde{\mathcal{V}}_{\text{Weyl, $S^4$ bk}} \to \tilde{\mathcal{V}}_{\text{Weyl, $S^4$ bk}} + \lambda \tilde{\mathcal{V}}_{\text{Weyl}}$, the action changes as
	\es{eq:weylst}{S[\tilde{\mathcal{V}}_{\text{Weyl, $S^4$ bk}} &+ \lambda \tilde{\mathcal{V}}_{\text{Weyl}}]  = S[\mathcal{V}_{\text{Weyl, $S^4$ bk}}] + \lambda  S_{\text{Weyl-ST}} + O(\lambda^2)\,,\\
	S_{\text{Weyl-ST}} =\int d^4 x \, \sqrt{g(x)} \, \Big(&-\frac{1}{2}h^{\mu\nu}T_{\mu\nu}+ \tensor{\overline{\psi}}{^\mu^i}\tensor{\kappa}{_\mu_i} +\tensor{\overline{\psi}}{^\mu_i}\tensor{\kappa}{_\mu^i}+Z^{-}_{ab}\tensor{T}{^{-ab}}+Z^{+}_{ab}\tensor{T}{^{+ab}}\\
	&+\tensor{j}{_{\mu j}^i} \tensor{V}{^\mu_i^j} +{j_\mu} A^{\mu}+\overline{\chi}_i\zeta^i+\overline{\chi}^i\zeta_i+ D \Phi\Big),}
	where $h_{\mu\nu}$ is the linear order change in the metric under this transformation $g_{\mu\nu} \to g_{\mu\nu} + \lambda h_{\mu\nu}$.\footnote{One can check that the coupling term $S_{\text{Weyl-ST}}$ is invariant under $\mathcal{N}=2$ SUSY transformations at order $O(\lambda)$ using the transformations of operators in the stress-tensor multiplet, given in \eqref{eq:STmultSUSY}, around a $S^4$ background, and the linearized transformations of the fields in $\tilde{\mathcal{V}}_{\text{Weyl}}$, which can be obtained by taking the transformations of the Weyl multiplet operators, given in \eqref{eq:WeylSUSY}, and keeping only the terms that are linear in $\lambda$. The SUSY transformation of the frame field $e^a_\mu$ can be used to get the SUSY transformation of $h_{\mu\nu}$. Note additionally, that while the Weyl multiplet fields listed in Table \ref{tab:WeylMultTable} contains the dilation gauge field $b_\mu$, it does not appear in \eqref{eq:weylst}. Indeed at linear order in $\lambda$, $b_\mu$ does not appear in the SUSY transformations of the other fields in the Weyl multiplet and hence all the Weyl multiplet fields excluding $b_\mu$ are closed under SUSY transformations at this order. While it seems like $b_\mu$ might appear in the SUSY transformation of $\psi_{\mu \, i},\tensor{\psi}{_{\mu}^{i}}$ in \eqref{eq:WeylSUSY}, the particular form of the $b_\mu$ dependence of $\omega^{ab}_\mu$, given in \eqref{eq:Curvs}, ensures that is not the case.} Note that this implies that the inverse metric $g^{\mu\nu}$ transforms as $g^{\mu\nu} \to g^{\mu\nu} - \lambda h^{\mu\nu}$, where $h^{\mu\nu} = g^{\mu\rho}g^{\nu\sigma}h_{\rho\sigma}$. While the coupling of the Weyl multiplet fields to the stress-tensor multiplet at linear order in $\lambda$ is universal, the particulars of the subleading terms in \eqref{eq:weylst} depends on the specific $\mathcal{N}=2$ SCFT being coupled to the Weyl multiplet. 
	
	By treating the squashing parameter $(b-1)$ as the deformation parameter $\lambda$ and substituting the non-zero linearized background fields in \eqref{bkgrndintermsofxi} into \eqref{eq:weylst}, we obtain the squashing deformation given in \eqref{eq:squashcoupl}.
	
	\section{\texorpdfstring{$\mathfrak{su}(4)$}{su(4)}-invariants and \texorpdfstring{$\mathcal{N}=4$}{N = 4} SUSY transformations}
	\label{sec:su4andN4conv}
	
	In this paper, we make extensive use of the $\mathcal{N}=4$ SUSY transformations and the Ward identities they imply. These transformations relate operators in different representations of $\mathfrak{su}(4)_R$, and so they are expressed using three-point $\mathfrak{su}(4)$-invariant tensors. 
	
	In this appendix, we first describe how these invariants can be constructed. In order to do so, we need to first identify the $\mathfrak{su}(4)$ generators in all relevant representations. We will use the letters $A,B,C,\ldots$ to represent the indices for the $\mathbf{4}$ and $\overline{\mathbf{4}}$ representations ($\mathbf{4}$ if raised, and $\overline{\mathbf{4}}$ if lowered). Additionally, we use $a,b,c,\ldots$ to represent adjoint ($\mathbf{15}$) indices and $I, J, \ldots$ to represent $\mathbf{6}$ indices. We will use the letters $K,L,M,\ldots$ to represent the indices for other representations. For complex representations, we will use raised indices for unbarred representations (like $\mathbf{10}$ and $\mathbf{20}$) and lowered indices for barred representations (like $\overline{\mathbf{10}}$ and $\overline{\mathbf{20}}$).

	The 15 $\mathfrak{su}(4)$ generators in the fundamental representation, $\tensor{(T^a_{\mathbf{4}})}{^A_B}$, span the space of $4\times 4$ traceless Hermitian matrices. Seven of these generators, which correspond to the generators of the $\mathfrak{su}(2)_R\times\mathfrak{su}(2)_F\times\mathfrak{u}(1)_R$ subalgebra of $\mathfrak{su}(4)_R$, are given explicitly in \eqref{eq:gensu2bas4}. If we start with a generic vector $v_\mathbf{4}$ in the fundamental ($\mathbf{4}$) representation whose components are $v^A_{\mathbf{4}}$, with $A =1,2,3,4$, then the $\SU(4)$ transformation $U(\alpha^a) = e^{-i\alpha^a T^a} $ takes this vector to
	\es{eq:delta4rep}{v'^{A}_{\mathbf{4}} =  v^A_{\mathbf{4}} -i \alpha^a \tensor{(T^a_{\mathbf{4}})}{^A_B}v^B_{\mathbf{4}} + O(\alpha^2)\,.}
	The complex conjugate vector $\tilde{v}_{\overline{\mathbf{4}}}$ with components $\tilde{v}_{\overline{\mathbf{4}}\,A} \equiv (v^A_{\mathbf{4}})^*$ is in the anti-fundamental ($\overline{\mathbf{4}}$) representation. The transformation of the $\tilde{v}_{\overline{\mathbf{4}}}$ vector under the $\SU(4)$ transformation $U(\alpha^a) = e^{-i\alpha^a T^a}$ is given by
	\es{eq:derivgen4barrep}{\tilde{v}'_{\overline{\mathbf{4}}\,A} &=   (v^A_{\mathbf{4}} -i \alpha^a \tensor{(T^a_{\mathbf{4}})}{^A_B}v^B_{\mathbf{4}} )^* + O(\alpha^2) \\
		&= \tilde{v}_{\overline{\mathbf{4}}\,A} +i \alpha^a (\tensor{(T^a_{\mathbf{4}})}{^A_B})^* \tilde{v}_{\overline{\mathbf{4}}\,A}  + O(\alpha^2) \, , }
	at linear order in $\alpha^a$, and hence, the anti-fundamental generators are \es{eq:gen4barrep}{ \tensor{(T^a_{\overline{\mathbf{4}}})}{_A^B} = -(\tensor{(T^a_{\mathbf{4}})}{^A_B})^* =-\tensor{(T^a_{\mathbf{4}})}{^B_A}  \,.}
	Note that we used the Hermiticity of the fundamental representation generators in the second equality. In general, if we know the generators in some complex representation $\mathbf{R}$, a manipulation similar to that in \eqref{eq:derivgen4barrep} shows that the generators in the conjugate representation $\overline{\mathbf{R}}$ are given by
	\es{eq:gen4barrep2}{ \tensor{(T^a_{\overline{\mathbf{R}}})}{_A^B} = -(\tensor{(T^a_{\mathbf{R}})}{^A_B})^* \,.}

	The norm of $v_{\mathbf{4}}$ is given by $\braket{v_{\mathbf{4}}|v_{\mathbf{4}}} \equiv \tilde{v}_{\overline{{\mathbf{4}}}\,A} v_{\mathbf{4}}^A$. In general, we choose the bilinear form between two conjugate representations, that defines the norm for vectors in those representations, to be Kronecker delta $\tensor{\delta}{^K_L}$. 
	
	A vector in the $\bold{6}$ rep, with components $v_{\mathbf{6}}^I$ ($I = 1,\ldots, 6$), can also be represented using two antisymmerized $\bold{4}$ indices as $v_{\mathbf{6}}^{[AB]}$. The relation between these two forms is given by
	\es{eq:basis6rep}{v^{I}_{\mathbf{6}} = \tensor{(C^{\mathbf{6}}_{\mathbf{4}\otimes \mathbf{4}})}{^I_{AB}} v^{[AB]}_{\mathbf{6}}\, , \qquad  v^{[AB]}_{\mathbf{6}} = \tensor{(C_{\mathbf{6}}^{\mathbf{4}\otimes \mathbf{4}})}{^{IAB}} v^{I}_{\mathbf{6}} \, , }
	where the components of the tensor $\tensor{(C^{\mathbf{6}}_{\mathbf{4}\otimes \mathbf{4}})}{^I_{AB}}$ are the Clebsch-Gordan coefficients $C^{I}_{AB}$ listed in \eqref{eq:clebsch644}, and $\tensor{(C_{\mathbf{6}}^{\mathbf{4}\otimes \mathbf{4}})}{^{IAB}} \equiv (\tensor{(C^{\mathbf{6}}_{\mathbf{4}\otimes \mathbf{4}})}{^I_{AB}})^*$. The tensors $\tensor{(C^{\mathbf{6}}_{\mathbf{4}\otimes \mathbf{4}})}{^I_{AB}}$ and $\tensor{(C_{\mathbf{6}}^{\mathbf{4}\otimes \mathbf{4}})}{^{IAB}}$ satisfy the relations
	\es{eq:clebsch644id}{\tensor{(C^{\mathbf{6}}_{\mathbf{4}\otimes \mathbf{4}})}{^I_{AB}}\tensor{(C_{\mathbf{6}}^{\mathbf{4}\otimes \mathbf{4}})}{^{JAB}} = \delta^{IJ}\,,\qquad \tensor{(C^{\mathbf{6}}_{\mathbf{4}\otimes \mathbf{4}})}{^I_{AB}}\tensor{(C_{\mathbf{6}}^{\mathbf{4}\otimes \mathbf{4}})}{^{ICD}} = \frac{1}{2}(\delta^A_C\delta^B_D-\delta^A_C\delta^B_D)\,,}
	which ensures that the two relations in \eqref{eq:basis6rep} are consistent. Alternatively, the components  $v_{\mathbf{6}}^I$ can be represented using two antisymmetrized $\overline{\bold{4}}$ indices as $v_{\mathbf{6} \, [AB]}$, via
	\es{eq:basis6rep2}{v^{I}_{\mathbf{6}} = \tensor{(C^{\mathbf{6}}_{\overline{\mathbf{4}}\otimes \overline{\mathbf{4}}})}{^{IAB}} v_{\mathbf{6}\,[AB]}\, , \qquad  v_{\mathbf{6}\,[AB]}  = \tensor{(C_{\mathbf{6}}^{\overline{\mathbf{4}}\otimes \overline{\mathbf{4}}})}{^{I}_{AB}} v^{I}_{\mathbf{6}}\, ,}
	where the tensors $\tensor{(C^{\mathbf{6}}_{\overline{\mathbf{4}}\otimes \overline{\mathbf{4}}})}{^{IAB}}$ and $\tensor{(C_{\mathbf{6}}^{\overline{\mathbf{4}}\otimes \overline{\mathbf{4}}})}{^{I}_{AB}}$ are related to the $C$-tensors in \eqref{eq:basis6rep} by
	\es{eq:defclebschother}{ \tensor{(C^{\mathbf{6}}_{\overline{\mathbf{4}}\otimes \overline{\mathbf{4}}})}{^{IAB}} = \tensor{(C_{\mathbf{6}}^{\mathbf{4}\otimes \mathbf{4}})}{^{IAB}} \, , \qquad \tensor{(C_{\mathbf{6}}^{\overline{\mathbf{4}}\otimes \overline{\mathbf{4}}})}{^{I}_{AB}}= \tensor{(C^{\mathbf{6}}_{\mathbf{4}\otimes \mathbf{4}})}{^I_{AB}}  \, .}
	The norm of $v_{\mathbf{6}}$ is given by $\braket{v_{\mathbf{6}}|v_{\mathbf{6}}} \equiv v^{I}_{\mathbf{6}}v^{I}_{\mathbf{6}} = v^{[AB]}_{\mathbf{6}}v_{\mathbf{6}\, [AB]}$, where the second equality follows from \eqref{eq:basis6rep}, \eqref{eq:clebsch644id} and \eqref{eq:basis6rep2}. The transformation of $v_{\mathbf{6}}$ under the $\SU(4)$ transformation $U(\alpha^a) = e^{-i\alpha^a T^a}$ is given by 
	\es{eq:delta6rep}{v'^{I}_{\mathbf{6}} &= \tensor{(C^{\mathbf{6}}_{\mathbf{4}\otimes \mathbf{4}})}{^I_{AB}} {v'}_{\mathbf{6}}^{[AB]} \\ 
		&= v^{I}_{\mathbf{6}} -i \alpha^a  (\tensor{(C^{\mathbf{6}}_{\mathbf{4}\otimes \mathbf{4}})}{^I_{AB}}  \tensor{(T^a_{\mathbf{4}})}{^A_C} v_{\mathbf{6}}^{[CB]} + \tensor{(C^{\mathbf{6}}_{\mathbf{4}\otimes \mathbf{4}})}{^I_{AB}}  \tensor{(T^a_{\mathbf{4}})}{^B_C} v_{\mathbf{6}}^{[AC]})+ O(\alpha^2) \\
		&= v^{I}_{\mathbf{6}} -i \alpha^a  (\tensor{(C^{\mathbf{6}}_{\mathbf{4}\otimes \mathbf{4}})}{^I_{CB}}  \tensor{(T^a_{\mathbf{4}})}{^C_A} v_{\mathbf{6}}^{[AB]} + \tensor{(C^{\mathbf{6}}_{\mathbf{4}\otimes \mathbf{4}})}{^I_{AC}}  \tensor{(T^a_{\mathbf{4}})}{^C_B} v_{\mathbf{6}}^{[AB]})+ O(\alpha^2) \\}
	Equating this expression with $v'^{I}_{\mathbf{6}} = v^{I}_{\mathbf{6}} -i \alpha^a  \tensor{(T^a_{\mathbf{6}})}{^I_J}   v^{J}_{\mathbf{6}}+ O(\alpha^2)$ and using \eqref{eq:basis6rep} to relate $v_{\mathbf{6}}^{[AB]}$ and $v_{\mathbf{6}}^{I}$, we can identify the $\mathfrak{su}(4)$ generators in the $\mathbf{6}$ representation $\tensor{(T^a_{\mathbf{6}})}{^I_J}$ via the relation
	\es{eq:gen6rep}{\tensor{(T^a_{\mathbf{6}})}{^I_J} \tensor{(C^{\mathbf{6}}_{\mathbf{4}\otimes \mathbf{4}})}{^J_{AB}} = \tensor{(C^{\mathbf{6}}_{\mathbf{4}\otimes \mathbf{4}})}{^I_{CB}} \tensor{(T^a_{\mathbf{4}})}{^C_A}  + \tensor{C}{^I_{AC}}\tensor{(T^a_{\mathbf{4}})}{^C_B} \, .}
	
	Similarly, a vector in the $\mathbf{10}$ rep with components $v^{K}_{\mathbf{10}}$ ($K = 1,\ldots, 10$) can be represented alternatively using two symmetrized $\mathbf{4}$ indices as $v^{(AB)}_{\mathbf{10}}$, via
	\es{eq:basis10reprels}{v^{K}_{\mathbf{10}} = \tensor{(C^{\mathbf{10}}_{\mathbf{4}\otimes \mathbf{4}})}{^K_{AB}} v^{(AB)}_{\mathbf{10}}\, ,  \qquad  v^{(AB)}_{\mathbf{10}} = \tensor{(C_{\mathbf{10}}^{\mathbf{4}\otimes \mathbf{4}})}{_K^{AB}} v^{K}_{\mathbf{10}} \, , }
	where the coefficients $\tensor{(C^{\mathbf{10}}_{\mathbf{4}\otimes \mathbf{4}})}{^K_{AB}}$ are the Clebsch-Gordan coefficients for $(\mathbf{4}\otimes\mathbf{4})_s=\mathbf{10}$, and $\tensor{(C_{\mathbf{10}}^{\mathbf{4}\otimes \mathbf{4}})}{_K^{AB}} \equiv (\tensor{(C^{\mathbf{10}}_{\mathbf{4}\otimes \mathbf{4}})}{^K_{AB}})^*$. In our choices of basis for $\mathfrak{su}(4)$ representations, the $\tensor{(C^{\mathbf{10}}_{\mathbf{4}\otimes \mathbf{4}})}{^K_{AB}}$ and $\tensor{(C_{\mathbf{10}}^{\mathbf{4}\otimes \mathbf{4}})}{_K^{AB}}$ are defined such that the components of the $4\times4$ matrix  $v^{(AB)}_{\mathbf{10}} \equiv \tensor{(C_{\mathbf{10}}^{\mathbf{4}\otimes \mathbf{4}})}{_K^{AB}} v^{K}_{\mathbf{10}}$ are
	\es{eq:basis10rep}{v^{(AB)}_{\mathbf{10}} = \begin{pmatrix}
			v^{1}_{\mathbf{10}} & \frac{1}{\sqrt{2}}v^{2}_{\mathbf{10}} & \frac{1}{\sqrt{2}}v^{3}_{\mathbf{10}} & \frac{1}{\sqrt{2}}v^{4}_{\mathbf{10}} \\
			\frac{1}{\sqrt{2}}v^{2}_{\mathbf{10}} & v^{5}_{\mathbf{10}} & \frac{1}{\sqrt{2}}v^{6}_{\mathbf{10}} & \frac{1}{\sqrt{2}}v^{7}_{\mathbf{10}} \\
			\frac{1}{\sqrt{2}}v^{3}_{\mathbf{10}}  & \frac{1}{\sqrt{2}}v^{6}_{\mathbf{10}} & v^{8}_{\mathbf{10}} & \frac{1}{\sqrt{2}}v^{9}_{\mathbf{10}}\\
			\frac{1}{\sqrt{2}}v^{4}_{\mathbf{10}} & \frac{1}{\sqrt{2}}v^{7}_{\mathbf{10}} & \frac{1}{\sqrt{2}}v^{9}_{\mathbf{10}} & v^{10}_{\mathbf{10}} \\
		\end{pmatrix} \,,
	}
	where the entries in the matrix have $\mathbf{10}$ indices. Note that the $\tensor{(C^{\mathbf{10}}_{\mathbf{4}\otimes \mathbf{4}})}{^K_{AB}}$ and $\tensor{(C_{\mathbf{10}}^{\mathbf{4}\otimes \mathbf{4}})}{_K^{AB}}$ tensors defined this way satisfy the identities
	\es{eq:clebsch1044id}{ \tensor{(C^{\mathbf{10}}_{\mathbf{4}\otimes \mathbf{4}})}{^K_{AB}} \tensor{(C_{\mathbf{10}}^{\mathbf{4}\otimes \mathbf{4}})}{_L^{AB}}   = \tensor{\delta}{^{K}_{L}}\,,\qquad \tensor{(C_{\mathbf{10}}^{\mathbf{4}\otimes \mathbf{4}})}{_K^{AB}} \tensor{(C^{\mathbf{10}}_{\mathbf{4}\otimes \mathbf{4}})}{^K_{CD}} = \frac{1}{2}(\delta^A_C\delta^B_D+\delta^A_C\delta^B_D)\,,}
	which are required for the two relations in \eqref{eq:basis10reprels} to be consistent. We can identify the $\mathfrak{su}(4)$ generators in the $\mathbf{10}$ representation $\tensor{(T^a_{\mathbf{10}})}{^K_L}$ from the relation
	\es{eq:gen10rep}{\tensor{(T^a_{\mathbf{10}})}{^K_L} \tensor{(C^{\mathbf{10}}_{\mathbf{4}\otimes \mathbf{4}})}{^L_{AB}} = \tensor{(C^{\mathbf{10}}_{\mathbf{4}\otimes \mathbf{4}})}{^K_{CB}} \tensor{(T^a_{\mathbf{4}})}{^C_A}  + \tensor{(C^{\mathbf{10}}_{\mathbf{4}\otimes \mathbf{4}})}{^K_{AC}}\tensor{(T^a_{\mathbf{4}})}{^C_B} \, ,}
	which is obtained by analogy with \eqref{eq:delta6rep} and \eqref{eq:gen6rep}. In general, one can identify the generators of a larger representation $\mathbf{R}_3$ using the generators of two smaller representations $\mathbf{R}_1$ and $\mathbf{R}_2$, if $\mathbf{R}_3$ lies in the decomposition of the tensor product $\mathbf{R}_1\otimes \mathbf{R}_2$ into irreps, using the relation
	\es{eq:genR3repfromR1R2}{\tensor{(T^a_{\mathbf{R}_3})}{^K_L} \tensor{(C^{\mathbf{R}_3}_{\mathbf{R}_1\otimes \mathbf{R}_2})}{^L_{MN}} = \tensor{(C^{\mathbf{R}_3}_{\mathbf{R}_1\otimes \mathbf{R}_2})}{^K_{LN}} \tensor{(T^a_{\mathbf{R}_1})}{^L_M}  + \tensor{(C^{\mathbf{R}_3}_{\mathbf{R}_1\otimes \mathbf{R}_2})}{^K_{ML}}\tensor{(T^a_{\mathbf{R}_2})}{^L_N} \, ,}
	where $\tensor{(C^{\mathbf{R}_3}_{\mathbf{R}_1\otimes \mathbf{R}_2})}{^K_{MN}}$ are the Clebsch-Gordan coefficients for the projection\footnote{In this paper, we will not need to address cases in which $\mathbf{R}_3$ appears with multiplicity in $\mathbf{R}_1\otimes\mathbf{R}_2$.} $\mathbf{R}_1\otimes \mathbf{R}_2 \to \mathbf{R}_3$.

	Components of vectors in the $\overline{\mathbf{10}}$ representation $\tilde{v}_{\overline{\mathbf{10}}\,K}$ can be expressed instead using two symmetrized $\overline{\mathbf{4}}$ indices $\tilde{v}_{\overline{\mathbf{10}}\,(AB)}$, via
	\es{eq:basis10brep}{\tilde{v}_{\overline{\mathbf{10}}\,K} = \tensor{(C^{\overline{\mathbf{10}}}_{\overline{\mathbf{4}}\otimes \overline{\mathbf{4}}})}{_K^{AB}} \tilde{v}_{\overline{\mathbf{10}}\,(AB)}\, ,  \qquad  \tilde{v}_{\overline{\mathbf{10}}\,(AB)} = \tensor{(C_{\overline{\mathbf{10}}}^{\overline{\mathbf{4}}\otimes \overline{\mathbf{4}}})}{^K_{AB}} \tilde{v}_{\overline{\mathbf{10}}\,K} \, , }
	where  $\tensor{(C^{\overline{\mathbf{10}}}_{\overline{\mathbf{4}}\otimes \overline{\mathbf{4}}})}{_K^{AB}} = \tensor{(C_{\mathbf{10}}^{\mathbf{4}\otimes \mathbf{4}})}{_K^{AB}}$ and $ \tensor{(C_{\overline{\mathbf{10}}}^{\overline{\mathbf{4}}\otimes \overline{\mathbf{4}}})}{^K_{AB}} = \tensor{(C^{\mathbf{10}}_{\mathbf{4}\otimes \mathbf{4}})}{^K_{AB}}$. 
	
	The components of vectors in the $\mathbf{15}$ and $\mathbf{20}'$ representations can be represented using the antisymmetric and traceless-symmetric combinations of two $\mathbf{6}$ indices, respectively. For the $\mathbf{15}$ rep, the relation between the components with $\mathbf{15}$ indices and with two $\mathbf{6}$ indices is given by
	\es{eq:basis15rep}{v^{K}_{\mathbf{15}} = \tensor{(C^{\mathbf{15}}_{\mathbf{6}\otimes \mathbf{6}})}{^{KIJ}} v^{[IJ]}_{\mathbf{15}}  \,, \qquad v^{[IJ]}_{\mathbf{15}} = \tensor{(C_{\mathbf{15}}^{\mathbf{6}\otimes \mathbf{6}})}{^{KIJ}} v^{K}_{\mathbf{15}} \,,}
	where the coefficients $\tensor{(C^{\mathbf{15}}_{\mathbf{6}\otimes \mathbf{6}})}{^{KIJ}}$ are chosen to satisfy $\tensor{(C^{\mathbf{15}}_{\mathbf{6}\otimes \mathbf{6}})}{^{KIJ}} = \tensor{(C_{\mathbf{15}}^{\mathbf{6}\otimes \mathbf{6}})}{^{KIJ}}$. We list these coefficients explicitly via the relations:
	\es{def:comp15rep}{\begin{aligned}
			&v^{[12]}_{\mathbf{15}} = -\frac{1}{2}v^{1}_{\mathbf{15}}-\frac{1}{2}v^{6}_{\mathbf{15}}\,, \qquad
			&v^{[13]}_{\mathbf{15}} = -\frac{1}{2}v^{7}_{\mathbf{15}}-\frac{1}{2}v^{12}_{\mathbf{15}}\,, \qquad
			&v^{[14]}_{\mathbf{15}} = -\frac{1}{2}v^{13}_{\mathbf{15}}+\frac{1}{2}v^{14}_{\mathbf{15}}\,,\\	
			&v^{[15]}_{\mathbf{15}} = -\frac{1}{2}v^{2}_{\mathbf{15}}+\frac{1}{2}v^{5}_{\mathbf{15}}\,, \qquad
			&v^{[16]}_{\mathbf{15}} = -\frac{1}{2}v^{8}_{\mathbf{15}}+\frac{1}{2}v^{11}_{\mathbf{15}}\,, \qquad
			&v^{[23]}_{\mathbf{15}} = \frac{1}{2}v^{13}_{\mathbf{15}}+\frac{1}{2}v^{14}_{\mathbf{15}}\,,\\	
			&v^{[24]}_{\mathbf{15}} = -\frac{1}{2}v^{7}_{\mathbf{15}}+\frac{1}{2}v^{12}_{\mathbf{15}}\,, \qquad
			&v^{[25]}_{\mathbf{15}} = -\frac{1}{2}v^{3}_{\mathbf{15}}-\frac{1}{2}v^{4}_{\mathbf{15}}\,,	\qquad
			&v^{[26]}_{\mathbf{15}} = -\frac{1}{2}v^{9}_{\mathbf{15}}-\frac{1}{2}v^{10}_{\mathbf{15}}\,,\\	
			&v^{[34]}_{\mathbf{15}} = \frac{1}{2}v^{1}_{\mathbf{15}}-\frac{1}{2}v^{6}_{\mathbf{15}}\,, \qquad	
			&v^{[35]}_{\mathbf{15}} = -\frac{1}{2}v^{9}_{\mathbf{15}}+\frac{1}{2}v^{10}_{\mathbf{15}}\,, \qquad	
			&v^{[36]}_{\mathbf{15}} = \frac{1}{2}v^{3}_{\mathbf{15}}-\frac{1}{2}v^{4}_{\mathbf{15}}\,,\\	
			&v^{[45]}_{\mathbf{15}} = \frac{1}{2}v^{8}_{\mathbf{15}}+\frac{1}{2}v^{11}_{\mathbf{15}}\,, \qquad	
			&v^{[46]}_{\mathbf{15}} = -\frac{1}{2}v^{2}_{\mathbf{15}}-\frac{1}{2}v^{5}_{\mathbf{15}}\,, \qquad	
			&v^{[56]}_{\mathbf{15}} = \frac{1}{\sqrt{2}}v^{15}_{\mathbf{15}} \,.
		\end{aligned}
	}
	For the $\mathbf{20}'$ rep, the relation between the components with $\mathbf{20}'$ indices and two $\mathbf{6}$ indices is given by
	\es{eq:basis20prep}{v^{K}_{\mathbf{20}'} = \tensor{(C^{\mathbf{20}'}_{\mathbf{6}\otimes \mathbf{6}})}{^{KIJ}} v^{(IJ)}_{\mathbf{15}}  \,, \qquad v^{(IJ)}_{\mathbf{15}} = \tensor{(C_{\mathbf{20}'}^{\mathbf{6}\otimes \mathbf{6}})}{^{KIJ}} v^{K}_{\mathbf{20}'} \,,}
	where the Clebsch-Gordan coefficients $\tensor{(C^{\mathbf{20}'}_{\mathbf{6}\otimes \mathbf{6}})}{^{KIJ}}$ are chosen such that $\tensor{(C^{\mathbf{20}'}_{\mathbf{6}\otimes \mathbf{6}})}{^{KIJ}} = \tensor{(C_{\mathbf{20}'}^{\mathbf{6}\otimes \mathbf{6}})}{^{KIJ}}$. These coefficients are given in our basis by the expressions
	\es{def:comp20prepp1}{
		v^{(11)}_{\mathbf{20}'} &=   \frac{\sqrt{6} v_{\mathbf{20}'}^1+\sqrt{2} v_{\mathbf{20}'}^7 +v_{\mathbf{20}'}^{12}+\sqrt{\frac{3}{5}} v_{\mathbf{20}'}^{16}+\sqrt{\frac{2}{5}} v_{\mathbf{20}'}^{19}}{2 \sqrt{3}} \, , \\
		v^{(22)}_{\mathbf{20}'} &=   \frac{-\sqrt{6} v_{\mathbf{20}'}^1+\sqrt{2} v_{\mathbf{20}'}^7 +v_{\mathbf{20}'}^{12}+\sqrt{\frac{3}{5}} v_{\mathbf{20}'}^{16}+\sqrt{\frac{2}{5}} v_{\mathbf{20}'}^{19}}{2 \sqrt{3}} \, ,\\ 
		v^{(33)}_{\mathbf{20}'} &=   \frac{-4\sqrt{\frac{3}{5}} v_{\mathbf{20}'}^{16}+\sqrt{\frac{2}{5}} v_{\mathbf{20}'}^{19}}{2 \sqrt{3}} \, ,\\
		v^{(44)}_{\mathbf{20}'} &=  \frac{-3v_{\mathbf{20}'}^{12}+\sqrt{\frac{3}{5}} v_{\mathbf{20}'}^{16}+\sqrt{\frac{2}{5}} v_{\mathbf{20}'}^{19}}{2 \sqrt{3}}\, ,\\
		v^{(55)}_{\mathbf{20}'} &= \frac{-2\sqrt{2} v_{\mathbf{20}'}^7 +v_{\mathbf{20}'}^{12}+\sqrt{\frac{3}{5}} v_{\mathbf{20}'}^{16}+\sqrt{\frac{2}{5}} v_{\mathbf{20}'}^{19}}{2 \sqrt{3}}\, , \\
		v^{(66)}_{\mathbf{20}'} &=  -\sqrt{\frac{5}{6}} v_{\mathbf{20}'}^{19}\, ,\\
	}
	and
	\es{def:comp20prepp2}{\begin{aligned}
			&v^{(12)}_{\mathbf{20}'} =  \frac{1}{\sqrt{2}} v_{\mathbf{20}'}^2\, , \qquad &v^{(13)}_{\mathbf{20}'} =  \frac{1}{\sqrt{2}} v_{\mathbf{20}'}^5 \, , \qquad &v^{(14)}_{\mathbf{20}'} =  \frac{1}{\sqrt{2}} v_{\mathbf{20}'}^4\,, \qquad &v^{(15)}_{\mathbf{20}'} =  \frac{1}{\sqrt{2}} v_{\mathbf{20}'}^3 \,, \\
			&v^{(16)}_{\mathbf{20}'} =  \frac{1}{\sqrt{2}} v_{\mathbf{20}'}^6\, , \qquad &v^{(23)}_{\mathbf{20}'} =  \frac{1}{\sqrt{2}} v_{\mathbf{20}'}^{10}\, , \qquad &v^{(24)}_{\mathbf{20}'} =  \frac{1}{\sqrt{2}} v_{\mathbf{20}'}^9\, ,\qquad &v^{(25)}_{\mathbf{20}'} =  \frac{1}{\sqrt{2}} v_{\mathbf{20}'}^8\,,\\
			&v^{(26)}_{\mathbf{20}'} =  \frac{1}{\sqrt{2}} v_{\mathbf{20}'}^{11}\,, \qquad &v^{(34)}_{\mathbf{20}'} =  \frac{1}{\sqrt{2}} v_{\mathbf{20}'}^{17}\,, \qquad &v^{(35)}_{\mathbf{20}'} =  \frac{1}{\sqrt{2}} v_{\mathbf{20}'}^{14}\,, \qquad &v^{(36)}_{\mathbf{20}'} =  \frac{1}{\sqrt{2}} v_{\mathbf{20}'}^{20}\,,\\
			&v^{(45)}_{\mathbf{20}'} =  \frac{1}{\sqrt{2}} v_{\mathbf{20}'}^{13}\,, \qquad &v^{(46)}_{\mathbf{20}'} =  \frac{1}{\sqrt{2}} v_{\mathbf{20}'}^{18} \,,  \qquad &v^{(56)}_{\mathbf{20}'} =  \frac{1}{\sqrt{2}} v_{\mathbf{20}'}^{15}\, . &
	\end{aligned}}
	
	A vector in the $\mathbf{20}$ rep with components $v^K_{\mathbf{20}}$ can also be expressed using a $\mathbf{6}$ index and a $\overline{\mathbf{4}}$ index $\tensor{(v_{\mathbf{20}})}{_A^I}$. These two forms are related by
	\es{eq:basis20rep}{v^{K}_{\mathbf{20}} = \tensor{(C^{\mathbf{20}}_{\overline{\mathbf{4}}\otimes\mathbf{6}})}{^{KA}_{I}} \tensor{(v_{\mathbf{20}})}{_A^I} \,, \qquad \tensor{(v_{\mathbf{20}})}{_A^I} = \tensor{(C_{\mathbf{20}}^{\overline{\mathbf{4}}\otimes\mathbf{6}})}{_{KA}^{I}} v^{K}_{\mathbf{20}}  \,,}
	where the coefficients $\tensor{(C_{\mathbf{20}}^{\overline{\mathbf{4}}\otimes\mathbf{6}})}{_{KA}^{I}}$ and $\tensor{(C^{\mathbf{20}}_{\overline{\mathbf{4}}\otimes\mathbf{6}})}{^{KA}_{I}} \tensor{(v_{\mathbf{20}})}{_A^I}$ can be identified from the relations
	\begin{align}\label{eq:comps20rep}
		\tensor{(v_{\mathbf{20}})}{_1^1} &= -i\sqrt{\frac{5}{6}}v^{1}_{\mathbf{20}}\,, \notag \\ 
		\tensor{(v_{\mathbf{20}})}{_1^2} &= -\frac{v^4_{\mathbf{20}}}{\sqrt{30}}+\frac{2 i v^5_{\mathbf{20}}}{\sqrt{5}} \,, \notag \\
		\tensor{(v_{\mathbf{20}})}{_1^3} &= \frac{v^3_{\mathbf{20}}}{\sqrt{30}}+\frac{2 i v^6_{\mathbf{20}}}{\sqrt{5}}\,, \notag\\ 
		\tensor{(v_{\mathbf{20}})}{_1^4} &= -\frac{v^1_{\mathbf{20}}}{\sqrt{30}}+\frac{2 i v^7_{\mathbf{20}}}{\sqrt{5}} \,, \notag\\ 
		\tensor{(v_{\mathbf{20}})}{_1^5} &= \frac{i \sqrt{\frac{2}{5}} v^4_{\mathbf{20}}-\sqrt{\frac{3}{5}} v^5_{\mathbf{20}}+v^9_{\mathbf{20}}+2 i \sqrt{2} v^{13}_{\mathbf{20}}}{2 \sqrt{3}}\,, \notag\\ 
		\tensor{(v_{\mathbf{20}})}{_1^6} &= \frac{-i \sqrt{\frac{2}{5}} v^3_{\mathbf{20}}-\sqrt{\frac{3}{5}} v^6_{\mathbf{20}}+v^{11}_{\mathbf{20}}+2 i \sqrt{2} v^{15}_{\mathbf{20}}}{2 \sqrt{3}}\,, \notag\\
		\tensor{(v_{\mathbf{20}})}{_2^1} &= \frac{\sqrt{10} v^4_{\mathbf{20}}+i \sqrt{15} v^5_{\mathbf{20}}-5 i v^9_{\mathbf{20}}-5 \sqrt{2} \left(v^{13}_{\mathbf{20}}+\sqrt{3} v^{17}_{\mathbf{20}}\right)}{10 \sqrt{3}}\,, \notag \\ 
		\tensor{(v_{\mathbf{20}})}{_2^2} &= \frac{1}{30} i \left(\sqrt{30} v^1_{\mathbf{20}}+3 i \sqrt{5} v^7_{\mathbf{20}}+5 i \sqrt{3} v^{10}_{\mathbf{20}}-10 \sqrt{6} v^{14}_{\mathbf{20}}\right)\,, \notag\\
		\tensor{(v_{\mathbf{20}})}{_2^3} &= -\frac{v^2_{\mathbf{20}}}{\sqrt{30}}+\frac{2 i v^8_{\mathbf{20}}}{\sqrt{5}}\,, \notag\\ 
		\tensor{(v_{\mathbf{20}})}{_2^4} &= \frac{-i \sqrt{10} v^4_{\mathbf{20}}+\sqrt{15} v^5_{\mathbf{20}}-5 v^9_{\mathbf{20}}+5 i \sqrt{2} \left(v^{13}_{\mathbf{20}}-\sqrt{3} v^{17}_{\mathbf{20}}\right)}{10 \sqrt{3}}\,, \notag\\ 
		\tensor{(v_{\mathbf{20}})}{_2^5} &= -\frac{\sqrt{10} v^1_{\mathbf{20}}+i \sqrt{15} v^7_{\mathbf{20}}+5 i v^{10}_{\mathbf{20}}+5 \sqrt{2} v^{14}_{\mathbf{20}}+5 \sqrt{6} v^{18}_{\mathbf{20}}}{10 \sqrt{3}}\,, \notag\\ 
		\tensor{(v_{\mathbf{20}})}{_2^6} &= \frac{i \sqrt{\frac{2}{5}} v^2_{\mathbf{20}}-\sqrt{\frac{3}{5}} v^8_{\mathbf{20}}+v^{12}_{\mathbf{20}}+2 i \sqrt{2} v^{16}_{\mathbf{20}}}{2 \sqrt{3}}\,, \\
		\tensor{(v_{\mathbf{20}})}{_3^1} &= -\frac{\sqrt{10} v^3_{\mathbf{20}}-i \sqrt{15} v^6_{\mathbf{20}}+5 i v^{11}_{\mathbf{20}}+5 \sqrt{2} v^{15}_{\mathbf{20}}+5 \sqrt{6} v^{19}_{\mathbf{20}}}{10 \sqrt{3}}\,, \notag\\ 
		\tensor{(v_{\mathbf{20}})}{_3^2} &= \frac{\sqrt{10} v^2_{\mathbf{20}}+i \sqrt{15} v^8_{\mathbf{20}}-5 i v^{12}_{\mathbf{20}}-5 \sqrt{2} \left(v^{16}_{\mathbf{20}}+\sqrt{3} v^{20}_{\mathbf{20}}\right)}{10 \sqrt{3}}\,, \notag\\
		\tensor{(v_{\mathbf{20}})}{_3^3} &= \frac{1}{30} i \left(\sqrt{30} v^1_{\mathbf{20}}+3 i \sqrt{5} v^7_{\mathbf{20}}+5 i \sqrt{3} v^{10}_{\mathbf{20}}+5 \sqrt{6} v^{14}_{\mathbf{20}}-15 \sqrt{2} v^{18}_{\mathbf{20}}\right)\,, \notag\\ 
		\tensor{(v_{\mathbf{20}})}{_3^4} &= \frac{i \sqrt{10} v^3_{\mathbf{20}}+\sqrt{15} v^6_{\mathbf{20}}-5 v^{11}_{\mathbf{20}}+5 i \sqrt{2} \left(v^{15}_{\mathbf{20}}-\sqrt{3} v^{19}_{\mathbf{20}}\right)}{10 \sqrt{3}}\,, \notag\\ 
		\tensor{(v_{\mathbf{20}})}{_3^5} &= \frac{-i \sqrt{10} v^2_{\mathbf{20}}+\sqrt{15} v^8_{\mathbf{20}}-5 v^{12}_{\mathbf{20}}+5 i \sqrt{2} \left(v^{16}_{\mathbf{20}}-\sqrt{3} v^{20}_{\mathbf{20}}\right)}{10 \sqrt{3}}\,, \notag\\ 
		\tensor{(v_{\mathbf{20}})}{_3^6} &= -\frac{v^1_{\mathbf{20}}}{\sqrt{30}}-\frac{1}{10} i \left(\sqrt{5} v^7_{\mathbf{20}}-5 \sqrt{3} v^{10}_{\mathbf{20}}\right)\,, \notag\\
		\tensor{(v_{\mathbf{20}})}{_4^1} &= -\frac{\sqrt{10} v^2_{\mathbf{20}}+i \left(\sqrt{15} v^8_{\mathbf{20}}+15 v^{12}_{\mathbf{20}}\right)}{10 \sqrt{3}}\,, \notag\\ 
		\tensor{(v_{\mathbf{20}})}{_4^2} &= \frac{i \left(i \sqrt{10} v^3_{\mathbf{20}}+\sqrt{15} v^6_{\mathbf{20}}+15 v^{11}_{\mathbf{20}}\right)}{10 \sqrt{3}}\,, \notag\\
		\tensor{(v_{\mathbf{20}})}{_4^3} &= -\frac{\sqrt{10} v^4_{\mathbf{20}}+i \left(\sqrt{15} v^5_{\mathbf{20}}+15 v^9_{\mathbf{20}}\right)}{10 \sqrt{3}} \,, \notag\\ 
		\tensor{(v_{\mathbf{20}})}{_4^4} &= i \sqrt{\frac{5}{6}} v^2_{\mathbf{20}} \,, \notag \\ 
		\tensor{(v_{\mathbf{20}})}{_4^5} &= i \sqrt{\frac{5}{6}} v^3_{\mathbf{20}} \,, \notag \\ 
		\tensor{(v_{\mathbf{20}})}{_4^6} &= i \sqrt{\frac{5}{6}} v^4_{\mathbf{20}}\,. \notag
	\end{align}
	
	After all the generators in the relevant representations are identified, we can compute three-point $\mathfrak{su}(4)$-invariants $\tensor{(C_{\mathbf{R}_1\mathbf{R}_2\mathbf{R}_3})}{^{LMN}}$ between any three $\mathfrak{su}(4)$ representations $\mathbf{R}_1$, $\mathbf{R}_2$ and $\mathbf{R}_3$, where $L$, $M$ and $N$ are indices in the representation $\mathbf{R}_1$, $\mathbf{R}_2$ and $\mathbf{R}_3$.\footnote{If any of these representations are barred complex representations, we lower their indices as usual; when the representations are generic we will keep their indices raised.} The three-point invariant tensor $\tensor{(C_{\mathbf{R}_1\mathbf{R}_2\mathbf{R}_3})}{^{LMN}}$ has dimensions $\dim \mathbf{R}_1 \times \dim \mathbf{R}_2 \times \dim \mathbf{R}_3$, and is non-zero only if the singlet representation appears in the $\mathbf{R}_1 \otimes \mathbf{R}_2 \otimes \mathbf{R}_3$ tensor product. It is determined uniquely\footnote{In this paper we will not deal with cases in which the singlet representation appears with multiplicity in $\mathbf{R}_1 \otimes \mathbf{R}_2 \otimes \mathbf{R}_3$. In those cases, \eqref{eq:defthreeptinv} has multiple solutions.} (up to an overall multiplicative constant) by the condition 
	\es{eq:defthreeptinv}{\tensor{(T^a_{\mathbf{R}_1})}{^{L}_{K}}\tensor{(C_{\mathbf{R}_1\mathbf{R}_2\mathbf{R}_3})}{^{KMN}}+ \tensor{(T^a_{\mathbf{R}_2})}{^{M}_{K}}\tensor{(C_{\mathbf{R}_1\mathbf{R}_2\mathbf{R}_3})}{^{LKN}}+ \tensor{(T^a_{\mathbf{R}_3})}{^{N}_{K}}\tensor{(C_{\mathbf{R}_1\mathbf{R}_2\mathbf{R}_3})}{^{LMK}} = 0 \,.}
	Comparing this with \eqref{eq:gen6rep}, and using the form of anti-fundamental generators given in \eqref{eq:gen4barrep}, we see that the $\tensor{(C^{\mathbf{6}}_{\mathbf{4}\otimes \mathbf{4}})}{^L_{MN}}$ tensor in \eqref{eq:basis6rep} is equivalent to the three-point invariant between the $\mathbf{6}$, $\overline{\mathbf{4}}$, and $\overline{\mathbf{4}}$ representations, $\tensor{(C_{\mathbf{6}\,\overline{\mathbf{4}}\,\overline{\mathbf{4}}})}{^{L}_{MN}}$. Indeed, it is generally the case that the Clebsch-Gordan coefficients for the projection $\mathbf{R}_1\otimes \mathbf{R}_2 \to \mathbf{R}_3$, which we have denoted $\tensor{(C^{\mathbf{R}_3}_{\mathbf{R}_1\otimes \mathbf{R}_2})}{^L_{MN}}$, is also a three-point invariant between the representations $\mathbf{R}_3$, $\mathbf{\overline{R}_1}$ and $\mathbf{\overline{R}_2}$, denoted $\tensor{(C_{\mathbf{R}_3 \mathbf{\overline{R}_1} {\mathbf{\overline{R}_2}}})}{^L_{MN}}$ .

	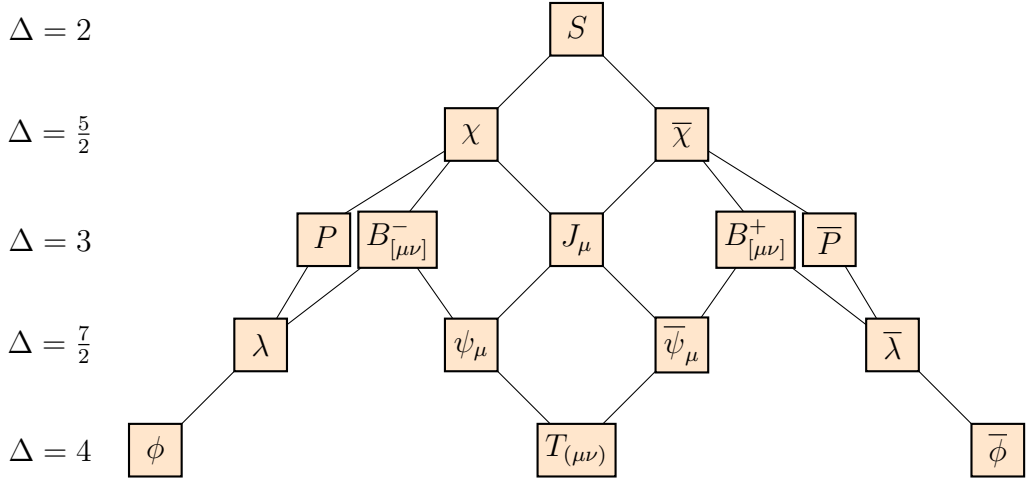
\begin{figure}[tb]
		\begin{tikzpicture}[box/.style = {draw,black,thick,rectangle,fill=orange!20,inner sep=.1cm,minimum height=.7cm,minimum width=.7cm},scale=1.4]
			\draw[->] (0, -4) -- (-1, -5);
			\draw (0, -4) -- (1, -5);
			\draw (-1, -5) -- (0, -6);
			\draw (-1, -5) -- (-2.6, -6);
			\draw (-1, -5) -- (-1.8, -6);
			\draw (1, -5) -- (0, -6);
			\draw (1, -5) -- (2.6, -6);
			\draw (1, -5) -- (1.8, -6);
			\draw (0, -6) -- (-1, -7);
			\draw (0, -6) -- (1, -7);
			\draw (-2.4, -6) -- (-3, -7);
			\draw (-1.7, -6) -- (-3, -7);
			\draw (-1.7, -6) -- (-1, -7);
			\draw (1.7, -6) -- (1, -7);
			\draw (1.7, -6) -- (3, -7);
			\draw (2.4, -6) -- (3, -7);
			\draw (-3, -7) -- (-4, -8);
			\draw (-1, -7) -- (0, -8);
			\draw (1, -7) -- (0, -8);
			\draw (3, -7) -- (4, -8);
			\node[box] at (0, -4) {$S_{\text{}}$};
			\node[box] at (1, -5) {$\overline{\chi }$};
			\node[box] at (-1, -5) {$\chi$};
			\node[box] at (0, -6) {$J_{\mu}$};
			\node[box] at (1.7, -6) {$B^{+}_{[\mu\nu]}$};
			\node[box] at (2.4, -6) {$\bar{P}$};
			\node[box] at (-1.7, -6) {$B^{-}_{[\mu\nu]}$};
			\node[box] at (-2.4, -6) {$P_{\text{}}$};
			\node[box] at (3, -7) {$\overline{\lambda }$};
			\node[box] at (1, -7) {$\overline{\psi }_{\mu}$};
			\node[box] at (-1, -7) {$\psi _{\mu}$};
			\node[box] at (-3, -7) {$\lambda$};
			\node[box] at (4, -8) {$\overline{\phi}$};
			\node[box] at (0, -8) {$T_{(\mu\nu)}$};
			\node[box] at (-4, -8) {$\phi$};
			\node[] at (-5, -4) {$\Delta = 2$};
			\node[] at (-5, -5) {$\Delta = \frac{5}{2}$};
			\node[] at (-5, -6) {$\Delta = 3$};
			\node[] at (-5, -7) {$\Delta = \frac{7}{2}$};
			\node[] at (-5, -8) {$\Delta = 4$};
			
		\end{tikzpicture}
		\caption{Conformal primary operators in the stress-tensor multiplet of an $\mathcal{N}=4$ SCFT\@. The lines joining operators indicate which operators are related to each other by a SUSY transformation.}
		\label{fig:STmultiplet}
	\end{figure}

	We can now use these three-point $\mathfrak{su}(4)$-invariants to fix the Poincar\'e SUSY transformations of the operators of the $\mathcal{N}=4$ stress-tensor multiplet, listed in Table \ref{tab:N4STOps}. This is done by first using the R-symmetry representation, Lorentz representation, and conformal dimension of each operator in the multiplet to write down an ansatz for the $\mathcal{N}=4$ SUSY transformations of the operators in the stress-tensor multiplet. Figure \ref{fig:STmultiplet} depicts which operators in the stress-tensor multiplet are related to each other by SUSY transformations.  For example, for the superprimary we have the ansatz
	\es{}{[\mathbf{Q}^{A}_{\alpha},S^{M}] = a_{S,1} \tensor{(C_{\mathbf{4}\,\mathbf{20}' \,\overline{\mathbf{20}}})}{^{AM}_{N}}  \chi^{N}_{\alpha}\,, \qquad [\bar{\mathbf{Q}}_{B\dot{\alpha}},S^{M}] = a_{S,2} \tensor{(C_{\overline{\mathbf{4}}\,\mathbf{20}' \,\mathbf{20}})}{_{B}^{MN}}  \bar{\chi}_{N\dot{\alpha}}\,,}
	where $\mathbf{Q}^A_{\alpha}$ and $\bar{\mathbf{Q}}_{B\dot{\alpha}}$ are the left-handed and right-handed SUSY generators in the $\mathbf{4}$ and $\overline{\mathbf{4}}$ representations respectively. The coefficients in this ansatz are constrained by the quantum SUSY algebra
	\es{eq:qSUSYalgN4}{[\{\mathbf{Q}^A_{\alpha},\bar{\mathbf{Q}}_{B\dot{\alpha}}\},\mathcal{O}] = -\frac{i \tensor{\delta}{^A_B}}{2}[\mathbf{P}_\mu \sigma^\mu_{\alpha\dot{\alpha}} , \mathcal{O}]\,, \qquad A,B = 1,2,3,4\,,}
	where $\mathbf{P}_\mu$ is the momentum generator. We can also derive additional constraints from $\mathcal{N} = 4$ Ward identities for two-point functions.
	
	These constraints cannot fix the coefficients in the transformation completely, because we are free to rescale the operators of the multiplet by any complex multiplicative factor. We can partially restrict this freedom by choosing normalizations for the two-point functions of each operator. We still retain some residual freedom because we are free to multiply any complex operator in the stress-tensor multiplet (i.e., all operators except the scalar superprimary, the conserved $\mathfrak{su}(4)_R$ current and the stress-tensor) by a phase factor; these free phases in the solution to the constraints from the SUSY algebra and the two-point Ward identities can be chosen arbitrarily.
	
	For example, our choice of the $\mathcal{N}=4$ SUSY transformations of the dimension-2 and dimension-$\frac{5}{2}$ operators that satisfy the SUSY algebra is given by
	\es{eq:N4SUSYexamples}{[\mathbf{Q}^{A}_{\alpha},S^{M}] &=  \tensor{(C_{\mathbf{4}\,\mathbf{20}' \,\overline{\mathbf{20}}})}{^{AM}_{N}}  \chi^{N}_{\alpha}\,, \\
		[\bar{\mathbf{Q}}_{A\dot{\alpha}},S^{M}] &=  -\tensor{(C_{\overline{\mathbf{4}}\,\mathbf{20}' \,\mathbf{20}})}{_{A}^{MN}}  \bar{\chi}_{N\dot{\alpha}}\,,\\
		\{\mathbf{Q}^{A}_{\alpha},\chi^{M}_{\beta}\} &=  \tensor{(C_{\mathbf{4}\,\mathbf{20} \,\mathbf{10}})}{^{AMN}}  P_{N} \epsilon_{\alpha\beta}+ \tensor{(C_{\mathbf{4}\,\mathbf{20} \,\mathbf{6}})}{^{AMN}}  B^{-N}_{\mu\nu} \sigma^{\mu\nu}_{\alpha\beta} \,, \\
		\{\bar{\mathbf{Q}}_{A\dot{\alpha}},\chi^{M}_{\alpha}\} &=  i\tensor{(C_{\overline{\mathbf{4}}\,\mathbf{20} \,\mathbf{15}})}{_A^{MN}}  J^{N}_{\alpha\dot{\alpha}} - \frac{i}{4} \tensor{(C_{\overline{\mathbf{4}}\,\mathbf{20}'\,\mathbf{20} })}{_A^{NM}} \partial_{\alpha\dot{\alpha}} S^{N} \,, \\
		\{\mathbf{Q}^A_{\alpha},\bar{\chi}_{M\dot{\alpha}}\} &=  -i\tensor{(C_{\mathbf{4}\,\overline{\mathbf{20}}\,\mathbf{15} })}{^{A}_{MN}}  J^{N}_{\alpha\dot{\alpha}}+ \frac{i}{4} \tensor{(C_{\mathbf{4}\,\mathbf{20}'\,\overline{\mathbf{20}} })}{^{AN}_{M}} \partial_{\alpha\dot{\alpha}} S^{N} \,, \\ \{\bar{\mathbf{Q}}_{A\dot{\alpha}},\bar{\chi}_{M\dot{\beta}}\} &=  \tensor{(C_{\overline{\mathbf{4}}\,\overline{\mathbf{20}} \,\overline{\mathbf{10}}})}{_{AMN}}  \bar{P}^{N} \epsilon_{\dot{\alpha}\dot{\beta}}+ \tensor{(C_{\overline{\mathbf{4}}\,\overline{\mathbf{20}} \,\mathbf{6}})}{_{AM}^{N}}  B^{+N}_{\mu\nu} \bar{\sigma}^{\mu\nu}_{\dot{\alpha}\dot{\beta}} \,, \\
	}
	where our the three-point $\mathfrak{su}(4)$-invariants are normalized such that
	\es{eq:threeptexnorm}{
		\begin{aligned}
			\tensor{(C_{\mathbf{4}\,\mathbf{20}' \,\overline{\mathbf{20}}})}{^{1,1}_{12}} &= -\sqrt{\frac{2}{3}}\,, \qquad &\tensor{(C_{\overline{\mathbf{4}}\,\mathbf{20}' \,\mathbf{20}})}{_1^{1,12}} &= -\sqrt{\frac{2}{3}}\,, \\
			\tensor{(C_{\mathbf{4}\,\mathbf{20} \,\mathbf{10}})}{^{1,1,4}} &=\sqrt{\frac{3}{10}}\,,  \qquad&
			\tensor{(C_{\mathbf{4}\,\mathbf{20} \,\mathbf{6}})}{^{1,1,1}} &= -\sqrt{\frac{5}{2}}\,,\\
			\tensor{(C_{\overline{\mathbf{4}}\,\overline{\mathbf{20}} \,\overline{\mathbf{10}}})}{_{1,1,4}} &=\sqrt{\frac{3}{10}}\,,  \qquad&
			\tensor{(C_{\overline{\mathbf{4}}\,\overline{\mathbf{20}} \,\mathbf{6}})}{_{1,1}^{1}} &= -\sqrt{\frac{5}{2}}\,,\\
			\tensor{(C_{\overline{\mathbf{4}}\,\mathbf{20} \,\mathbf{15}})}{_1^{1,3}}&=-\frac{1}{\sqrt{5}}\,,  \qquad&
			\tensor{(C_{\mathbf{4}\,\overline{\mathbf{20}}\,\mathbf{15} })}{^{1}_{1,3}} &= -\frac{1}{\sqrt{5}}\,.\\
	\end{aligned} }
	
	We use these $\mathcal{N}=4$ SUSY transformations as well as suitably chosen SUSY transformations for the rest of the stress-tensor multiplet to solve the chain of Ward identities listed in Table \ref{tab:N4STWard} in order to obtain the correlation functions that are relevant for the computation of integral constraints on the squashed sphere. The relevant correlation functions are summarized in Appendix~\ref{sec:corrs}.

	\section{Identifying \texorpdfstring{$\mathcal{N}=2$}{N=2} deformation operators as \texorpdfstring{$\mathcal{N}=4$}{N=4} stress-tensor operators}
	
	\label{sec:N4toN2decomp}

	In Section~\ref{sec:background}, we described the mass, squashing and $\tau$ deformation operators as operators in an $\mathcal{N}=2$ SCFT\@. In Section~\ref{N2asN4}, we identified how these operators lie in the $\mathcal{N}=4$ stress-tensor multiplet. In this appendix, we elaborate on how the $\mathcal{N}=4$ stress-tensor multiplet can be decomposed into $\mathcal{N}=2$ multiplets in order to identify the mass and squashing deformation operators. (The superprimary of the $\mathcal{N}=2$ chiral multiplet is sufficient to describe the $\tau$ deformation using \eqref{eq:taupole}, and so we do not need to identify the other operators in the $\mathcal{N}=2$ chiral multiplet.) 
	
	We can determine how the $\mathcal{N}=4$ stress-tensor multiplet decomposes into $\mathcal{N}=2$ multiplets by using their SUSY transformations. We have described in Appendix~\ref{sec:su4andN4conv} how these $\mathcal{N}=4$ SUSY transformations can be derived. When the $\mathcal{N}=4$ theory is described in $\mathcal{N}=2$ language, the SUSY algebra \eqref{eq:qSUSYalgN4} is restricted to $A,B = 1,2$. This is equivalent to the algebra in the first line of \eqref{eq:SUSYAlg}, after accounting for a factor of $i$ that arises when converting the Poisson bracket relation in \eqref{eq:SUSYAlg} into a quantum commutator. Note that our conventions are such that the left-handed $\mathcal{N}=2$ SUSY generator has a lowered $\mathfrak{su}(2)_R$ index, while the left-handed $\mathcal{N}=4$ SUSY generator has a raised  $\mathfrak{su}(4)_R$ index; and likewise, the right-handed $\mathcal{N}=2$ SUSY generator has a raised $\mathfrak{su}(2)_R$ index, while the right-handed $\mathcal{N}=4$ SUSY generator has a lowered  $\mathfrak{su}(4)_R$ index.
	
	We will start by describing how the $\mathcal{N}=2$ flavor current multiplet satisfying the SUSY algebra in \eqref{eq:SUSYflavor} can be identified. The operator $J_{12}$ in \eqref{eq:masscoupl} lies in the Cartan of the $\mathfrak{su}(2)_R$ as well as the $\mathfrak{su}(2)_F$ in the $(\mathbf{3},\mathbf{3})_0$ subrepresentation in the decomposition of the stress-tensor superprimary $S$ in \eqref{eq:Ssplit}. In our chosen basis for the $\bold{20}'$ representation (see Appendix~\ref{sec:su4andN4conv}), $J_{12}$ can thus be identified, up to overall normalization, as the following piece of the $\mathcal{N} = 4$ stress tensor superprimary $S$:
	\es{eq:J12compsansatz}{
		J_{12} &=  iN_m\left(\frac{1}{\sqrt{2}}S_1 - \frac{\sqrt{3}}{4}S_{12}+\frac{\sqrt{5}}{4}S_{16}\right)\, .
	}
	Here $N_m$ is a numerical prefactor and the subscript on $S$ is an index for the $\bold{20}'$ representation and runs from 1 to 20.
	
	Acting on the expression for $J_{12}$ in \eqref{eq:J12compsansatz} with the left-handed $\mathcal{N}=4$ SUSY generator $\mathbf{Q}^1$ gives 
	\es{eq:J12toxi2}{[\mathbf{Q}^1, J_{12}]&= iN_m \left(\frac{1}{\sqrt{2}}[\mathbf{Q}^1,S_1] - \frac{\sqrt{3}}{4}[\mathbf{Q}^1,S_{12}]+\frac{\sqrt{5}}{4}[\mathbf{Q}^1,S_{16}]\right) \\ 
		&= \frac{iN_m}{60} \left(6i\sqrt{30}\chi^2 +3\sqrt{5}\chi^8-25\sqrt{3}\chi^{12} - 10i\sqrt{6}\chi^{16}\right) \, ,\\}
	where we used the $\mathcal{N}=4$ SUSY transformation rules for $S$ given in \eqref{eq:N4SUSYexamples} to derive the second line.
	Additionally, the $\mathcal{N}=2$ SUSY transformation rules in \eqref{eq:SUSYflavor} tells us that we should have
	\es{eq:J12toxi22}{[\mathbf{Q}^1, J_{12}]&= -\frac{i}{2}\xi^2\, .}
	where we emphasize again that $\mathbf{Q}^1$ is a left-handed $\mathcal{N}=4$ SUSY generator. We thus have 
	\es{eq:xi2downdef}{\xi^2 =\frac{N_m}{30} \left(-6i\sqrt{30}\chi^2 -3\sqrt{5}\chi^8+25\sqrt{3}\chi^{12} + 10i\sqrt{6}\chi^{16}\right)\,, }
	where the superscripts on $\chi$ run from 1 to 20 and label the $\bold{20}$ representation.
	A similar calculation using the other left-handed $\mathcal{N}=2 \subset \mathcal{N}=4$ generator $\mathbf{Q}^2$ as well as the right-handed generators $\mathbf{Q}_{1,2}$ gives us
	\es{eq:xiotherdef}{\xi^1 &=\frac{N_m}{30} \left(6i\sqrt{30}\chi^3 -3\sqrt{5}\chi^6+25\sqrt{3}\chi^{11} + 10i\sqrt{6}\chi^{15}\right)\,,\\ 
		\xi_1 &=\frac{N_m}{30} \left(-6i\sqrt{30}\overline{\chi}_3 -3\sqrt{5}\overline{\chi}_6+25\sqrt{3}\overline{\chi}_{11} - 10i\sqrt{6}\overline{\chi}_{15}\right)\,,\\
		\xi_2 &=\frac{N_m}{30} \left(6i\sqrt{30}\overline{\chi}_2 -3\sqrt{5}\overline{\chi}_8+25\sqrt{3}\overline{\chi}_{12} - 10i\sqrt{6}\overline{\chi}_{16}\right)\,, }
	where the superscripts on the first line label the $\bold{20}$ representation, while the subscripts in the other two lines label the $\overline{\bold{20}}$ representation. 
	Acting on both sides of \eqref{eq:xi2downdef} and \eqref{eq:xiotherdef} with the two left-handed ($\mathbf{Q}^{1,2}$) and two right-handed generators ($\mathbf{Q}_{1,2}$) again, using \eqref{eq:N4SUSYexamples} and comparing to the $\mathcal{N}=2$ SUSY transformations in \eqref{eq:SUSYflavor} would give the expressions for the $\mathfrak{su}(2)_F$ flavor current in terms of the $\mathfrak{su}(4)_R$-symmetry current as well as the dimension-three scalars in the flavor current multiplet ($K$ and $\overline{K}$) in terms of components of the dimension-three scalars of the stress-tensor multiplet ($P$ and $\overline{P}$). These scalars are given by
	\es{eq:KintermsofP}{K &= 4N_m P_9\,, \qquad
		\overline{K} = 4N_m \overline{P}^9\, ,}
	where the subscript in the first line is a label for the $\overline{\bold{10}}$ representation while the superscript in the second line is a label for the $\bold{10}$ representation.
	
	We can repeat a similar exercise for the $\mathcal{N}=2$ stress-tensor multiplet by starting with its superprimary $\Phi$, which lies in the $(\bold{1},\bold{1})_0$ representation in the decomposition \eqref{eq:Ssplit}. Using this fact, $\Phi$ can be identified as
	\es{eq:PhiintermsofS}{\Phi = N_b\left(\frac{1}{\sqrt{2}}S_{7}-\frac{1}{4}S_{12}-\frac{1}{4}\sqrt{\frac{3}{5}}S_{16} + \sqrt{\frac{2}{5}}S_{19}\right)\,,} 
	where $N_b$ is a numerical prefactor. We can then use the $\mathcal{N}=2$ SUSY transformations in \eqref{eq:STmultSUSY} and $\mathcal{N}=4$ SUSY transformations in \eqref{eq:N4SUSYexamples} to identify the relevant squashing deformation operators in \eqref{eq:squashcoupl}. The result is
	\es{eq:otherN2STmultope}{B^+_{\mu\nu} &= N_b((Z^+_{6})_{\mu\nu}- i (Z^{+}_5)_{\mu\nu}) \\
		B^-_{\mu\nu} &= N_b((Z^-_{6})_{\mu\nu}+ i (Z^{-}_5)_{\mu\nu}) \\
		\tensor{(j_\mu)}{_1^1}=-\tensor{(j_\mu)}{_2^2}&= 2i\sqrt{2}N_b J_{13\mu} \,,\\}
	where the subscripts in the first and second lines are $\mathbf{6}$ indices while the subscript in the third line is a label for the $\mathbf{15}$ rep.

	We can fix the coefficient $N_m$ by requiring that the $\mathcal{N} = 2$ flavor current operator is conventionally normalized. Likewise, we can fix $N_b$ by requiring that the $\mathfrak{su}(2)_R$ current in the $\mathcal{N}=2$ stress-tensor multiplet is conventionally normalized. The result is
	\es{eq:normfactssu4tosu2}{N_m = \sqrt{\frac{c}{8 \pi^4}}\,,\qquad N_b = \sqrt{\frac{3c}{8\pi^4}}\,.}
	Using these prefactors, one can check that the the operators listed in Section~\ref{N2asN4} are the same as the operators in \eqref{eq:J12compsansatz}, \eqref{eq:KintermsofP} and \eqref{eq:otherN2STmultope}, after using \eqref{eq:basis10rep}, \eqref{eq:basis10brep}, \eqref{eq:basis15rep} and \eqref{eq:basis20prep} to relate the different ways of representing the components of vectors in various reps of $\mathfrak{su}(4)$.
	
	\section{Compact expressions for correlators}
	\label{sec:corrs}
	
	In Sections~\ref{sec:nosquash} and \ref{sec:squashintcorr}, we compute squashing, mass, $\tau$, and $\bar \tau$ derivatives of the partition function $Z$ of a general $\mathcal{N}=4$ superconformal field theory on $S^4$. These observables receive contributions from integrals of the four-point functions of various operators in the $\mathcal{N}=4$ stress-tensor multiplet, as discussed in Sections \ref{sec:background} and \ref{N2asN4}. In this appendix, we provide compact expressions relating these correlators to the four-point function of the $\mathcal{N}=4$ stress\nobreakdash-tensor superprimary, given in \eqref{eq:gssssform}. Note that \eqref{eq:gssssform} contains two parts, a universal free theory part and a part that is theory-specific and determined by the conformally-invariant reduced correlator function $\mathcal{T}(u,v)$. In this appendix, we provide just the $\mathcal{T}(u,v)$-dependent part of the relevant four-point functions computed in flat space.

	The general approach we follow, based on the component field method of \cite{Dolan:2001tt}, is discussed in Section~\ref{WardSolve}. Briefly, we expand the relevant four-point functions in a basis of R-symmetry and conformal structures, with coefficients that are linear combinations of $\mathcal{T}(u,v)$ and its derivatives. The R-symmetry structures are fixed by the $\mathfrak{su}(4)$ representation of the operators involved, while the conformal structures are fixed by conformal symmetry and the Lorentz representation of the operators. The coefficients of these structures can be determined by solving the $\mathcal{N}=4$ SUSY Ward identities that relate the various four-point functions to each other.

	However, the number of independent conformal structures can become large for correlators of spinning operators (e.g. 70 structures for the four-point function of the R-symmetry current $J_\mu$), making it cumbersome to present the full expressions. Instead, we will present compact expressions for the relevant four-point functions in terms of differential operators acting on $\mathcal{T}(u,v)$. These compact expressions are inspired by the master formula presented in \cite{Belitsky:2014zha} and \cite{Korchemsky:2015ssa} for four-point functions of operators in the $\mathcal{N}=4$ stress-tensor multiplet, expressed there using the harmonic superspace formalism. More concretely, we conjecture that all four-point functions of operators in the $\mathcal{N}=4$ stress-tensor multiplet are of the form\footnote{Note that operators with spinor indices are related to their forms with coordinate or frame indices using $\sigma$-matrices, as shown in \eqref{eq:frametospinor}.}
	\es{eq:N4ansatz}{&\braket{\tensor{\mathcal{O}}{_{1,\{\alpha_{1}\}\{\dot{\alpha}_{1}\}}}(x_1) \tensor{\mathcal{O}}{_{2,\{\alpha_{2}\}\{\dot{\alpha}_{2}\}}}(x_2)\tensor{\mathcal{O}}{_{3,\{\alpha_{3}\}\{\dot{\alpha}_{3}\}}}(x_3)\tensor{\mathcal{O}}{_{4,\{\alpha_{4}\}\{\dot{\alpha}_{4}\}}}(x_4)} \\
		{}={}& \left(\prod_{i=1}^4\tensor{({\mathcal{D}_{\mathcal{O}_i}})}{_{\{\dot{\alpha}_{i}\}}^{\{\beta_{i}\}}}\right) \bigg[\mathcal{V}^{\mathcal{O}_1\mathcal{O}_2\mathcal{O}_3\mathcal{O}_4}_{\{\alpha_1\}\ldots \{\alpha_4\}\{\beta_1\}\ldots \{\beta_4\}}(x_1,x_2,x_3,x_4) \mathcal{T}(u,v)\bigg] \, .}
	Here the operators $\tensor{\mathcal{O}}{_{i,\{\alpha_{i}\}\{\dot{\alpha}_{i}\}}}$, with their dotted and undotted spinor indices symmetrized separately, correspond to components of the operators listed in Table \ref{tab:N4STOps}. The differential operators $\mathcal{D}_{\mathcal{O}_i}$ are defined as\footnote{We define derivatives with spinor indices by
	\es{eq:spinderiv}{\partial_{\alpha\dot{\alpha}} \equiv \sigma^{\mu}_{\alpha\dot{\alpha}}\frac{\partial}{\partial x^{\mu}}\, , \qquad
		\tensor{{\partial}}{^{\alpha}_{\dot{\alpha}}}\equiv \epsilon^{\alpha\beta}\partial_{\beta\dot{\alpha}}\, , \qquad
		\tensor{{\partial}}{_{\alpha}^{\dot{\alpha}}}\equiv \epsilon^{\dot{\alpha}\dot{\beta}}\partial_{\alpha\dot{\beta}} \, .  }
	Our conventions for $\epsilon_{\alpha\beta}$ and $\epsilon_{\dot{\alpha}\dot{\beta}}$ are given in \eqref{eq:defepsRaiseLower}.}
	\es{eq:defDN4ansatz}{
		\tensor{({\mathcal{D}_{\overline{\chi}}})}{_{ \dot{\alpha}}^\beta} = \tensor{({\mathcal{D}_{J}})}{_{ \dot{\alpha}}^\beta} = \tensor{({\mathcal{D}_{\psi}})}{_{ \dot{\alpha}}^\beta}  &\equiv \tensor{\partial}{_{ \dot{\alpha}}^\beta}\, , \\
		\mathcal{D}_{\bar{P}} &\equiv \Box  \, , \\
		 \tensor{({\mathcal{D}_{\bar{\lambda}}})}{_{\dot{\alpha}}^\beta} &\equiv \Box \tensor{\partial}{_{\dot{\alpha}}^\beta}\, ,  \\
		\tensor{({\mathcal{D}_{{B^{+}}}})}{_{(\dot{\alpha_1}\dot{\alpha_2})}^{(\beta_1\beta_2)}} = \tensor{({\mathcal{D}_{\bar{\psi}}})}{_{(\dot{\alpha_1}\dot{\alpha_2})}^{(\beta_1\beta_2)}} =\tensor{({\mathcal{D}_{T}})}{_{(\dot{\alpha_1}\dot{\alpha_2})}^{(\beta_1\beta_2)}} &\equiv \tensor{\partial}{_{(\dot{\alpha_1}}^{(\beta_1}} \tensor{\partial}{_{\dot{\alpha_2})}^{\beta_2)}}\, ,  \\
		\mathcal{D}_{\bar{\phi}} &\equiv \Box^2\, ,\\
		\mathcal{D}_{S}= \mathcal{D}_{\chi} =\mathcal{D}_{P} =\mathcal{D}_{{B^{-}}}= \mathcal{D}_{\phi} &\equiv 1 \, ,  \\}
	where the derivatives are taken at the point corresponding to the associated operator. The tensor $\mathcal{V}^{\mathcal{O}_1\mathcal{O}_2\mathcal{O}_3\mathcal{O}_4}_{\{\alpha_1\}\ldots \{\alpha_4\}\{\beta_1\}\ldots \{\beta_4\}}(x_1,x_2,x_3,x_4)$ must have the appropriate Lorentz structure, and it is additionally constrained by translation invariance along with the conformal Ward identities satisfied by the correlator in \eqref{eq:N4ansatz}. However, this does not fully fix the specific form of the $\mathcal{V}^{\mathcal{O}_1\mathcal{O}_2\mathcal{O}_3\mathcal{O}_4}_{\{\alpha_1\}\ldots \{\alpha_4\}\{\beta_1\}\ldots \{\beta_4\}}(x_1,x_2,x_3,x_4)$; we fix it by comparison with our results from the component field method.

	It is helpful to illustrate how the ansatz in \eqref{eq:N4ansatz} is constructed with an example. Consider the $\mathcal{N} = 4$ stress-tensor multiplet four-point function $\braket{SP B^{+}_{\dot{\alpha_1}\dot{\alpha_2}}J_{\alpha_3\dot{\alpha}_3}}$. The dimensions and R-symmetry representations of the operators in this correlator are provided in Table \ref{tab:N4STOps}; we have suppressed R-symmetry indices to avoid clutter. From \eqref{eq:defDN4ansatz}, we find that the differential operators for the operators in this correlator are
	\es{}{\tensor{{\mathcal{D}_{{B^{+}}}}}{_{(\dot{\alpha_1}\dot{\alpha_2})}^{(\beta_1\beta_2)}}=  \tensor{\partial}{_{(\dot{\alpha_1}}^{(\beta_1}} \tensor{\partial}{_{\dot{\alpha_2})}^{\beta_2)}}\, , \qquad \tensor{{\mathcal{D}_{J}}}{_{ \dot{\alpha_3}}^{\beta_3}} =  \tensor{\partial}{_{ \dot{\alpha_3}}^{\beta_3}} \, , \qquad \mathcal{D}_{S} = \mathcal{D}_{\bar{P}}  = 1 \, .}
	Hence, this correlator must be of the form
	\es{}{\braket{SP B^{+}_{\dot{\alpha_1}\dot{\alpha_2}}J_{\alpha_3\dot{\alpha}_3}} = \tensor{{\partial_2}}{_{(\dot{\alpha_1}}^{(\beta_1}} \tensor{\partial}{_{\dot{\alpha_2})}^{\beta_2)}} \tensor{{\partial_4}}{_{ \dot{\alpha_3}}^{\beta_3}}\bigg[\mathcal{V}^{{S\bar{P}B^+J}}_{(\beta_1\beta_2)\alpha_3\beta_3}\mathcal{T}(u,v)\bigg] \, .}
	Note that the undotted spinor indices $\{\alpha_i\}$ of the correlator correspond to indices of the tensor $\mathcal{V}$, while the dotted spinor indices $\{\dot{\alpha}_i\}$ of the correlator correspond to indices on the derivatives acting on $\mathcal{V}$. The undotted indices $\{\beta_i\}$ on the derivatives are all contracted with additional undotted indices of $\mathcal{V}$.

	To construct the tensors $\mathcal{V}$ for various correlators, we will use the tensors $\bold{K}^{\{i,j\}}_{k}$ and $\tilde{\bold{K}}^{\{i,j\}}_{k}$, defined as
	\es{eq:Kdefn}{
		\bold{K}^{\{i,j\}}_{k\,\alpha\beta} &\equiv  2 x^{\mu}_{ik}  x^{\nu}_{jk} \sigma_{\mu \alpha\dot{\alpha}}  \sigma_{\nu \beta\dot{\beta}}\epsilon^{\dot{\beta}\dot{\alpha}}\,,\\
		\tilde{\bold{K}}^{\{i,j\}}_{k\,\dot{\alpha}\dot{\beta}} &\equiv  2 x^{\mu}_{ik}  x^{\nu}_{jk} \sigma_{\mu \alpha\dot{\alpha}}  \sigma_{\nu \beta\dot{\beta}}\epsilon^{\beta\alpha}\,,\\
	}
	as building blocks to write down the $\mathcal{V}^{\mathcal{O}_1\mathcal{O}_2\mathcal{O}_3\mathcal{O}_4}_{\{\alpha_1\}\ldots \{\alpha_4\}\{\beta_1\}\ldots \{\beta_4\}}$ tensors for the correlators relevant to our computation.\footnote{The two tensor structures in \eqref{eq:Kdefn} are related to two of the building blocks for 4D CFT correlators given in Section D of \cite{Cuomo:2017wme}.}
	
	\subsection{Correlators for \texorpdfstring{$\partial_\tau \partial_{\bar{\tau}} \partial^2_m \log Z$}{tau² m² derivative}}
	
	The $\mathcal{N}=2$ operators $\mathcal{A}$ and $\bar{\mathcal{A}}$ generate the $\tau$ and $\bar{\tau}$ deformations, defined in \eqref{eq:taupole}, while the $\mathcal{N}=2$ operators $K$, $\overline{K}$, and $J_{12}$ are involved in the mass deformation, given in \eqref{eq:masscoupl}. We identified $\mathcal{A}$, $\bar{\mathcal{A}}$, and $J_{12}$ as particular components of the $\mathcal{N}=4$ stress-tensor superprimary $S$  in \eqref{eq:AAbarcomps} and \eqref{eq:J12comps}, and in \eqref{eq:KKbarcomps} we identified $K$ and $\overline{K}$ as specific components of the $\mathcal{N}=4$ dimension-three scalar operators $P$ and $\bar{P}$ respectively. The correlators contributing to the $\partial_\tau \partial_{\bar{\tau}} \partial^2_m \log Z$ integrated correlator are
	\es{eq:corrmmtt}{
		\braket{\mathcal{A}(x_1)\bar{\mathcal{A}}(x_2)J_{12}(x_3)J_{12}(x_4)}&=  -\frac{c^2v\mathcal{T}(u,v)}{16\pi^4 \tau^2_2 x^4_{12} x^4_{34}}\,, \\ 
		\braket{\mathcal{A}(x_1)\bar{\mathcal{A}}(x_2)K(x_3)\overline{K}(x_4)}&= \Box_4\left[ \frac{c^2x^2_{14} v\mathcal{T}(u,v)}{8\pi^4 \tau^2_2 x^2_{13} x^4_{12} x^4_{34}} \right]\,. \\ 
	}	
	
	\subsection{Correlators for \texorpdfstring{$\partial^4_m \log Z$}{m⁴ derivative}}
	
	The correlators contributing to $\partial^4_m \log Z$ are
	\es{eq:corrmmmm}{
		\braket{J_{12}(x_1)J_{12}(x_2)J_{12}(x_3)J_{12}(x_4)}&=  \frac{c^2 (1+u+v)^2\mathcal{T}(u,v)}{256\pi^8 x^4_{12}  x^4_{34}}\,, \\
		\braket{J_{12}(x_1)J_{12}(x_2)K(x_3)\overline{K}(x_4)}&= \Box_4\left[ \frac{-c^2 x^2_{14} (1+u+v)\mathcal{T}(u,v)}{64\pi^8 x^2_{13} x^4_{12}  x^4_{34}} \right]\,, \\ 
		\braket{K(x_1) K(x_2) \overline{K}(x_3) \overline{K}(x_4)}&= \Box_3\,\Box_4\left[ \frac{c^2 (1+u+v)\mathcal{T}(u,v)}{32 \pi^8 x^2_{24} x^2_{13} x^4_{12}} \right]\,, \\
	}
	where we use \eqref{eq:AAbarcomps} and \eqref{eq:KKbarcomps} to identify the appropriate $\mathcal{N}=4$ stress-tensor multiplet operators appearing in these correlators.
	
	\subsection{Correlators for \texorpdfstring{$\partial_{\tau} \partial_{\bar{\tau}} \partial^2_b \log Z$}{tau² b² derivative}}
	\label{sec:corrbbtt}
	
	The squashing deformation, defined in \eqref{eq:squashcoupl}, involves the spinning operators $\tensor{j}{_{a\,1}^1}$, $Z^{-}_{ab}$, and $Z^+_{ab}$ of the $\mathcal{N} = 2$ stress-tensor multiplet. In \eqref{eq:jdiagcomps} we identified $\tensor{j}{_{a\,1}^1}$ as particular components of the $\mathcal{N}=4$ $\mathfrak{su}(4)_R$-symmetry current operator $J_\mu$, and in \eqref{eq:ZZbarcomps} we identified $Z^{-}_{ab}$ and  $Z^+_{ab}$ with specific components of the $\mathcal{N}=4$ tensor operators. The correlators contributing to $\partial_{\tau} \partial_{\bar{\tau}} \partial^2_b \log Z$ are 
	\es{eq:corrbbtt}{
		\braket{\mathcal{A}(x_1)\bar{\mathcal{A}}(x_2)\tensor{j}{_{1}^1_{\alpha_1\dot{\alpha}_1}}(x_3)\tensor{j}{_{1}^1_{\alpha_2\dot{\alpha}_2}}(x_4)} &= \frac{c^2}{32\pi^4\tau^2_2}\tensor{{\partial_3}}{^{\beta_1}_{\dot{\alpha}_1}}\tensor{{\partial_4}}{^{\beta_2}_{\dot{\alpha}_2}}\left[\mathcal{P}_{(\alpha_1\beta_1)(\alpha_2\beta_2)}\frac{x^2_{14}\mathcal{T}(u,v)}{x^4_{12}x^4_{34}x^2_{13}x^4_{24}}\right]\, ,\\
		\braket{\mathcal{A}(x_1)\bar{\mathcal{A}}(x_2) Z^{-}_{\alpha_1\beta_1}(x_3)Z^{+}_{\dot{\alpha}_2\dot{\beta}_2}(x_4)} &= -\frac{c^2}{256\pi^4\tau^2_2}\tensor{{\partial_4}}{^{\alpha_2}_{\dot{\alpha}_2}}\tensor{{\partial_4}}{^{\beta_2}_{\dot{\beta}_2}}\left[\mathcal{P}_{(\alpha_1\beta_1)(\alpha_2\beta_2)}\frac{x^4_{14}\mathcal{T}(u,v)}{x^4_{24}x^4_{13}x^4_{12}x^4_{34}}\right]\, ,\\} where we define the tensor
	\es{eq:Ptensdefn}{
		\mathcal{P}_{(\alpha_1\beta_1)(\alpha_2\beta_2)}&= \frac{1}{2}\left[\bold{K}^{\{3,4\}}_{2\,\alpha_1\alpha_2} \bold{K}^{\{3,4\}}_{2\,\beta_1\beta_2}+ \bold{K}^{\{3,4\}}_{2\,\beta_1\alpha_2} \bold{K}^{\{3,4\}}_{2\,\alpha_1\beta_2}\right]\, .  \\ }
	For large $|x_2|$, this tensor behaves as
	\es{eq:Plim}{
		\lim_{|x_2|\to \infty}\frac{\mathcal{P}_{(\alpha_1\beta_1)(\alpha_2\beta_2)}}{|x_2|^4}&= 2 (\epsilon_{\alpha_1\alpha_2} \epsilon_{\beta_1\beta_2} +\epsilon_{\alpha_1\beta_2} \epsilon_{\beta_1\alpha_2}).}
	We can also express this tensor in frame indices as
	\es{eq:Ptensdefn2}{
		\mathcal{P}_{(\alpha_1\beta_1)(\alpha_2\beta_2)} & \equiv 
		\mathcal{{P}}_{[a_1 b_1][a_2b_2]}\sigma^{a_1b_1}_{\alpha_1\beta_1}\sigma^{a_2b_2}_{\alpha_2\beta_2}\, , \\
		\mathcal{{P}}^{[a_1b_1][a_2b_2]}&= \, \tensor{{P^-}}{^{[a_1b_1]}_{[a_3b_3]}}\tensor{{P^-}}{^{[a_2b_2]}_{[a_4b_4]}}\left(16 x^{a_3}_{32}x^{b_3}_{42}x^{a_4}_{32}x^{b_4}_{42} + \eta^{a_3a_4} \left(16 x_{32}\cdot x_{42} x^{b_3}_{32}x^{b_4}_{42}  \right.\right. \\ 
		&\left.\left.{} -4 x^2_{42}x^{b_3}_{32}x^{b_4}_{32}-4 x^2_{32}x^{b_3}_{42}x^{b_4}_{42} \right)\right)\, . \\
	}
	Here the tensor $\tensor{{P^-}}{^{[ab]}_{[cd]}}$, defined as
	\es{eq:projasddef}{\tensor{{P^-}}{^{[ab]}_{[cd]}}=  \frac{1}{2}\left[\frac{1}{2}(\tensor{\delta}{^a_c}\tensor{\delta}{^b_d}-\tensor{\delta}{^a_d}\tensor{\delta}{^b_c}) -\frac{1}{2}\tensor{\epsilon}{^{ab}_{cd}}\right]\, ,}
	projects out the antisymmetric anti-self-dual part of a tensor.

	\subsection{Correlators for \texorpdfstring{$\partial^2_b\partial^2_m \log Z$}{b² m² derivative}}
	
	The correlators that contribute to the $\bbmmZ$ integrated correlator are
	\es{eq:corrbbmm}{
		\braket{J_{12}(x_1)J_{12}(x_2)\tensor{j}{_{1}^1_{\alpha_1\dot{\alpha}_1}}(x_3)\tensor{j}{_{1}^1_{\alpha_2\dot{\alpha}_2}}(x_4)} &= \frac{-c^2}{256\pi^8}\tensor{{\partial_3}}{^{\beta_1}_{\dot{\alpha}_1}}\tensor{{\partial_4}}{^{\beta_2}_{\dot{\alpha}_2}}\left[\mathcal{K}_{\text{flat}\,(\alpha_1\beta_1)(\alpha_2\beta_2)}\frac{u(1+u+v)\mathcal{T}(u,v)}{x^6_{12}x^6_{34}}\right]\, ,\\
		\braket{K(x_1)\overline{K}(x_2)\tensor{j}{_{1}^1_{\alpha_1\dot{\alpha}_1}}(x_3)\tensor{j}{_{1}^1_{\alpha_2\dot{\alpha}_2}}(x_4)} &= \frac{c^2}{64\pi^8}\Box_2\,\tensor{{\partial_3}}{^{\beta_1}_{\dot{\alpha}_1}}\tensor{{\partial_4}}{^{\beta_2}_{\dot{\alpha}_2}}\left[\mathcal{K}_{\text{flat}\,(\alpha_1\beta_1)(\alpha_2\beta_2)}\frac{x^2_{23}\mathcal{T}(u,v)}{x^2_{24}x^4_{13}x^4_{12}x^4_{34}}\right]\, ,\\
		\braket{J_{12}(x_1)J_{12}(x_2)Z^{-}_{\alpha_1\beta_1}(x_3)Z^{+}_{\dot{\alpha}_2\dot{\beta}_2}(x_4)} &= \frac{c^2}{1024\pi^8}\tensor{{\partial_4}}{^{\alpha_2}_{\dot{\alpha}_2}}\tensor{{\partial_4}}{^{\beta_2}_{\dot{\beta}_2}}\left[\mathcal{K}_{\text{flat}\,(\alpha_1\beta_1)(\alpha_2\beta_2)}\frac{x^2_{14}\mathcal{T}(u,v)}{x^2_{24}x^4_{13}x^4_{12}x^4_{34}}\right]\, ,\\
		\braket{K(x_1)\overline{K}(x_2)Z^{-}_{\alpha_1\beta_1}(x_3)Z^{+}_{\dot{\alpha}_2\dot{\beta}_2}(x_4)} &= \frac{-c^2}{512\pi^8}\Box_2\tensor{{\partial_4}}{^{\alpha_2}_{\dot{\alpha}_2}}\tensor{{\partial_4}}{^{\beta_2}_{\dot{\beta}_2}}\left[\mathcal{K}_{\text{flat}\,(\alpha_1\beta_1)(\alpha_2\beta_2)}\frac{\mathcal{T}(u,v)}{x^4_{13}x^4_{12}x^4_{34}}\right]\, ,\\
		\braket{\overline{K}(x_1)\overline{K}(x_2)Z^{-}_{\alpha_1\beta_1}(x_3)Z^{-}_{\alpha_2\beta_2}(x_4)} &= \frac{c^2}{512\pi^8}\Box_1\Box_2\left[\mathcal{K}_{\text{flat}\,(\alpha_1\beta_1)(\alpha_2\beta_2)}\frac{\mathcal{T}(u,v)}{x^4_{13}x^4_{24}x^4_{34}}\right]\, ,\\
		`    \braket{K(x_1)K(x_2)Z^{+}_{\dot{\alpha}_1\dot{\beta}_1}(x_3)Z^{+}_{\dot{\alpha}_2\dot{\beta}_2}(x_4)} &= \frac{c^2}{512\pi^8}\tensor{{\partial_3}}{^{\alpha_1}_{\dot{\alpha}_1}}\tensor{{\partial_3}}{^{\beta_1}_{\dot{\beta}_1}}\tensor{{\partial_4}}{^{\alpha_2}_{\dot{\alpha}_2}}\tensor{{\partial_4}}{^{\beta_2}_{\dot{\beta}_2}}\left[\mathcal{K}_{\text{flat}\,(\alpha_1\beta_1)(\alpha_2\beta_2)}\frac{\mathcal{T}(u,v)}{x^4_{13}x^4_{24}x^4_{12}}\right]\, ,\\
		\braket{J_{12}(x_1)K(x_2)Z^{+}_{\dot{\alpha}_1\dot{\beta}_1}(x_3)\tensor{j}{_{1}^1_{\alpha_2\dot{\alpha}_2}}(x_4)} &= \frac{ic^2}{256\pi^8}\tensor{{\partial_3}}{^{\alpha_1}_{\dot{\alpha}_1}}\tensor{{\partial_3}}{^{\beta_1}_{\dot{\beta}_1}}\tensor{{\partial_4}}{^{\beta_2}_{\dot{\alpha}_2}} \left[\mathcal{K}_{\text{flat}\,(\alpha_1\beta_1)(\alpha_2\beta_2)}\frac{\mathcal{T}(u,v)}{x^2_{13}x^2_{34}x^4_{12}x^4_{24}}\right]\, ,\\
		\braket{J_{12}(x_1)\overline{K}(x_2)Z^{-}_{\alpha_1\beta_1}(x_3)\tensor{j}{_{1}^1_{\alpha_2\dot{\alpha}_2}}(x_4)} &= -\frac{ic^2}{256\pi^8}\Box_2 \tensor{{\partial_4}}{^{\beta_2}_{\dot{\alpha}_2}} \left[\mathcal{K}_{\text{flat}\,(\alpha_1\beta_1)(\alpha_2\beta_2)}\frac{\mathcal{T}(u,v)}{x^2_{12}x^2_{24}x^4_{13}x^4_{34}}\right]\, .\\
	}
	Here we have used \eqref{eq:J12comps}, \eqref{eq:KKbarcomps}, \eqref{eq:ZZbarcomps}, and \eqref{eq:jdiagcomps}  to identify the relevant $\mathcal{N}=4$ operators in these correlators, and we define $\mathcal{K}_{\text{flat}\,(\alpha_1\beta_1)(\alpha_2\beta_2)}$ as
	\es{eq:Ktensdefn}{
		\mathcal{K}_{\text{flat}\,(\alpha_1\beta_1)(\alpha_2\beta_2)}&= \frac{1}{4}\left(\bold{K}^{\{3,4\}}_{2\,\alpha_1\alpha_2} \bold{K}^{\{3,4\}}_{1\,\beta_1\beta_2}+\bold{K}^{\{3,4\}}_{2\,\beta_1\alpha_2} \bold{K}^{\{3,4\}}_{1\,\alpha_1\beta_2}+\bold{K}^{\{3,4\}}_{2\,\alpha_1\beta_2} \bold{K}^{\{3,4\}}_{1\,\beta_1\alpha_2}+\bold{K}^{\{3,4\}}_{2\,\beta_1\beta_2} \bold{K}^{\{3,4\}}_{1\,\alpha_1\alpha_2}\right)\, ,  \\ }
	where $\bold{K}^{\{i,j\}}_{k\,\alpha\beta} $ is defined in \eqref{eq:Kdefn}. We can also write the $\mathcal{K}_{\text{flat}}$ tensor as
	\es{eq:Ktenscorrindex}{\mathcal{K}_{\text{flat}\,(\alpha_1\beta_1)(\alpha_2\beta_2)} &= \mathcal{K}^{[a_1b_1][a_2b_2]}_{\text{flat}} \tensor{({\sigma_{a_1b_1}})}{_{\alpha_1\beta_1}}\tensor{({\sigma_{a_2b_2}})}{_{\alpha_2\beta_2}}, \\ 
		\mathcal{K}_{\text{flat}}^{[a_1b_1][a_2b_2]} & = 8\tensor{{P^{-}}}{^{[a_1 b_1]}_{[a_3 b_3]}} \tensor{{P^{-}}}{^{[a_2b_2]}_{[a_4b_4]}}\left(  x^{a_3}_{31}x^{b_3}_{32}x^{a_4}_{41}x^{b_4}_{42} +  x^{a_3}_{31}x^{b_3}_{42}x^{a_4}_{32}x^{b_4}_{41}+ x^{a_3}_{31}x^{b_3}_{41}x^{a_4}_{32}x^{b_4}_{42} \right.\\
		&\left. + \eta^{a_3a_4}(-(x_{42}\cdot x_{41})x^{b_3}_{31}x^{b_4}_{32} +(x_{32}\cdot x_{41})x^{b_3}_{31}x^{b_4}_{42} + (x_{32}\cdot x_{42})x^{b_3}_{31}x^{b_4}_{41} ) \right). \\  }
	where  $\tensor{{P^{-}}}{^{[ab]}_{[cd]}}$, defined in \eqref{eq:projasddef}, projects out the antisymmetric anti-self-dual part of a tensor. Note that in Section~\ref{sec:bbmm}, we will use these identities with $\mathcal{K}_\text{flat}$ replaced with a version multiplied by conformal factors to convert from $\mathbb{R}^4$ to $S^4$:
	\es{eq:Kcaltensdefn}{\mathcal{K}_{\,[a_1b_1][a_2b_2]}(x_1,x_2,x_3,x_4) = \left(\prod^4_{i=1} \Omega(x_i)\right) \mathcal{K}_{\text{flat}\,[a_1b_1][a_2b_2]}\, , }
	where $\Omega(x)$ is defined as in \eqref{eq:S4metric}.
	
	Note also that $\braket{J_{12}(x_1)K(x_2)Z^{+}_{\dot{\alpha}_1\dot{\beta}_1}(x_3)\tensor{j}{_{1}^1_{\alpha_2\dot{\alpha}_2}}(x_4)}$ and $\braket{J_{12}(x_1)\overline{K}(x_2)Z^{-}_{\alpha_1\beta_1}(x_3)\tensor{j}{_{1}^1_{\alpha_2\dot{\alpha}_2}}(x_4)}$ are conjugate to each other, although this is not apparent from the expressions in \eqref{eq:corrbbmm}. The same is true for the  $\braket{K(x_1)K(x_2)Z^{+}_{\dot{\alpha}_1\dot{\beta}_1}(x_3)Z^{+}_{\dot{\alpha}_2\dot{\beta}_2}(x_4)}$ and $\braket{\overline{K}(x_1)\overline{K}(x_2)Z^{-}_{\alpha_1\beta_1}(x_3)Z^{-}_{\alpha_2\beta_2}(x_4)}$ correlators. These relations are made manifest in the alternate expressions
	\es{eq:corrbbmm2}{
		\braket{J_{12}(x_1)K(x_2)Z^{+}_{\dot{\alpha}_1\dot{\beta}_1}(x_3)\tensor{j}{_{1}^1_{\alpha_2\dot{\alpha}_2}}(x_4)} &= \frac{ic^2}{128\pi^8}\Box_2 \tensor{{\partial_4}}{_{\alpha_2}^{\dot{\beta}_2}} \left[\tilde{\mathcal{K}}_{\text{flat}\,(\dot{\alpha}_1\dot{\beta}_1)(\dot{\alpha}_2\dot{\beta}_2)}\frac{\mathcal{T}(u,v)}{x^2_{12}x^2_{24}x^4_{13}x^4_{34}}\right],\\
		\braket{K(x_1)K(x_2)Z^{+}_{\dot{\alpha}_1\dot{\beta}_1}(x_3)Z^{+}_{\dot{\alpha}_2\dot{\beta}_2}(x_4)} &= \frac{c^2}{512\pi^8}\Box_1\Box_2\left[\tilde{\mathcal{K}}_{\text{flat}\,(\dot{\alpha}_1\dot{\beta}_1)(\dot{\alpha}_2\dot{\beta}_2)}\frac{\mathcal{T}(u,v)}{x^4_{13}x^4_{24}x^4_{34}}\right],\\}
	where, using $\tilde{\bold{K}}^{\{i,j\}}_{k\,\dot{\alpha}\dot{\beta}} $ defined in \eqref{eq:Kdefn}, we have
	\es{eq:Ktenstildefn}{	\tilde{\mathcal{K}}_{\text{flat}\,(\dot{\alpha}_1\dot{\beta}_1)(\dot{\alpha}_2\dot{\beta}_2)}&= \frac{1}{4}\left(\tilde{\bold{K}}^{\{3,4\}}_{2\,\dot{\alpha}_1\dot{\alpha}_2} \tilde{\bold{K}}^{\{3,4\}}_{1\,\dot{\beta}_1\dot{\beta}_2}+\tilde{\bold{K}}^{\{3,4\}}_{2\,\dot{\beta}_1\dot{\alpha}_2} \tilde{\bold{K}}^{\{3,4\}}_{1\,\dot{\alpha}_1\dot{\beta}_2}+\tilde{\bold{K}}^{\{3,4\}}_{2\,\dot{\alpha}_1\dot{\beta}_2} \tilde{\bold{K}}^{\{3,4\}}_{1\,\dot{\beta}_1\dot{\alpha}_2}+\tilde{\bold{K}}^{\{3,4\}}_{2\,\dot{\beta}_1\dot{\beta}_2} \tilde{\bold{K}}^{\{3,4\}}_{1\,\dot{\alpha}_1\dot{\alpha}_2}\right)\, .}

	\subsection{Correlators for \texorpdfstring{$\bbbbZ$}{b⁴ derivative}}
	
	The correlators  relevant for computing $\bbbbZ$ are
	\es{eq:corrbbbb}{
		&\braket{\tensor{j}{_{1}^1_{\alpha_1\dot{\alpha}_1}}(x_1)\tensor{j}{_{1}^1_{\alpha_2\dot{\alpha}_2}}(x_2)\tensor{j}{_{1}^1_{\alpha_3\dot{\alpha}_3}}(x_3)\tensor{j}{_{1}^1_{\alpha_4\dot{\alpha}_4}}(x_4)} \\
		&= {}-\tensor{{\partial_1}}{^{\beta_1}_{\dot{\alpha}_1}} \tensor{{\partial_2}}{^{\beta_2}_{\dot{\alpha}_2}} \tensor{{\partial_3}}{^{\beta_3}_{\dot{\alpha}_3}} \tensor{{\partial_4}}{^{\beta_4}_{\dot{\alpha}_4}}\left[\mathcal{M}_{\text{flat}\,(\alpha_1\beta_1)(\alpha_2\beta_2)(\alpha_3\beta_3)(\alpha_4\beta_4)}  \frac{u^2(1+u+v)}{v}\frac{c^2\mathcal{T}(u,v)}{512\pi^8 x^8_{12}x^8_{34}}\right], \\
		&\braket{Z^{-}_{{\alpha_1\beta_1}}(x_1)\tensor{j}{_{1}^1_{\alpha_2\dot{\alpha}_2}}(x_2)\tensor{j}{_{1}^1_{\alpha_3\dot{\alpha}_3}}(x_3)Z^{+}_{{\dot{\alpha}_4\dot{\beta}_4}}(x_4)} \\
		&=  \tensor{{\partial_2}}{^{\beta_2}_{\dot{\alpha}_2}} \tensor{{\partial_3}}{^{\beta_3}_{\dot{\alpha}_3}} \tensor{{\partial_4}}{^{\alpha_4}_{\dot{\alpha}_4}} \tensor{{\partial_4}}{^{\beta_4}_{\dot{\beta}_4}}\left[\mathcal{M}_{\text{flat}\,(\alpha_1\beta_1)(\alpha_2\beta_2)(\alpha_3\beta_3)(\alpha_4\beta_4)}  \frac{u^2}{v}\frac{c^2\mathcal{T}(u,v)}{2048\pi^8 x^8_{12}x^2_{13}x^6_{34}}\right], \\
		&\braket{Z^{-}_{{\alpha_1\beta_1}}(x_1)Z^{-}_{{\alpha_2\beta_2}}(x_2)Z^{+}_{{\dot{\alpha}_3\dot{\beta}_3}}(x_3)Z^{+}_{{\dot{\alpha}_4\dot{\beta}_4}}(x_4)}\\
		&= {}-\tensor{{\partial_3}}{^{\alpha_3}_{\dot{\alpha}_3}} \tensor{{\partial_3}}{^{\beta_3}_{\dot{\beta}_3}}\tensor{{\partial_4}}{^{\alpha_4}_{\dot{\alpha}_4}} \tensor{{\partial_4}}{^{\beta_4}_{\dot{\beta}_4}}\left[\mathcal{M}_{\text{flat}\,(\alpha_1\beta_1)(\alpha_2\beta_2)(\alpha_3\beta_3)(\alpha_4\beta_4)}  \frac{u^3}{v}\frac{c^2\mathcal{T}(u,v)}{16384\pi^8 x^{10}_{12}x^6_{34}}\right], \\	
	}
	where we use \eqref{eq:ZZbarcomps} and \eqref{eq:jdiagcomps} to identify the relevant components of $\mathcal{N}=4$ operators and we define
	\es{eq:Mtensdefn}{
		\mathcal{M}_{\text{flat}\,(\alpha_1\beta_1)(\alpha_2\beta_2)(\alpha_3\beta_3)(\alpha_4\beta_4)} &= \frac{1}{16} \Bigl[\left(\bold{K}^{\{1,2\}}_{3\,\alpha_1\alpha_2} \bold{K}^{\{1,3\}}_{4\,\beta_1\alpha_3} \bold{K}^{\{2,4\}}_{1\,\beta_2\alpha_4} \bold{K}^{\{3,4\}}_{2\,\beta_3\beta_4} + \bold{K}^{\{1,2\}}_{4\,\alpha_1\alpha_2} \bold{K}^{\{1,3\}}_{2\,\beta_1\alpha_3} \bold{K}^{\{2,4\}}_{3\,\beta_2\alpha_4} \bold{K}^{\{3,4\}}_{1\,\beta_3\beta_4} \right.\\
		& {} -  \left. \bold{K}^{\{1,2\}}_{3\,\alpha_1\alpha_2} \bold{K}^{\{1,2\}}_{4\,\beta_1\beta_2} \bold{K}^{\{3,4\}}_{1\,\alpha_3\alpha_4} \bold{K}^{\{3,4\}}_{2\,\beta_3\beta_4} -\bold{K}^{\{1,3\}}_{2\,\alpha_1\alpha_3} \bold{K}^{\{1,3\}}_{4\,\beta_1\beta_3} \bold{K}^{\{2,4\}}_{1\,\alpha_2\alpha_4} \bold{K}^{\{2,4\}}_{3\,\beta_2\beta_4}\right)\\ 
		&{}+ (\text{15 other terms given by symmetrizing } (\alpha_i \leftrightarrow\beta_i)  \text{ pairs})\Bigr]\, ,\\
	}
	using $\bold{K}^{\{i,j\}}_{k\,\alpha\beta} $ given in \eqref{eq:Kdefn}. Note that in Section~\ref{sec:bbbb}, we use these identities with $\mathcal{M}_\text{flat}$ replaced by a version multiplied by conformal factors to convert from $\mathbb{R}^4$ to $S^4$:
	\es{eq:Mtenscorrindex}{\mathcal{M}_{[a_1b_1][a_2b_2][a_3b_3][a_4b_4]}(x_1,x_2,x_3,x_4) = \left(\prod^4_{i=1} \Omega(x_i)\right)^2 \mathcal{M}_{\text{flat}\,[a_1b_1][a_2b_2][a_3b_3][a_4b_4]}\, , }
	where $\Omega(x)$ is defined as in \eqref{eq:S4metric}. 
	
	The tensor structure 
	\es{}{\mathcal{M}\equiv \frac{1}{16}\mathcal{M}_{[\mu_1 \nu_1][\mu_2 \nu_2][\mu_3 \nu_3][\mu_4 \nu_4]} (dx_1^{\mu_1}\wedge dx_1^{\nu_1}) \, (dx_2^{\mu_2}\wedge dx_2^{\nu_2}) \, (dx_3^{\mu_3}\wedge dx_3^{\nu_3}) \, (dx_4^{\mu_4}\wedge dx_4^{\nu_4})}
	can also be written as 
	\es{eq:MtensHarmonic}{ 
		\mathcal{M}(x_1,x_2,x_3,x_4) =& 16 y^2_{23}y^2_{14} T^{\mathcal{M}}_{[p_1q_1][p_2q_2][p_3q_3][p_4q_4]} \\
		&\times  \mathcal{Y}^{-[p_1q_1]}(y_1)\mathcal{Y}^{-[p_2q_2]}(y_2)\mathcal{Y}^{-[p_3q_3]}(y_3)\mathcal{Y}^{-[p_4q_4]}(y_4)\, ,\\
	}
	where $p_i,q_i$ are $\SO(5)$ indices that we sum over, $y^2_{ij}=\Omega(x_i)\Omega(x_j)x^2_{ij}$ is the chordal distance on $S^4$, $\mathcal{Y}^{-[p_4q_4]}$ are $\bold{10}$ anti-self-dual tensor harmonics defined as 
	\es{}{\mathcal{Y}^{-[pq]} = \frac{1}{4}\left(dy^p\wedge dy^q -\frac{1}{2}\epsilon^{pqklm}y^k dy^l\wedge dy^m\right)\,,} 
	 and $T^{\mathcal{M}}$ is an $\SO(5)$-invariant tensor that can be written in terms of the basis \eqref{eq:Tso5basis} of  $\SO(5)$-invariant tensors as
	\es{eq:TMasTsum}{T^{\mathcal{M}} = \frac{1}{2}(T^1 + T^2 + T^3)-2(T^4 + T^5 + T^6)\, .}

	\section{Converting to conformally-invariant form}
	\label{sec:AdSavg}
	
	In this appendix, we will describe how an integral of an $\SO(5)$\nobreakdash-invariant function of four points on $S^4$ can be written in a conformally-invariant form. Following the discussion in Section~\ref{sec:embed6d}, the integrand can be written as a function in the embedding space $\mathbb{R}^{5,1}$. Therefore, we can express this integral in the form
	\es{eq:intI1}{\mathcal{I}[f] =  \int  \frac{\prod^4_{i=1} d^4\Omega^i_{S^4}}{(y^2_{12}y^2_{34})^4} \, \left.f\left(\frac{\bold{X}_i\cdot \bold{X}_j}{(\bold{X}_i\cdot \bold{Y}_*)(\bold{X}_j\cdot \bold{Y}_*)}\right)\right|_{\bold{X}_i \in \mathbb{L}_{\vec{y}_i}} \, ,}
	where $d^4\Omega^i_{S^4} = d^5 y_i \,\delta(y^2_i - 1)$ is the measure on $S^4$, the point $y_i$ on $S^4$ is embedded in $\mathbb{R}^{5,1}$ as the lightray $\mathbb{L}_{\vec{y}_i}$  defined in \eqref{eq:embed}, and $\bold{Y}_* = (0,0,0,0,0,1)$ is a point on the AdS hyperboloid, defined as the locus of points satisfying $\bold{Y}\cdot\bold{Y} = -1$ and $Y_0 > 0$. Note that the factors $\frac{\prod_{i=1}^2 d^4\Omega^i_{S^4}}{(y^2_{12})^4}$ and $\frac{\prod_{i=3}^4 d^4\Omega^i_{S^4}}{(y^2_{34})^4}$ are both conformally invariant.\footnote{ This follows from the same argument as the footnote following \eqref{eq:mmmm5}. } 
	
	We begin simplifying \eqref{eq:bbmm5} by first using a $\SO(5)$ rotation followed by a $\SO(4)$ rotation to fix
	\es{eq:lcset1}{\vec{y}_3 = (0,0,0,0,-1)\, , \qquad \vec{y}_4 = (0,0,0,\sin\lambda,\cos\lambda)\, ,}
	which gives
	\es{eq:intI2}{\mathcal{I}[f] =  \frac{\vol(S^4) \vol(S^3)}{16} \int  \frac{\prod^2_{i=1} d^4\Omega^i_{S^4} }{(y^2_{12})^4}\, dt \, t^3 \, \left.f\left(\frac{\bold{X}_i\cdot \bold{X}_j}{(\bold{X}_i\cdot \bold{Y}_*)(\bold{X}_j\cdot \bold{Y}_*)}\right)\right|_{\bold{X}_i \in \mathbb{L}_{\vec{y}_i}} \, ,}
	where $t = \tan \frac{\lambda}{2}$. We then perform an $\SO(5,1)$ transformation $\bold{\Lambda}_1$ given by the matrix 
	\es{eq:conftransf1}{\tensor{\Lambda}{_1^M_N} =  \begin{pmatrix}
			\mathbbm{1}_3 & 0 & 0 & 0\\
			0  & 1 &  -t & -t\\
			0 &  t & 1 - \frac{t^2}{2} & - \frac{t^2}{2} \\
			0 & - t &  \frac{t^2}{2} & 1+\frac{t^2}{2}\\
		\end{pmatrix}}
	to set $\vec{y}_4 =(0,0,0,0,1)$. This transformation sends $\bold{Y}_*$ to $\bold{Y}_1 = \left(0,0,0,- t ,-\frac{t^2}{2},1+\frac{t^2}{2}\right)$, while $\vec{y}_3$ is left unchanged. In \eqref{eq:intI2}, $y_1$ and $y_2$ are generic points on $S^4$. Since the measure $\frac{\prod^2_{i=1} d^4\Omega^i_{S^4} }{(y^2_{12})^4}$ in \eqref{eq:intI2} is invariant under conformal transformations, the conformal transformation of $y_1$ and $y_2$ under $\bold{\Lambda}_1$ can be absorbed conveniently by a change of integration variables. Hence, following this conformal transformation, the integral \eqref{eq:intI2} is reduced to the form
	\es{eq:bbmm7}{\mathcal{I}[f] =  \frac{\vol(S^4) \vol(S^3)}{16} \int  \frac{\prod^2_{i=1} d^4\Omega^i_{S^4} }{(y^2_{12})^4}\, dt \, t^3 \, \left.f\left(\frac{\bold{X}_i\cdot \bold{X}_j}{(\bold{X}_i\cdot \bold{Y}_1)(\bold{X}_j\cdot \bold{Y}_1)}\right) \right|_{\bold{X}_i \in \mathbb{L}_{\vec{y}_i}} \, ,}
	where 
	\es{eq:fixpts1}{\bold{Y}_1 &=\left(0,0,0,- t ,-\frac{t^2}{2},1+\frac{t^2}{2}\right)\, ,\\
		\vec{y}_3 &= \vec{y}^{\,*}_3 = (0,0,0,0,-1)\, , \\
		\vec{y}_4 &= \vec{y}^{\,*}_4 = (0,0,0,0,1)\, .\\}
	In terms of the stereographic coordinates \eqref{eq:varsR4toS4}, this corresponds to sending $\vec{x}_3$ to infinity while fixing $\vec{x}_4$ at the origin. Simplifying \eqref{eq:bbmm7} further becomes convenient if we switch to stereographic coordinates, defined in \eqref{eq:varsR4toS4}, for the points $y_1$ and $y_2$. The measure factor $\frac{\prod^2_{i=1} d^4\Omega^i_{S^4} }{(y^2_{12})^4}$ in \eqref{eq:bbmm7} changes to $\frac{\prod^2_{i=1} d^4 x_i}{(x^2_{12})^4}$ with this choice of parametrization. 
	
	We can now do a dilation in these stereographic coordinates to fix $|x_2| = 1$. The dilation, given by the $\SO(5,1)$ matrix
	\es{eq:conftransf2}{\tensor{\Lambda}{_2^M_N} = \begin{pmatrix}
			\mathbbm{1}_4 & 0 & 0 \\
			0 & \frac{|x_2|+|x_2|^{-1}}{2}	&  \frac{|x_2|-|x_2|^{-1}}{2} \\
			0 & \frac{|x_2|-|x_2|^{-1}}{2}	&  \frac{|x_2|+|x_2|^{-1}}{2} \\
		\end{pmatrix}\, ,  }
	sends $\vec{x}_2$ to $\vec{x}'_2=\frac{\vec{x}_2}{|x_2|}$, $\vec{x}_1$ to $\vec{x}'_1=\frac{\vec{x}_1}{|x_2|}$, and $\bold{Y}_1$ to $\bold{Y}_2=\left(0,0,0,-t,\frac{|x_2|^2-t^2-1}{2|x_2|},\frac{|x_2|^2+t^2+1}{2|x_2|}\right)$, while leaving $\vec{x}_3$ and $\vec{x}_4$ fixed. Switching integration variables in \eqref{eq:bbmm7} gives us
	\es{eq:bbmm8}{\mathcal{I}[f] =  \frac{\vol(S^4) \vol(S^3)}{16} \int  \frac{d^4 x_2}{|x_2|^4} \frac{ d^4 x'_1 \,  dt \, t^3}{|\vec{x}'_1 - \vec{x}'_2|^8}\, \, \left.f\left(\frac{\bold{X}_i\cdot \bold{X}_j}{(\bold{X}_i\cdot \bold{Y}_2)(\bold{X}_j\cdot \bold{Y}_2)}\right) \right|_{\bold{X}_i \in \mathbb{L}_{\vec{y}_i}} \, ,}
	where $\vec{y}_1$ and $\vec{y}_2$ are related to $\vec{x}'_1$ and $\vec{x}'_2$ via \eqref{eq:varsR4toS4} and $\vec{y}_3$ and $\vec{y}_4$ are given in \eqref{eq:fixpts1}. Next, we can perform an $\SO(4)$ rotation that sends $\vec{x}'_2$ to
	\es{eq:lcset3}{ \vec{x}'_2 = \vec{x}_2^* &=  (0,0,0,1) \\}
	and sends $\bold{Y}_2$ to
	\es{eq:lcset4}{ \bold{Y}&=\left(t \frac{\vec{w}}{|w|},\frac{|w|^2-t^2-1}{2|w|},\frac{1+t^2+|w|^2}{2|w|}\right) \, ,\\}
	where $\vec{w} = (x_{2,1},x_{2,2},x_{2,3},-x_{2,4})$ is related to $\vec{x}_2$ by a parity transformation. Once again, the transformation of $\vec{x}'_1$ can be absorbed by a change of integration variables.  Following this $\SO(4)$ rotation, \eqref{eq:bbmm8} becomes
	\es{eq:bbmm9}{\mathcal{I}[f] =  \frac{\vol(S^4) \vol(S^3)}{16} \int  \frac{ d^4 w}{|w|^4}\frac{d^4 x'_1\, dt \, t^3}{|\vec{x}'_1 - \vec{x}^{\, *}_2|^8}\, \left. f\left(\frac{\bold{X}_i\cdot \bold{X}_j}{(\bold{X}_i\cdot \bold{Y})(\bold{X}_j\cdot \bold{Y})}\right)\right|_{\bold{X}_i \in \mathbb{L}_{\vec{y}_i}} \, .}
	Using the residual $\SO(3)$ symmetry to set $\vec{x}'_1$ to $\vec{x}^*_1 =(0,0,r\sin\theta, r\cos\theta)$ and changing integration variables to $\vec{z}$ and $z_0$, which are related to $\vec{w}$ and $t$ by $\vec{z}=\frac{t\vec{w}}{|w|^2}, z_0 = \frac{1}{|w|}$, we obtain the following conformally-invariant form for the integral \eqref{eq:bbmm5}:
	\es{eq:bbmm10}{\mathcal{I}[f] =  \frac{\vol(S^4) \vol(S^3) \vol(S^2)}{16} \int dr\, d\theta \, \frac{ r^3 \sin^2\theta}{u^4} \mathcal{F}(u,v)\, .}
	Here the conformally-invariant function $\mathcal{F}(u,v)$ is related to $f$ by the functional $\mathcal{J}[\ldots]$ which we define as the integral over the AdS hyperboloid:
	\es{eq:defFcal}{ \mathcal{F}(u,v)  =   \mathcal{J}[f] \equiv \int \frac{d^4 \vec{z} \, dz_0}{z^5_0} \, \left. f\left(\frac{\bold{X}_i\cdot \bold{X}_j}{(\bold{X}_i\cdot \bold{Y})(\bold{X}_j\cdot \bold{Y})}\right)\right|_{\bold{X}_i \in \mathbb{L}_{\vec{y}^*_i}} \, ,}
	where the point $\bold{Y}$ on the hyperboloid is parametrized as
	\es{}{\bold{Y}=\left(\frac{\vec{z}}{z_0},\frac{1-z^2_0-|z|^2}{2z_0},\frac{1+z^2_0+|z|^2}{2z_0}\right) \,,} 
	the kinematic configuration of the four-points on $S^4$ is
	\es{eq:conffix}{
		\begin{aligned}
			\vec{y}^{\,*}_1 &= \left(0,0,\frac{2r\sin\theta}{1+r^2},\frac{2r\cos\theta}{1+r^2},\frac{1-r^2}{1+r^2}\right)\, ,  &\vec{y}^{\,*}_2 = (0,0,0,1,0) \, , \\  
			\vec{y}^{\,*}_3 &= (0,0,0,0,-1)\, , &\vec{y}^{\,*}_4 = (0,0,0,0,1)\, ,	\end{aligned}}
	and the conformal cross-ratios $u$ and $v$ are related to $r$ and $\theta$ by $u = 1 + r^2 -2 r\cos\theta$ and $v = r^2$, using the definition \eqref{eq:uvR5} and the configuration \eqref{eq:conffix}.

	\section{Useful identities of \texorpdfstring{$\bar{D}$}{Dbar}-functions}
	\label{sec:Dfunc}

	The $n$-point $D_{r_1,\ldots,r_n}$-function in four dimensions is defined by the integral \cite{Witten:1998qj}
	\begin{equation}\label{eq:Ddefnalt}
		D_{r_1,\ldots,r_n}(x_1,\ldots,x_n) = \int \frac{dz_0\, d^4 \vec{z}}{z_0^5} \, \prod_{i=1}^n\left(\frac{z_0}{(z_0^2+(\vec{z}-\vec{x})^2)}\right)^{r_i} \, .
	\end{equation}
	The conformally-invariant $\bar{D}$-functions are related to four-point $D$-functions by
	\begin{equation}\label{eq:DtoDbar2}
		D_{r_1,r_2,r_3,r_4}(x_1,x_2,x_3,x_4) = \frac{\pi^2}{2}\frac{\Gamma\left(\bar{r} -2\right)}{\prod_{i=1}^4\Gamma\left(r_i\right)}\frac{(x^2_{14})^{\bar{r}-r_1-r_4}(x^2_{34})^{\bar{r}-r_3-r_4}}{(x^2_{13})^{\bar{r}-r_4}(x^2_{24})^{r_2}}\bar{D}_{r_1,r_2,r_3,r_4}(u,v),
	\end{equation}
	where $\bar{r} = \frac{\sum_{i=1}^4 r_i}{2}$. The simplest $\bar{D}$-function with integer coefficients, $\bar{D}_{1,1,1,1}(u,v)$, can be written in terms of the dilog function $\text{Li}$ as
	\es{eq:D1111func}{\bar{D}_{1,1,1,1}(u,v) = \frac{1}{z-\bar{z}}\left( \log(z\bar{z})\log\left(\frac{1-z}{1-\bar{z}}\right) + 2\text{Li}(z)-2\text{Li}(\bar{z}) \right) \, , } 
	where the variables $z$ and $\bar{z}$ are related to the conformal cross-ratios $u$ and $v$ via $ u = z\bar{z}$ and $v = (1-z)(1-\bar{z})$.
	
	The $\bar{D}$-functions satisfy the following differential relations \cite{Dolan:2000ut,Arutyunov:2002fh}:
	\es{eq:Drecurse}{
		\bar{D}_{r_1+1,r_2+1,r_3,r_4}&={}-\partial_u \bar{D}_{r_1,r_2,r_3,r_4}\,,\\
		\bar{D}_{r_1,r_2,r_3+1,r_4+1}&=\left(\frac{r_3+r_4-r_1-r_2}{2}-u\partial_u\right) \bar{D}_{r_1,r_2,r_3,r_4}\,,\\
		\bar{D}_{r_1,r_2+1,r_3+1,r_4}&={}-\partial_v \bar{D}_{r_1,r_2,r_3,r_4}\,,\\
		\bar{D}_{r_1+1,r_2,r_3,r_4+1}&= \left(\frac{r_1+r_4-r_2-r_3}{2}-v\partial_v\right) \bar{D}_{r_1,r_2,r_3,r_4}\,,\\
		\bar{D}_{r_1,r_2+1,r_3,r_4+1}&=\left(r_2+u\partial_u+v\partial_v\right) \bar{D}_{r_1,r_2,r_3,r_4}\,,\\
		\bar{D}_{r_1+1,r_2,r_3+1,r_4}&=\left(\frac{r_1+r_2+r_3-r_4}{2}+v\partial_v+u\partial_u\right) \bar{D}_{r_1,r_2,r_3,r_4}\,.
	}
	The $D$-functions satisfying the condition $ \max r_i < \frac{\sum_j r_j}{2}$ are called non-extremal $D$-functions, while $D$-functions saturating the inequality are called extremal $D$-functions. Extremal and non-extremal $\bar{D}$-functions are defined similarly. One can check using conformal symmetry that all non-extremal two-point $D$-functions, i.e. those with $r_1\neq r_2$, vanish.

	\subsection{Basis for non-extremal \texorpdfstring{$\bar{D}$}{Dbar}-functions}
	\label{sec:Dbarbasis}
	
	Using the argument-raising differential operators given in \eqref{eq:Drecurse} recursively and the expression for $\bar{D}_{1111}$ given in \eqref{eq:D1111func}, we can compute any non-extremal $\bar{D}$-functions with positive integer arguments. In fact, it is possible to express any such $\bar{D}$-functions as a linear combination of four chosen $\bar{D}$-functions with coefficients that are rational functions of $u$ and $v$. We demonstrate this fact in this appendix. 
	
	First, we use the first and third identities in \eqref{eq:Drecurse} to eliminate the derivatives in the other four. This leaves us with the following linear relations between $\bar{D}$-functions with different arguments:
	\es{eq:Dbarbasiseqns}{(\bar{r}-r_1-r_2) \bar{D}_{r_1,r_2,r_3,r_4}-\bar{D}_{r_1,r_2,r_3+1,r_4+1}+u \bar{D}_{r_1+1,r_2+1,r_3,r_4}&= 0,\\
		(\bar{r}-r_2-r_3) \bar{D}_{r_1,r_2,r_3,r_4}+v \bar{D}_{r_1,r_2+1,r_3+1,r_4}-\bar{D}_{r_1+1,r_2,r_3,r_4+1}&=0\,,\\
		r_2 \bar{D}_{r_1,r_2,r_3,r_4}-\bar{D}_{r_1,r_2+1,r_3,r_4+1}-v \bar{D}_{r_1,r_2+1,r_3+1,r_4}-u \bar{D}_{r_1+1,r_2+1,r_3,r_4}&=0\,,\\
		(\bar{r}-r_4)\bar{D}_{r_1,r_2,r_3,r_4}- v \bar{D}_{r_1,r_2+1,r_3+1,r_4}-\bar{D}_{r_1+1,r_2,r_3+1,r_4}-u \bar{D}_{r_1+1,r_2+1,r_3,r_4}&= 0 \,,
	}
	where $\bar{r} = \frac{\sum^4_{i=1}r_i}{2}$. We then build a set of equations by evaluating \eqref{eq:Dbarbasiseqns} at arguments $(r_1,r_2,r_3,r_4)$ satisfying the conditions $ r_i >0$, $\max r_i < \frac{\sum_{j=1}^4 r_j}{2}$, and $\sum_{i=1}^4 r_i \leq 8$. There are 26 such sets of arguments, giving a system of 104 linear equations containing 70 different $\bar{D}$\nobreakdash-functions (namely, all the non-extremal $\bar{D}_{r_1,r_2,r_3,r_4}$-functions satisfying $r_i >0$ and $\sum_{i=1}^4 r_i \leq 10$). Solving this system of equations, we find that the dimension of the vector space of this set of non-extremal $\bar{D}_{r_1,r_2,r_3,r_4}$-functions over the field of rational functions of $u$ and $v$ is 4. 
	
	We will choose our basis for this vector space to be \es{eq:Dbarbasis}{\mathcal{W}=\{\bar{D}_{1,1,1,1},\bar{D}_{1,2,2,1},\bar{D}_{1,1,2,2},\bar{D}_{2,2,2,2}\}\,.}  
	For example, in this basis, we find that $\bar{D}_{2,2,3,3}$ is given by
	\es{eq:Dbar2233}{\bar{D}_{2,2,3,3} &=  \frac{1}{(u^2 + (-1 + v)^2 - 2 u (1 + v))}\big[ \left(4 u^2-3 u (v+1)-(v-1)^2\right) \bar{D}_{2,2,2,2} \\ 
		{}&+(u-v+1) \bar{D}_{1,1,1,1}(u,v)+(-3 u+v-1) \bar{D}_{1,1,2,2}+2 (v-1) v \bar{D}_{1,2,2,1} \big] \, ,}
	and $\bar{D}_{2,3,2,1}$ is given by
	\es{eq:Dbar2321}{\bar{D}_{2,3,2,1} &= \frac{1}{2uv}\left(-\bar{D}_{1,1,1,1}+2 \bar{D}_{1,1,2,2}+2 v \bar{D}_{1,2,2,1}+(1-u-v) \bar{D}_{2,2,2,2}\right) \, .}

	We can now show inductively that any non-extremal $\bar{D}$-function with positive integer arguments can be written in this basis with coefficients that are rational functions of $u$ and $v$. Let us assume that this is true for all such $\bar{D}_{r_1,r_2,r_3,r_4}$ functions satisfying $\sum_{i=1}^4 r_i \leq \Lambda$, where $\Lambda \geq 10$. Any non-extremal $\bar{D}_{s_1,s_2,s_3,s_4}$ function satisfying $\sum_{i=1}^4 s_i =\Lambda+2$, can be written in the form 
	\es{eq:Dexp1}{ \bar{D}_{s_1,s_2,s_3,s_4}(u,v) = \mathcal{D}_{(u,v)} \bar{D}_{r'_1,r'_2,r'_3,r'_4}(u,v) \, , }
	for some $\{r'_1,r'_2,r'_3,r'_4\}$ satisfying $\sum_{i=1}^4 r'_i = \Lambda$, where $\mathcal{D}_{(u,v)}$ is one of the argument-raising differential operators appearing on the right-hand side of the relations in \eqref{eq:Drecurse}. Using our inductive hypothesis, we can write $\bar{D}_{s_1,s_2,s_3,s_4}(u,v)$ in the form
	\es{eq:Dexp2}{ \bar{D}_{s_1,s_2,s_3,s_4}(u,v) = \sum_{f\in \mathcal{W}} c_{f}^{(0,0)}(u,v) f + c_{f}^{(1,0)}(u,v)\partial_u f + c_{f}^{(0,1)}(u,v) \partial_v  f  \, , } 
	where $\mathcal{W}$ is our chosen basis for non-extremal $\bar{D}$-functions \eqref{eq:Dbarbasis}, and the $c_{f}^{(a,b)}$ functions in \eqref{eq:Dexp2} are rational functions of $u$ and $v$. Using the first and third relations in \eqref{eq:Drecurse}, one can write the $\partial_u f$ and $\partial_v f$ terms in \eqref{eq:Dexp2} in terms of other $\bar{D}_{r'_1,r'_2,r'_3,r'_4}$ functions, which also satisfy $\sum_{i=1}^4 r'_i \leq 10$. The expressions for these $\bar{D}_{r'_1,r'_2,r'_3,r'_4}$ functions in the $\bar{D}$-function basis can now be substituted in giving us the final expression for $\bar{D}_{s_1,s_2,s_3,s_4}(u,v)$ in our chosen basis. Thus any non-extremal $\bar{D}_{s_1,s_2,s_3,s_4}$ function satisfying $\sum_{i=1}^4 s_i =\Lambda+2$, can be written as a linear combination of the $\bar{D}$-functions in the basis \eqref{eq:Dbarbasis}. Our proof by induction is thus complete.
	
	We illustrate this process with an example. Using the first relation in \eqref{eq:Drecurse} and then substituting in the expression for $\bar{D}_{2,2,3,3}$ provided in \eqref{eq:Dbar2233}, we have
	\es{eq:dbar3333deriv}{ \bar{D}_{3,3,3,3}&= {} -\partial_u \bar{D}_{2,2,3,3} \\ 
		&= \frac{1}{(u^2 + (-1 + v)^2 - 2 u (1 + v))} \Bigl[ \left(-4 u^2+3 u (v+1)+(v-1)^2\right) \partial_u\bar{D}_{2,2,2,2} \\
		& {}+ (-u+v-1) \partial_u\bar{D}_{1,1,1,1}+(3 u-v+1) \partial_u\bar{D}_{1,1,2,2}-2 (v-1) v \partial_u\bar{D}_{1,2,2,1} \Bigr] \\
		& {}+  \frac{1}{(u^2 + (-1 + v)^2 - 2 u (1 + v))^2}\Bigl[\left(u^2-2 u (v-1)+v^2+2 v-3\right) \bar{D}_{1,1,1,1} \\
		& {}+ \left(-3 u^2+2 u (v-1)+v^2-6 v+5\right) \bar{D}_{1,1,2,2} -4 (v-1) v (1-u+v) \bar{D}_{1,2,2,1} \\
		& {}+ 5 \left(u^2 (v+1)-2 u (v-1)^2+(v-1)^2 (v+1)\right) \bar{D}_{2,2,2,2} \Bigr]\,.\\ }	
	Notice that we can use second relation in \eqref{eq:Drecurse} to replace $\partial_u\bar{D}_{2,2,2,2}$ with $\bar{D}_{2,2,3,3}$ and $\partial_u\bar{D}_{1,1,1,1}$ with $\bar{D}_{1,1,2,2}$. Additionally, we can use the first relation in \eqref{eq:Drecurse} to replace $\partial_u\bar{D}_{1,1,2,2}$ with $\bar{D}_{2,2,2,2}$ and $\partial_u\bar{D}_{1,2,2,1}$ with $\bar{D}_{2,3,2,1}$. Plugging in the expressions in \eqref{eq:Dbar2233} and \eqref{eq:Dbar2321}, we get the expression for $\bar{D}_{3,3,3,3} $ in the basis \eqref{eq:Dbarbasis}:
	\es{eq:dbar3333exp}{ \bar{D}_{3,3,3,3}  &= \frac{1}{(u^2 + (-1 + v)^2 - 2 u (1 + v))^2}\Bigl[ \left(5 u^2+u (4-10 v)+5 v^2+4 v-9\right) \bar{D}_{1,1,1,1} \\
		& {}+\left(13 u^3-13 u^2 (v+1)+u \left(-13 v^2+62 v-13\right)+13 (v-1)^2 (v+1)\right) \bar{D}_{2,2,2,2}  \\
		& {} -14 (u-1) (u-v+1) \bar{D}_{1,1,2,2}+14 (v-1) v (u-v-1) \bar{D}_{1,2,2,1}\Bigr] \, .} 
	
	In addition, we note that the space of functions spanned by \eqref{eq:Dbarbasis} contains the constant function, and hence it is possible to express one of the basis elements as a linear combination of the other three basis elements plus an additional rational function of $u$ and $v$. Indeed, we find
	\es{eq:dbar2222rel}{  \bar{D}_{2,2,2,2} = \frac{(1-u-v) \bar{D}_{1,1,1,1}+2 (u-1) \bar{D}_{1,1,2,2}+2 (v-1) v \bar{D}_{1,2,2,1}+2}{u^2-2 u (v+1)+(v-1)^2} \,.}
	One can check that this is true using \eqref{eq:Drecurse} to write the four $\bar{D}$-functions above as derivatives acting on $\bar{D}_{1,1,1,1}$ and then substituting in the functional form of $\bar{D}_{1,1,1,1}$ given in \eqref{eq:D1111func}. 
	
	\subsection{Identities for extremal \texorpdfstring{$\bar{D}$}{Dbar}-functions}
	
	We now move onto discussing how the identities in \eqref{eq:Dbarbasiseqns} can be used to manipulate extremal $\bar{D}$-functions. These identities imply the following relations which raise a particular argument of a $\Dbar$-function:
	\es{eq:DbarRaising}{\bar{D}_{r_1,r_2,r_3,r_4}&=\frac{u \bar{D}_{r_1+1,r_2+1,r_3,r_4}+\bar{D}_{r_1+1,r_2,r_3,r_4+1}+\bar{D}_{r_1+1,r_2,r_3+1,r_4}}{r_1}\,,\\
		\bar{D}_{r_1,r_2,r_3,r_4}&=\frac{u \bar{D}_{r_1+1,r_2+1,r_3,r_4}+v \bar{D}_{r_1,r_2+1,r_3+1,r_4}+\bar{D}_{r_1,r_2+1,r_3,r_4+1}}{r_2}\,,\\
		\bar{D}_{r_1,r_2,r_3,r_4}&=\frac{v \bar{D}_{r_1,r_2+1,r_3+1,r_4}+\bar{D}_{r_1,r_2,r_3+1,r_4+1}+\bar{D}_{r_1+1,r_2,r_3+1,r_4}}{r_3}\,,\\
		\bar{D}_{r_1,r_2,r_3,r_4}&=\frac{\bar{D}_{r_1,r_2,r_3+1,r_4+1}+\bar{D}_{r_1,r_2+1,r_3,r_4+1}+\bar{D}_{r_1+1,r_2,r_3,r_4+1}}{r_4}\,.
	}
	In the simplification discussed in Appendix~\ref{sec:extrmsimp}, these are used when any argument in a $\Dbar$\nobreakdash-function goes to zero in the $\epsilon\to 0 $ limit.
	
	The identities in \eqref{eq:Dbarbasiseqns} can also be manipulated to give the following identities in which a particular argument is held fixed and the others are raised:
	\es{eq:DbarRaisingOthers}{\bar{D}_{r_1,r_2,r_3,r_4}&=\frac{2 \left(v \bar{D}_{r_1,r_2+1,r_3+1,r_4}+\bar{D}_{r_1,r_2,r_3+1,r_4+1}+\bar{D}_{r_1,r_2+1,r_3,r_4+1}\right)}{-r_1+r_2+r_3+r_4}\,, \\
		\bar{D}_{r_1,r_2,r_3,r_4}&=\frac{2 \left(\bar{D}_{r_1,r_2,r_3+1,r_4+1}+\bar{D}_{r_1+1,r_2,r_3,r_4+1}+\bar{D}_{r_1+1,r_2,r_3+1,r_4}\right)}{r_1-r_2+r_3+r_4}\, ,\\
		\bar{D}_{r_1,r_2,r_3,r_4}&= \frac{2 \left(u \bar{D}_{r_1+1,r_2+1,r_3,r_4}+\bar{D}_{r_1,r_2+1,r_3,r_4+1}+\bar{D}_{r_1+1,r_2,r_3,r_4+1}\right)}{r_1+r_2-r_3+r_4}\,,\\
		\bar{D}_{r_1,r_2,r_3,r_4}&= \frac{2 \left(u \bar{D}_{r_1+1,r_2+1,r_3,r_4}+v \bar{D}_{r_1,r_2+1,r_3+1,r_4}+\bar{D}_{r_1+1,r_2,r_3+1,r_4}\right)}{r_1+r_2+r_3-r_4}\,.
	}
	The four identities given above are used in in Appendix~\ref{sec:extrmsimp} when a particular  $\bar{D}$-function becomes extremal in the $\epsilon \to 0$ limit.
	
	\subsection{Basis for \texorpdfstring{$\epsilon$}{ϵ}-deformed \texorpdfstring{$\bar{D}$}{Dbar}-functions}
	\label{sec:Dbarbasiseps}
	In the simplification in Section~\ref{sec:extrmsimp}, we use  \eqref{eq:DbarRaising} and \eqref{eq:DbarRaisingOthers} to ensure that we are working with $\Dbar_{r_1+\epsilon,r_2+\epsilon,r_3+\epsilon,r_4+\epsilon}$-functions satisfying $ r_i>0$ and $\max r_i < \frac{\sum^4_{j=1}r_j}{2}$ for $1\leq i \leq 4$. In the $\epsilon \to 0$ limit, these $\bar{D}$-functions are non-extremal with positive integer arguments, and hence finite. As was the case for $\bar{D}$-functions without any $\epsilon$-dependence, the identities in \eqref{eq:Dbarbasiseqns} ensure that the vector space of all such $\bar{D}$-functions over the field of rational functions of $u$ and $v$ is spanned by a basis of four $\bar{D}$-functions. The proof of this statement follows the same logic as the discussion in Appendix~\ref{sec:Dbarbasis}.
	
	For our simplification, we choose the basis of $\Dbar$-functions to be 
	\begin{equation} \label{eq:Dbarbasiseps}
		\mathcal{W}_{\epsilon} = \{\Dbar_{1+\epsilon,1+\epsilon,1+\epsilon,1+\epsilon}, \Dbar_{1+\epsilon,2+\epsilon,2+\epsilon,1+\epsilon}, \Dbar_{1+\epsilon,1+\epsilon,2+\epsilon,2+\epsilon}, \Dbar_{2+\epsilon,2+\epsilon,2+\epsilon,2+\epsilon}\}.
	\end{equation}
	We can write other $\epsilon$-deformed $\bar{D}$-functions in this basis; for instance,
	\es{eq:dbar2233eqneps}{    \Dbar_{2+\epsilon,2+\epsilon,3+\epsilon,3+\epsilon} &=\frac{1}{(u^2 + (-1 + v)^2 - 2 u (1 + v))}\left[(1 + \epsilon)^3 (u-v+1) \bar{D}_{1+\epsilon ,1+\epsilon ,1+\epsilon,1+\epsilon} \right. \\
		&\left.{}-(1 + \epsilon)^2 (3 u-v+1) \bar{D}_{1+\epsilon,1+\epsilon,2+\epsilon ,2+\epsilon}+2 (v-1) v (1 + \epsilon)^2 \bar{D}_{1+\epsilon , 2+ \epsilon , 2+\epsilon, 1+\epsilon }\right.\\
		&\left.{}+\left(u^2 (4 + 3\epsilon)-u (v+1) (3 + 2  \epsilon)-(v-1)^2 (1 + \epsilon)\right) \bar{D}_{2+\epsilon,2+\epsilon,2+\epsilon,2+\epsilon }\right] .}

	\section{Computing \texorpdfstring{$\mathcal{F}_{b^2m^2}(u,v)$}{F\_b²m²(u,v)}}
	\label{sec:FAdSsimp}
	
	The $\bbmmZT$ integrated correlator is given as an integral of the conformally-invariant function $\mathcal{F}_{b^2m^2}(u,v)$ in \eqref{eq:bbmm5}. The function $\mathcal{F}_{b^2m^2}(u,v)$, itself defined as an integral in \eqref{eq:defFcalbbmm}, can be written as a sum of two-point, three-point and four-point $D$-functions  using \eqref{eq:Ddefnalt}. However, applying \eqref{eq:Ddefnalt} directly to \eqref{eq:defFcalbbmm} would result in an expression involving divergent terms. These divergences come from several extremal three-point $D$-functions, such as $D_{4,2,2}(x_1,x_2,x_3)$, as well as the extremal two-point $D$-function $D_{4,4}(x_1,x_2)$ and its crossed versions.\footnote{As described in Appendix~\ref{sec:Dfunc}, non-extremal $D_{r_1,\ldots, r_n}$-functions satisfy the condition $2 \max r_i < \sum_j r_j$, while extremal $D$-functions saturate the inequality. The same condition holds for extremal and non-extremal $\bar{D}$-functions.}

	In order to understand the nature of this divergence, we separate out $\mathcal{F}_{b^2m^2}(u,v)$ into two parts,
	\es{eq:Fcalbbmmsplit}{ \mathcal{F}_{b^2m^2}(u,v)  = \mathcal{F}^{\text{conv}}_{b^2m^2}(u,v) +  \mathcal{F}^{\text{div}}_{b^2m^2}(u,v) \, ,}
	where $\mathcal{F}_{b^2m^2}^{\text{div}}(u,v)$ consists of the divergent $D$-functions that would arise from a naive application of \eqref{eq:Ddefnalt} to \eqref{eq:defFcalbbmm}, while $\mathcal{F}_{b^2m^2}^{\text{conv}}(u,v)$ contains the remaining finite terms.
	We can similarly split $F_{b^2m^2}$ and $N_{b^2m^2}(y^2_{ij})$, defined in \eqref{eq:defFbbmm} and \eqref{eq:defNbbmm} respectively, as 
	\es{eq:NFbbmmsplit}{ F_{b^2m^2}  = F^{\text{conv}}_{b^2m^2} +  F^{\text{div}}_{b^2m^2} \, , \qquad N_{b^2m^2} = N^{\text{conv}}_{b^2m^2} + N^{\text{div}}_{b^2m^2} \, , }
	such that they reproduce $\mathcal{F}_{b^2m^2}^{\text{div}}(u,v) $ and $\mathcal{F}_{b^2m^2}^{\text{conv}}(u,v)$ using the substitution in \eqref{eq:defFbbmm} and the integral in \eqref{eq:defFcalbbmm} respectively. The function $N_{b^2m^2}^{\text{div}}$ is given explicitly by the expression
	\es{eq:Nbbmmdiv}{N^{ \text{div}}_{b^2m^2}= \frac{c^2}{320\pi^8}\bigg[&	y^4_{12}\left(3y^4_{12} - 4y^2_{12} y^2_{13} -2y^4_{13}\right) \\
		&+ (\text{23 other  permutations of $\{y_1,y_2,y_3,y_4\}$})\bigg]\, .}
	We shall first discuss how the finite expression $\mathcal{F}_{b^2m^2}^{\text{conv}}(u,v)$ can be simplified before we discuss the apparently divergent expression $F_{b^2m^2}^{\text{div}}(u,v)$.

	\subsection{Simplifying \texorpdfstring{$\mathcal{F}_{b^2m^2}^{\mathrm{conv}}(u,v)$}{the convergent part}}
	\label{sec:nonextrmsimp}
	
	The function $\mathcal{F}_{b^2m^2}^{\text{conv}}(u,v)$, identified as the convergent part of $\mathcal{F}_{b^2m^2}(u,v)$ in \eqref{eq:Fcalbbmmsplit}, is a sum of finite three-point and four-point $D$-functions. Three-point $D$-functions are rational functions of the coordinates of the points, 
	\es{eq:3ptDfunc}{
		D_{r_1,r_2,r_3}(x_1,x_2,x_3) &= \frac{\pi^2 \prod_{i=1}^3\Gamma\left(\bar{r}-r_i\right)}{2 \prod_{i=1}^3\Gamma(r_i)}\frac{ \Gamma\left(\bar{r}-2\right) }{ (x^2_{23})^{\bar{r}-r_1}(x^2_{13})^{\bar{r}-r_2}(x^2_{12})^{\bar{r}-r_3}} \, ,
	}
	where $\bar{r}=\frac{\sum_{i=1}^3 r_i}{2} $. The four-point $D$-functions can be written in terms of conformally-invariant $\bar{D}$-functions, defined in \eqref{eq:DtoDbar2}. Using \eqref{eq:3ptDfunc} and \eqref{eq:DtoDbar2}, we can write $F_{b^2m^2}^{\text{conv}}$ as a sum of $\bar{D}$-functions along with a rational function of $u$ and $v$. By construction, all $\bar{D}$-functions appearing in $F_{b^2m^2}^{\text{conv}}$ are non-extremal and have positive integer arguments. As we explain in Appendix~\ref{sec:Dbarbasis}, any non-extremal $\bar{D}$-function with positive integer arguments can be written as a linear combination of the four $\bar{D}$-functions:  $\bar{D}_{1,1,1,1}$, $\bar{D}_{1,2,2,1}$, $\bar{D}_{1,1,2,2}$, and $\bar{D}_{2,2,2,2}$, with coefficients that are rational functions of $u$ and $v$. Substituting in the expressions for various $\bar{D}$-functions, $\mathcal{F}_{b^2m^2}^{\text{conv}}(u,v)$ becomes
	\es{eq:exprFbbmmnondiv1}{\mathcal{F}_{b^2m^2}^{\text{conv}}(u,v) &= \frac{4 c^2 u^2}{5 \pi ^6} \Bigl( (-9 u-9 v+29) \bar{D}_{1,1,1,1}+38 (u-1) \bar{D}_{1,1,2,2}\\
		& {}+38 (v-1) v \bar{D}_{1,2,2,1} -19 \left(u^2-2 u (v+1)+(v-1)^2\right) \bar{D}_{2,2,2,2} \Bigr) \, .}
	Furthermore, using the relation \eqref{eq:dbar2222rel}, $\mathcal{F}_{b^2m^2}^{\text{conv}}$ reduces to the simple expression
	\es{eq:exprFbbmmnondiv2}{\mathcal{F}_{b^2m^2}^{\text{conv}}(u,v) &= \frac{8c^2 u^2}{5 \pi ^6} \left(5 (u+v+1) \bar{D}_{1,1,1,1}(u,v)-19\right) \, .}
	We now discuss how the divergent $\mathcal{F}_{b^2m^2}^{\text{div}}(u,v)$ term in  \eqref{eq:Fcalbbmmsplit} can be computed and simplified.

	\subsection{Simplifying \texorpdfstring{$\mathcal{F}_{b^2m^2}^{\mathrm{div}}(u,v)$}{the divergent part}}
	\label{sec:extrmsimp}
	
	As we discussed at the beginning of this appendix, if we attempt to express $\mathcal{F}_{b^2m^2}^{\text{div}}(u,v)$ in terms of $D$-functions we will find an expression with divergent extremal two-point and three-point $D$-functions. These divergent terms show up only because we are separating the integral in \eqref{eq:defFcalbbmm} into a sum of several integrals, some of which are individually divergent when the integral over AdS space is computed. One can see from the definition of two\nobreakdash-point and three-point $D$-functions that these divergences arise from the region of the AdS boundary in the direction of the point associated with the largest argument of the divergent $D$-function being considered. For example, for the 3-point $D$-function $D_{2,2,4}(x_1,x_2,x_4)$, using the definition \eqref{eq:Ddefnalt}, we have a divergent contribution from the region with $\vec{z} = \vec{x}_4 + \vec{\zeta}$,  $|\zeta|^2 < \epsilon^2$, and $z_0 <\epsilon$:
	\es{eq:div3ptp1}{&  D_{2,2,4}(x_1,x_2,x_4) \\ 
		= {}&   \int \frac{d^4\vec{z}\, dz_0}{z^5_0} \left( \frac{z_0}{z^2_0+(\vecx_1-\vec{z})^2}\right)^2 \left( \frac{z_0}{z^2_0+(\vecx_2-\vec{z})^2}\right)^2 \left( \frac{z_0}{z^2_0+(\vecx_4-\vec{z})^2}\right)^4 \\ 
		={} & \frac{1}{x^4_{14} x^4_{24}}\int_{_{\substack{|\zeta|^2<\epsilon^2 \\ z_0<\epsilon}}}\, d^4\vec{\zeta}\, dz_0  \frac{z^3_0}{(z^2_0+|\zeta|^2)^4} + \text{finite contributions}\, .\\}
	A careful analysis of the integrand $F_{b^2m^2}$ in \eqref{eq:defFcalbbmm} near the AdS boundary (in the small $z_0$ limit) in the direction of the lightrays $\mathbb{L}_{\vec{y}^*_i}$, where the vectors $\vec{y}^*_i$ are given in \eqref{eq:conffix}, reveals that $\mathcal{F}_{b^2m^2}(u,v)$ is finite. We illustrate how to investigate such apparent divergences using a simpler example. Consider the linear combination of extremal $D$-functions,  $2D_{2,2,4}(x_1,x_3,x_4)x^4_{14}x^4_{34} - D_{1,3,4}(x_1,x_3,x_4) x^2_{14}x^6_{34} - D_{1,3,4}(x_2,x_3,x_4)x^2_{24}x^6_{34}$, which might arise from a similarly naive application of \eqref{eq:Ddefnalt} to some integral. Each $D$-function in this expression receives divergent contributions from the region with $\vec{z} = \vec{x}_4 + \vec{\zeta}$, $|\zeta|^2 < \epsilon^2$, and $z_0 <\epsilon$, of the integral in \eqref{eq:Ddefnalt}. We can separate out the divergent part of the integral, as was done in \eqref{eq:div3ptp1} above, and check that
	\es{eq:sample3ptD}{&2D_{2,2,4}(x_1,x_3,x_4)x^4_{14}x^4_{34} - D_{1,3,4}(x_1,x_3,x_4) x^2_{14}x^6_{34} - D_{1,3,4}(x_2,x_3,x_4)x^2_{24}x^6_{34}\\
		={}& (2-1-1)\int_{_{\substack{|\zeta|^2<\epsilon^2 \\ z_0<\epsilon}}}\, d^4\vec{\zeta}\, dz_0  \frac{z^3_0}{(z^2_0+|\zeta|^2)^4} + \text{finite contributions}\,;\\ }
	that is, the divergent parts from each $D$-function cancel each other out and the full expression is finite.
	
	In order to regulate such apparent divergences in \eqref{eq:defFcalbbmm}, we introduce an $\epsilon$-dependent regulator to define the $\epsilon$-deformed conformally-invariant function $\mathcal{F}_{b^2m^2 }^{\text{div},\epsilon}(u,v) $ as 
	\es{eq:defFcalbbmmdiveps}{ \mathcal{F}_{b^2m^2}^{\text{div},\epsilon}(u,v) = \mathcal{J}[F_{b^2m^2}^{\text{div},\epsilon}] \, , }
	where the functional $\mathcal{J}[\ldots]$ is defined in \eqref{eq:defFcal} and the $\epsilon$-deformed function $F_{b^2m^2}^{\text{div},\epsilon}$ is given by
	\es{eq:defFbbmmdiveps}{F_{b^2m^2}^{\text{div},\epsilon}\left(\frac{\bold{X}_i\cdot \bold{X}_j}{(\bold{X}_i\cdot \bold{Y})(\bold{X}_j\cdot \bold{Y})}\right) &=  \left. (y^2_{12}y^2_{34})^\epsilon\frac{y^4_{12}y^4_{34}}{y^4_{13}y^4_{24}}N_{b^2m^2}^{\text{div}}(y^2_{ij})\right|_{y^2_{ij}\to \frac{-2 \bold{X}_i\cdot \bold{X}_j}{\bold{X}_i\cdot\bold{Y}\, \bold{X}_j\cdot\bold{Y}}} \, ,}
	with $\bold{Y}=\left(\frac{\vec{z}}{z_0},\frac{1-z^2_0-|z|^2}{2z_0},\frac{1+z^2_0+|z|^2}{2z_0}\right) $ and $\bold{X}_i \in \mathbb{L}_{\vec{y}^{*}_i}$, for the $\vec{y}^{*}_i$ vectors given in \eqref{eq:conffix}. 
	
	Following this $\epsilon$-deformation and using \eqref{eq:Ddefnalt}, $\mathcal{F}_{b^2m^2}^{\text{div},\epsilon}$ can be written as a sum of various four-point $D$-functions. For $\epsilon>0$, these $D$-functions are not extremal. Taking the $\epsilon \to 0$ limit naively would reproduce the same extremal two-point and three-point $D$-functions discussed earlier. Instead, we will use \eqref{eq:DtoDbar2} to rewrite $\mathcal{F}_{b^2m^2}^{\text{div},\epsilon}$ in terms of $\bar{D}_{r_1 + \epsilon,r_2+\epsilon,r_3+\epsilon,r_4+\epsilon}$-functions satisfying $r_i \geq 0$ and $\max r_i = \frac{\sum_{k=1}^4 r_k}{2}$ for $1\leq i \leq 4$. By construction, all of these $\bar{D}$-functions are divergent in the $\epsilon\to 0$ limit. We will then use the $\bar{D}$-identities discussed in Appendix~\ref{sec:Dfunc} to write $\mathcal{F}_{b^2m^2}^{\text{div},\epsilon}$ as a linear combination of a basis of $\bar{D}$-functions that are finite in the $\epsilon\to 0$ limit so that the $\epsilon\to 0$ limit can be taken smoothly.
	
	Using the four argument-raising identities in \eqref{eq:DbarRaising}, we can express $\mathcal{F}_{b^2m^2}^{\text{div},\epsilon}$ as a sum of $\bar{D}_{r_1 + \epsilon,r_2+\epsilon,r_3+\epsilon,r_4+\epsilon}$-functions with $r_i >0$. For example, using the first identity in \eqref{eq:DbarRaising}, we can rewrite $\bar{D}_{\epsilon ,1+\epsilon ,3+\epsilon, 4+ \epsilon } $ as the sum
	\es{eq:fb2m2divexp1}{\bar{D}_{\epsilon ,1+\epsilon ,3+\epsilon, 4+ \epsilon } = \frac{\bar{D}_{1 + \epsilon,1 + \epsilon,3 + \epsilon,5+ \epsilon}+\bar{D}_{1 + \epsilon,1 + \epsilon,4+ \epsilon,4+ \epsilon}+u \bar{D}_{1 + \epsilon,2 + \epsilon,3 + \epsilon,4+ \epsilon}}{\epsilon } \, .}
	The next step is to use another set of argument-raising identities in \eqref{eq:DbarRaisingOthers} to express $\mathcal{F}_{b^2m^2}^{\text{div},\epsilon}$ in terms of $\bar{D}_{r_1 + \epsilon,r_2+\epsilon,r_3+\epsilon,r_4+\epsilon}$-functions satisfying  $r_i > 0$ and $\max r_i < \frac{\sum_{k=1}^4 r_k}{2}$. For example, using the fourth identity in \eqref{eq:DbarRaisingOthers}, $\bar{D}_{1 + \epsilon,1 + \epsilon,3 + \epsilon,5+ \epsilon}$ in \eqref{eq:fb2m2divexp1} can be changed to
	\es{eq:fb2m2divexp2}{\bar{D}_{1 + \epsilon,1 + \epsilon,3 + \epsilon,5+ \epsilon} =  \frac{v \bar{D}_{1+ \epsilon,2+ \epsilon,4+ \epsilon,5+ \epsilon}+\bar{D}_{2+ \epsilon,1+ \epsilon,4+ \epsilon,5+ \epsilon} + u \bar{D}_{2+ \epsilon,2+ \epsilon,3+ \epsilon,5+ \epsilon}}{\epsilon } \, .}
	Following these two steps, we have an expression for $\mathcal{F}_{b^2m^2}^{\text{div},\epsilon}$ as a sum of $\bar{D}_{r_1 + \epsilon,r_2+\epsilon,r_3+\epsilon,r_4+\epsilon}$ that are finite in the $\epsilon \to 0$ limit. In Appendix~\ref{sec:Dbarbasiseps}, we discuss how all such $\bar{D}$-functions can be written as a linear combination of the basis of four $\bar{D}$-functions given in \eqref{eq:Dbarbasiseps}. Following this procedure, $\mathcal{F}_{b^2m^2 }^{\text{div},\epsilon}$ can be written in terms of these basis $\bar{D}$-functions. The $\bar{D}$-functions in this chosen basis are finite in the $\epsilon \to 0$ limit and hence we can now take the $\epsilon \to 0$ limit easily. The result is
	\es{eq:exprFbbmmdiv1}{\mathcal{F}_{b^2m^2}^{\text{div}} &=  \frac{4c^2 u^2}{5 \pi ^6} \bigl[-\left(u^2-2 u (v+1)+(v-1)^2\right) \bar{D}_{2,2,2,2}-(u+v-1) \bar{D}_{1,1,1,1}\\
		{}&+2 (u-1) \bar{D}_{1,1,2,2}+2 (v-1) v \bar{D}_{1,2,2,1}\bigr]\, .}
	Using the identity \eqref{eq:dbar2222rel}, this simplifies to
	\es{eq:exprFbbmmdiv2}{\mathcal{F}_{b^2m^2}^{\text{div}} = -\frac{8c^2u^2}{5 \pi ^6}\, .}
	Combining the two components $\mathcal{F}_{b^2m^2}^{\text{conv}}$ and  $\mathcal{F}_{b^2m^2}^{\text{div}}$, given in \eqref{eq:exprFbbmmnondiv2} and \eqref{eq:exprFbbmmdiv2} respectively, we have
	\es{eq:Fbbmmfinal}{\mathcal{F}_{b^2m^2} = \frac{8c^2 u^2}{\pi^6} \left((u+v+1) \bar{D}_{1,1,1,1}-4\right) \, .}

	\bibliographystyle{ssg}
	\bibliography{squash}

\end{document}